%% file: MS_arXiv.tex
\documentclass[journal,onecolumn]{IEEEtran}
\input{preambleT.tex}
\usepackage[export]{adjustbox}
\usepackage{threeparttable}
\usepackage{multirow}

\begin{document}

	\title{Control-Theoretic Modeling of Multi-Species Water Quality Dynamics in Drinking Water Networks: \\ Survey, Methods, and Test Cases}

	\author{Salma M. Elsheri$\text{f}^{\dagger, \P}$, Shen Wan$\text{g}^{\ast}$, Ahmad F. Tah$\text{a}^{\dagger, \ast \ast}$, Lina Sel$\text{a}^{\spadesuit}$, Marcio H. Giacomon$\text{i}^{\ddagger}$, and Ahmed Abokif$\text{a}^{\diamond}$
	
		\thanks{$^\dagger$Department of Civil and Environmental Engineering, Vanderbilt University, Nashville, TN, USA. Emails: salma.m.elsherif@vanderbilt.edu, ahmad.taha@vanderbilt.edu}
		\thanks{$\P$Secondary: Department of Irrigation and Hydraulics Engineering, Faculty of Engineering, Cairo University, Giza, Egypt.}
		\thanks{$^\ast$School of Cyberspace Security, Beijing University of Posts and Telecommunications. Email: shen.wang@bupt.edu.cn}
		\thanks{$^{\spadesuit}$Department of Civil, Architecture, and Environmental Engineering, The University of Texas at Austin. Email: linasela@utexas.edu}
		\thanks{$^\ddagger$School of Civil \& Environmental Engineering, and Construction Management, The University of Texas at San Antonio. Email: marcio.giacomoni@utsa.edu}
		\thanks{$^{\diamond}$Department of Civil, Materials, and Environmental Engineering, The University of Illinois Chicago. Email: abokifa@uic.edu}
		\thanks{$^{\ast \ast}$Corresponding author. }
		\thanks{This work is partially supported by National Science Foundation under grants 1728629, 2015603, 2015671, 2151392, and 2015658.}}
	
	\maketitle
	
	\begin{abstract} 
		\normalsize	 Chlorine is a widely used disinfectant and proxy for water quality (WQ) monitoring in water distribution networks (WDN). Chlorine-based WQ regulation and control aims to maintain pathogen-free water. Chlorine residual evolution within WDN is commonly modeled using the typical single-species decay and reaction dynamics that account for network-wide, spatiotemporal chlorine concentrations only. Prior studies have proposed more advanced and accurate descriptions via multi-species dynamics. This paper presents a host of novel state-space, control-theoretic representations of multi-species water quality dynamics. These representations describe decay, reaction, and transport of chlorine and a fictitious reactive substance to reflect realistic complex scenarios in WDN. Such dynamics are simulated over space- and time-discretized grids of the transport partial differential equation and the nonlinear reaction ordinary differential equation. To that end, this paper \textit{(i)} provides a full description on how to formulate a high fidelity model-driven state-space representation of the multi-species water quality dynamics and \textit{(ii)} investigates the applicability and performance of different Eulerian-based schemes (Lax-Wendroff, backward Euler, and Crank-Nicolson) and Lagrangian-based schemes (method of characteristics) in contrast with EPANET and its EPANET-MSX extension. Numerical case studies reveal that the Lax-Wendroff scheme and method of characteristics outperform other schemes with reliable results under reasonable assumptions and limitations. 
	\end{abstract}


	\section{\large Introduction and Literature Review}~\label{sec:Into-Lit}
	\IEEEPARstart{A}{s} water travels through water distribution networks (WDN), water quality deteriorates due to the physical-chemical processes (e.g., growing water age and microbial load increasing), malicious attacks, natural hazards, or lack of maintenance leading to potential contaminant intrusion through leaks. To that end, disinfection has been introduced as an important treatment process for drinking water to ensure meeting the public health requirements (e.g., waterborne disease prevention). To point out the great impact of disinfection, studies \cite{OhanianHealth, richardson2007occurrence} state that the occurrence of multiple waterborne diseases (e.g., cholera, typhoid) in the U.S. has decreased remarkably since its application at the start of the 20\textsuperscript{th} century \cite{constable2003century}. The National Academy of Engineering (NAE) listed \textit{Safe and Abundant Water} as one of the greatest engineering achievements. Yet, the disinfection process is not straightforward as there are challenges to overcome. The objective is to have a sufficient disinfectant residual all over the network to maintain pathogen-free water while limiting the formation of the health-threatening disinfection by-products (DBPs), while not causing undesirable odor, taste, or color \cite{demarini2020review, ngwenya2013recent}.

	There are different disinfection methods which are listed and compared in \cite{tsitsifli2018disinfection} according to their composition (chemical/non-chemical), effectiveness, and limitations. Among these disinfectants, chlorine is the most widely used one in drinking water distribution networks. Modeling the fate of chlorine concentrations within the WDN has been an effective proxy for monitoring real-time water quality. Consequently, various studies have focused on modeling chlorine as it leaves the treatment plant, and travels through the network, until it reaches water users \cite{grayman2018history}.
	
	To model chlorine evolution in WDN, a full understanding of its behavior is required. As chlorine travels within the water, it decays and reacts with other substances through the processes of Reaction-Transport-Diffusion. However, chlorine evolution is widely modeled by the one-dimension advection-reaction (1-D AR) partial differential equation (PDE). Several numerical methods are applied to solve the PDE, that are Eulerian, Lagrangian, and hybrid Eulerian-Lagrangian based schemes  \cite{rossman1996numerical,rossman1993discrete,liou1987modeling,basha2007eulerian}, which discretize the PDE in both time and space over a specific domain. The main distinction between these approaches is the formulation of either fixed or variable-sized segments in the pipes to form the numerical grid over which the equations are solved. Additionally, chlorine decay and reaction processes are characterized by nonlinear functions, which include interactions between multiple species. Yet, due to the modeling complexity the most frequently used model is the single-species first-order equations \cite{jonkergouw2009variable}. This simplified model depicts the chlorine decay and reaction as a function of the chlorine concentration with a constant decay rate without accounting for reactions with other species. Whilst, the decay and reaction models in realistic WDN are much more complicated and the simplified single-species model fails to capture the actual performance, especially under high variability in hydraulics and water quality states. On the other hand, a more advanced and complex description is provided by the multi-species second-order model \cite{jadas1992chlorine,clark1998chlorine, boccelli2003reactive}, which produces a more accurate description. The multi-species model can be used to model chlorine interactions with other species (e.g., microbial, chemical) in the bulk flow, attached to the pipe walls, or contamination events \cite{monteiro2014modeling}. The simulation can be limited to consider one specific substance (in addition to organic matter) or it can be upgraded to include more substances that consume chlorine interdependently or in form of a fictitious reactant  \cite{jadas1992chlorine, jonkergouw2009variable}.
	
	Building reasonably good models to predict the evolution of water quality states (e.g., concentrations of chemicals) in WDN is the key to perform model-driven feedback control---a mainstream control approach in various infrastructure. The aim is to control and maintain the desired chlorine residual that meets water quality standards. This has been expressed as an interesting research and applied problem in several studies \cite{ohar2014optimal, ostfeld2006conjunctive}: 
	\begin{quote}
		\textit{Is it more effective to apply primary chlorination with a high dose at the treatment plant, the very start of the water distribution networks, or distribute the injection locations over the network with smaller doses? Would the answer to this previous question change under abnormal conditions (e.g., contamination intrusion)?}
	\end{quote}
	Yet, to the best of our knowledge, there are virtually no model-, system-, and network-theoretic studies on the control of water quality in WDN in the existence of other reactive substances (i.e., multi-species dynamics), let alone, to solve the aforementioned research problem. Therefore, this paper's main objective is to investigate and build a model that traces chlorine concentrations in presence of such events. This model is built to be easily integrated in model-based control frameworks (e.g., model-predictive control, state-feedback control, output-feedback control, Lyapunov-based control) while integrating measurement models from WQ sensors. The proposed models can also be used to study control-theoretic and optimal WQ sensors and booster station placement formulations.  
	
	\noindent \textbf{Paper's objectives.} this paper's objectives are \textit{(i)} to formulate high fidelity water quality model depicting multi-species reaction dynamics of chlorine in the entire network, and \textit{(ii)} to investigate the performance, scalability, and complexity of various discretization-based techniques to simulate the multi-species water quality dynamics. This paper's results pave the way to build novel control algorithms of multi-species dynamics---the algorithms are outside the scope of this paper. Henceforward, we survey the literature to highlight the gap bridged by this paper.  
	
	\subsection{Literature review}
	
	The literature review covers two main aspects of water quality dynamics that are chlorine transport modeling and decay and reaction modeling. Study \cite{boccelli2003reactive} states that the system composition and how diverse and case-oriented the reactants with which chlorine reacts establish the existence of huge uncertainty and assumptions. Thus, simulating chlorine transport and reaction has been covered by several studies with different approaches with a wide range of applicability and limitation domains.  
	
	Chemicals transport model is derived using the advection-reaction equation in several studies \cite{boulos1994event,blokker2008importance,rossman1993discrete}. Study \cite{rossman1996numerical} applies different numerical schemes to discretize the advection-reaction equation. Two of these schemes are Eulerian-based: finite-difference and discrete-volume methods, and the other two are Lagrangian-based: time-driven and event-driven methods. Eulerian methods solve the equation on a fixed grid system while Lagrangian methods employ a moving coordinate system using particle tracking algorithms. Each scheme's stability is sensitive to one parameter or another including water quality time-step, segments sizes, and allowed simulation tolerance. Studies \cite{tzatchkov2002advection, basha2007eulerian} propose Lagrangian based-solution to the advection and reaction using the explicit method of characteristics to solve the advection and reaction terms. Study \cite{munavalli2005multi} uses a multi-step Eulerian scheme that solves the transport equations by an explicit Eulerian scheme, the MacCormack scheme, while solving the kinetic dynamics using the explicit Runge-Kutta method. On the other hand, the latest version of EPANET, a standard software widely used to model hydraulics and water quality of water distribution systems \cite{rossman2020epanet}, applies the Lagrangian time-driven method where it works within fixed time intervals and variable-sized segments. 
	
	Moreover, the effect of dispersion in chemicals transport has been investigated in several studies \cite{tzatchkov2002advection, abokifa2020investigating, shang2021lagrangian}. Results have shown that dispersion dominates over advection in dead-end branches where the flow velocities are low and approaching a laminar state. That is, it is an acceptable assumption to rely on the AR dynamics neglecting the dispersion term in networks with limited dead-end branches, higher velocities, and changing demands leading to a turbulent flow state.
	
	In addition to the transport modeling, decay and reaction modeling plays an important role in water quality dynamics. The most widely used model for chlorine decay is the first-order decay model which is based on the assumption that chlorine reacts at a constant rate with excess chlorine-demanding reactants \cite{vasconcelos1996characterization, jonkergouw2009variable}. Moreover, this first-order decay model has been used in  EPANET for water quality modeling in water networks \cite{rossman2020epanet}. However, the study \cite{fisher2011evaluation} verifies that the first-order decay model that accounts for only chlorine is not accurate under high variability in concentrations and rechlorination scenarios. Not only the reaction dynamics are affected by the chlorine-consuming reactant concentration in many cases, but also depend on the contact time needed for the reaction to be completed. Thus, other studies  \cite{fisher2012suitable,wang2019quantifying, fisher2017comprehensive} categorize the reactants to slow and fast dynamics and accordingly model the chlorine decay with two different second-order reaction models simultaneously. These models outperform the first-order model in the rechlorination scenarios. Studies \cite{jadas1992chlorine, jonkergouw2009variable} introduce a multi-species model of chlorine with every reactive component that presents in the water modeled with its specific reaction rate.  Yet, this model requires pre-categorization for each substance according to their reaction rates and to know their initial concentrations at all the network components \cite{monteiro2014modeling}.

	Furthermore, in \cite{munavalli2004dynamic}, a multicomponent (organic carbon, biomass, and chlorine) reaction transport model is developed. Although this model does not include the reaction dynamics with the microbial contaminant itself, it is proposed to be an alternative approach by tracking the organic carbon and bacterial growth. Yet, it has various parameters to be predetermined/measured through a complex approach. However, this study uses one explicit Eulerian discretization scheme and the time-driven Lagrangian method with no reference to compare to and with no brief description of the implementation of the method on different networks. In addition, the authors in \cite{clark1998chlorine, boccelli2003reactive} investigate the performance of a reaction model between chlorine and fictitious reactive species in a form of a second-order model to test its advantages over the typically used first-order model. In comparison, their models show a more accurate description, particularly for scenarios with chlorine injected into the network by chlorine booster stations. Note that both decay and reaction of chlorine are compromised in this fictitious reactant. That is, the parameters of their models have to be predetermined for the specific water characteristics of the network to reflect organic matter and substances reacting with chlorine. Yet, the inclusion of this fictitious reactant in the chlorine reaction and decay can play an important role to represent a specific substance/contaminant reacting with chlorine in addition to chlorine decay due to organic matter and bacterial growth. As the study \cite{helbling2009modeling} states, along with tracing chlorine concentrations throughout the network to maintain the desired levels under the normal conditions, it is important to model the chlorine with different special cases. These cases include contaminated water due to several reasons (e.g., microbial/chemical contamination intrusion, poorly treated water, substances derived from pipe materials) and due to the fact that WDN are considered vulnerable to contamination and health violation attacks which affect millions of people annually \cite{allaire2018national, furst2018tradeoffs}.
	
	Contamination sources that can affect water quality parameters vary from microbial contaminants, heavy metals, natural organic matter, etc. \cite{palansooriya2020occurrence}. Study \cite{helbling2009modeling} investigates chlorine residual modeling's response to microbial contamination events. Results from this study show how chlorine residual is affected by the contamination species, event location, and network topology and characteristics. Study \cite{mohan2018modeling} addresses uncontrolled microbial contamination caused by sewage intrusion and how chlorine is consumed as a result. Study \cite{burkhardt2017modeling} states that arsenic (As) is a naturally occurring toxic substance from pipe materials that is highly soluble and often found in water sources, while study \cite{sharma2017drinking} considers it the biggest mass poisoning case in the world. However, the authors in \cite{abhijith2021modeling} show that maintaining residual chlorine is recognized as an effective strategy to control levels of soluble arsenic. These aforementioned contaminates/substances are examples that react with chlorine at various rates leading to different scenarios, hence, controlling chlorine throughout the network follows a case-oriented approach.    
		
	On the other hand, several studies \cite{ohar2014optimal, ostfeld2006conjunctive, munavalli2003optimal} investigate the water quality control (WQC) problem and apply different optimization algorithms to reach the optimum scheduling/locations of the injection boosters while minimizing the system cost (i.e., minimizing the cost of chlorine injections and maintaining minimal deviations from chlorine setpoint concentrations) and minimizing the formation of the undesired DBPs. A framework for optimization chlorine and by-product concentrations is presented in \cite{fisher2018framework}. However, such studies do not include a closed-form, network- and control-theoretic representation of all system inputs, variables, and output measurements under normal and abnormal operation conditions thereby limiting their ability to perform network-wide WQC.  
	
	Recently, a novel state-space water quality control model based on the conversation of mass and a single-species reaction model has been developed in \cite{wang2021effective}. This state-space model captures the spatio-temporal evolution of chlorine concentrations throughout all the components of any network (e.g., reservoirs, tanks, junctions, pipes, pumps, valves, and booster stations). However, this state-space model does not simulate or include multi-species interactions between chlorine and any other reactants that are critical in water quality modeling as mentioned---and only uses the Lax-Wendroff scheme without further discussion on other discretization schemes. 
	
	\subsection{Paper Contributions and Organization}
	The paper contributions are three-fold:
	\begin{itemize}
		\item We present a state-space water quality model to include the decay and reaction of chlorine with a fictitious reactant resulting in a multi-species dynamic model. This model is a step forward to depict the existence of a specific reactant(s) with the chlorine in the system and to reflect various real-time water quality scenarios (e.g., water characteristics-, pipes material-, hazard-related). With which, a closed-form and network-theoretic representation is included of all system inputs and variables. 
		\item  Although the Lagrangian method has been used within EPANET for decades, other methods are less utilized, understood, or derived in the context of the nonlinear multi-species water quality dynamics. In this context, this paper investigates the performance, scalability, and complexity of various discretization-based techniques used in building this state-space water quality model. These techniques are Lax-Wendroff, Backward Euler, Crank-Nicolson, and Method of Characteristics. The derivation of these techniques is provided through a detailed example and applied on various scaled case studies. This adds a novel educational and theoretical value to the paper as it fills the gap of how different discrete methods perform for multi-species dynamics, how to apply them, and to add a comprehensive framework of the dynamics in a system theoretic way that empowers applying feedback control.
		\item To compare techniques' validity and simulation results with EPANET and its multi-species extension EPANET-MSX which allows the modeling of multi-species dynamics. EPANET-MSX extension gives the user free control to define the chemical reactions to be included in his model \cite{rossman2020epanet,shang2008epanet}. Comparison is considered reliable as the covering laws and equations are the same for all network components in both models.  
		
	\end{itemize}
	
	\begin{figure}[h!] 
		\centering
		\includegraphics[width=0.8\linewidth]{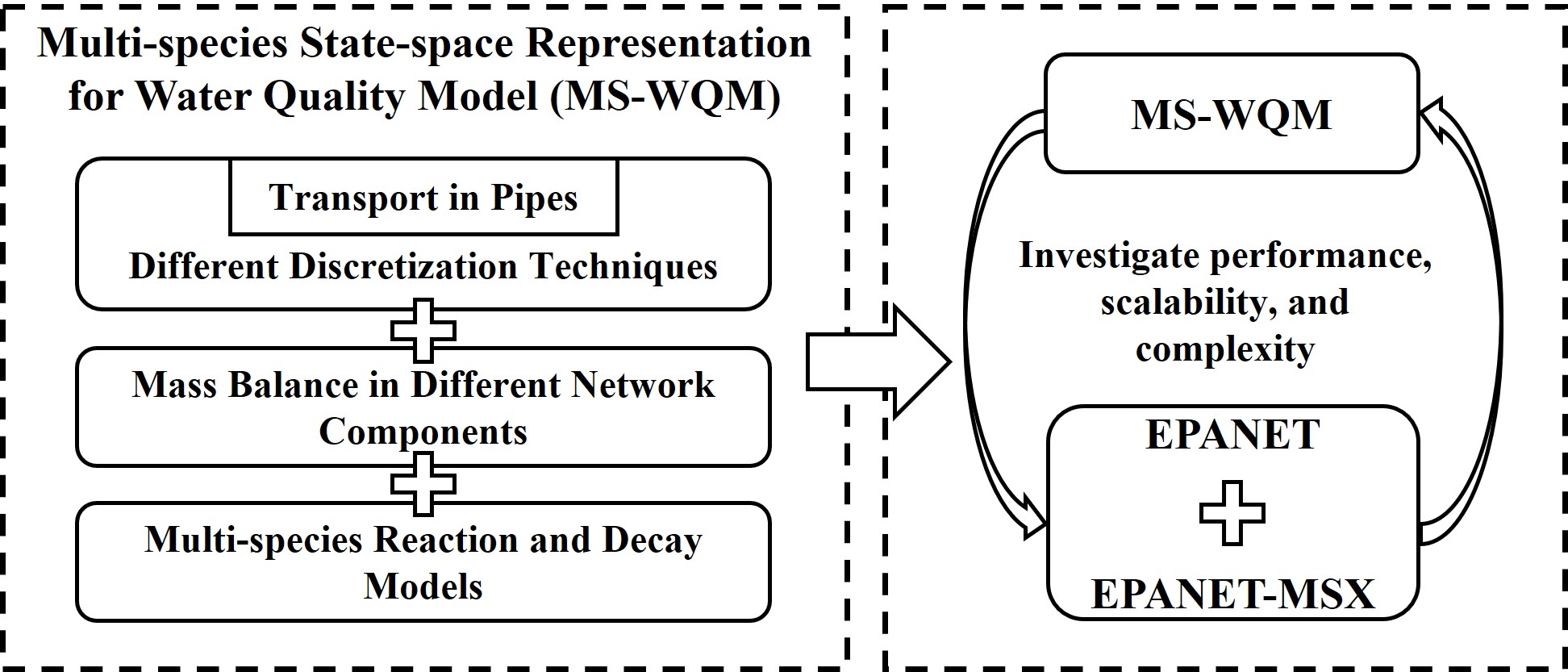}
		\caption{Conceptual framework of the paper.}
		\label{fig:PaperFC}
	\end{figure}
	
	The conceptual framework of the paper is illustrated in Fig. \ref{fig:PaperFC}. The formulation of the state-space representation for multi-species water quality model (MS-WQM) is based on transport in pipes, mass balance at the rest of the network components, and multi-species reaction and decay dynamics. In this paper, we adopt different discretization methods to simulate the transport and reaction in pipes. These methods are compared with EPANET and its extension EPANET-MSX to test their performance, scalability, and complexity.
	
	The rest of the paper is organized as follows. Section \ref{sec:WNF-SS} describes the problem formulation based on the principles of water quality in WDN. Section \ref{sec:TransPipes} delineates the transport and reaction model in pipes in detail and how pipes can be discretized using different schemes. Section \ref{sec:MassBal} lists the governing equations for different network components by applying mass balance. Section \ref{sec:reaction} breaks down the reaction models into decay and mutual reaction dynamics and their implementations. Section \ref{sec:SSForm} provides the final formulation of the state-space representation of the system. Section \ref{sec:Case-Studies} showcases the model implementation results on different networks. Based on these results, conclusion, paper's limitations, and recommendations for future work are all given in Section \ref{sec:Conc-Rec}. 
	
	\section{\large Water Network Fundamentals and Problem Formulation}~\label{sec:WNF-SS}
	The WDN is modeled by a directed graph $\mathcal{G} = (\mathcal{N},\mathcal{L})$.  The set $\mathcal{N}$ defines the nodes and is partitioned as $\mathcal{N} = \mathcal{J} \cup \mathcal{T} \cup \mathcal{R}$ where sets $\mathcal{J}$, $\mathcal{T}$, and $\mathcal{R}$ are the collections of junctions, tanks, and reservoirs, respectively. Let $\mathcal{L} \subseteq \mathcal{N} \times \mathcal{N}$ be the set of links, and define the partition $\mathcal{L} = \mathcal{P} \cup \mathcal{M} \cup \mathcal{V}$, where sets $\mathcal{P}$, $\mathcal{M}$, and $\mathcal{V}$ represent the collection of pipes, pumps, and valves.
	
	We simulate the interaction and fate of two chemical substances in our model, thus, the governing equations may differ in one component according to their characteristics. Therefore, general statements of laws and equations for each network component are briefly formulated followed by their application for different substances. In particular, herein we showcase a high-level state-space representation of the water quality model which captures the transport and evolution of the chemicals' concentrations in every component of the network (e.g., reservoirs, junctions, tanks, pipes, pumps, and valves). This representation is a generalized multi-species model that traces the evolution of chlorine and a fictitious reactant that can represent one specific chemical substance or multiples which react at a relatively close rate with chlorine. Furthermore, we showcase how this model can be appended to include more than two distinct chemicals by introducing different reactions dynamics representations.
	
	In general, for each component the water quality model is a function of time-dependent parameters and variables such as velocities, and flow rates, and might depend on the model of another component. The problem formulation in this paper is to obtain a state-space representation for both chlorine and the fictitious reactant concentrations evolution and their mutual reaction as a nonlinear difference equation (NDE). This representation is formulated for various discretization methods of the transport advection-reaction PDE. Eventually, we formulate a system representation of two-species that is able to capture chemicals evolution, booster stations injections, and sensors measurements, is given by an NDE as

	\begin{subequations}~\label{equ:NDE}
		\begin{align}
			\underbrace{	\begin{bmatrix}
					E_{11}(t) & 0 \\ 0 & E_{22}(t)
			\end{bmatrix}}_{{E}(t)} \underbrace{	\begin{bmatrix}
					x_1(t+\Delta t) \\ x_2(t+\Delta t)
			\end{bmatrix}}_{{x}(t+\Delta t)} &= 
			\underbrace{	\begin{bmatrix}
					A_{11}(t) & 0 \\ 0 & A_{22}(t)
			\end{bmatrix}}_{{A}(t)}
			\underbrace{ \begin{bmatrix}
					x_1(t) \\ x_2(t)
			\end{bmatrix}}_{{x}(t)}  + \underbrace{	\begin{bmatrix}
					B_{11}(t) & 0 \\ 0 & B_{22}(t)
			\end{bmatrix}}_{{B}(t)} 
			\underbrace{\begin{bmatrix}
					u_1(t) \\ u_2(t)
			\end{bmatrix}}_{{u}(t)} + f(x_1,x_2,t), \label{equ:NDE1} 
			\\ \underbrace{	\begin{bmatrix}
					y_1(t) \\ y_2(t)
			\end{bmatrix}}_{{y}(t)} &= 
			\underbrace{	\begin{bmatrix}	C_{11}(t) & 0 \\
					0 &  C_{22}(t)
			\end{bmatrix}}_{{C}(t)}
			\underbrace{ \begin{bmatrix}
					x_1(t) \\ x_2(t)
			\end{bmatrix}}_{x(t)} \label{equ:NDE2}
		\end{align}
	\end{subequations}
	where variable $t$ represents specific time index in a simulation period $[0,T_s]$; $\Delta t$ is the time-step or sampling time; vectors $x_1(t)$ and $x_2(t) \in \mathbb{R}^{n_x}$ depict  the concentrations of chlorine and the fictitious reactant (two species model) in the entire network; vector $u_1(t) \in \mathbb{R}^{n_{u_1}}$ represents the dosages of injected chlorine; vector $u_2(t) \in \mathbb{R}^{n_{u_2}}$ accounts for planned or unplanned injection of the fictitious component; vector $f(x_1,x_2,t)$ encapsulates the nonlinear part of the equations representing the mutual nonlinear reaction between the two chemicals; vector $y_1(t) \in \mathbb{R}^{n_{y_1}}$ denotes the sensor measurements of chlorine concentrations at specific locations in the network while $y_2(t) \in \mathbb{R}^{n_{y_2}}$ captures the fictitious reactant measurements by sensors in the network if they exist. The state-space matrices $E_{11}, E_{22}, A_{11}, A_{22}, B_{11}, B_{22}, C_{11}$ and $C_{22}$ are all time-varying matrices that depend on the network topology, hydraulic parameters, and disinfectant decay rate coefficients. It is customary to assume that these matrices evolve at a slower pace than the states $x(t)$ and control inputs $u(t)$. 
	
	The model, built on the conservation of mass law, transport, decay, and reaction models of the substances, captures the concentrations in all network components (junctions, tanks, pipes, valves, pumps, and reservoirs). Some of these models are in a form of ordinary differential equations (ODEs) or PDEs and converted to difference algebraic equations by applying different schemes and approaches (e.g., Eulerian-based and Lagrangian-based schemes for pipes, continuously stirred tank reactor model for tanks). On the other hand, the decay and reaction models are combined in a form of a multi-species model. Notice that the $A$ matrices are placed in a form of a block-diagonal big matrix, which indicates that the state-space modeling is mainly decoupled for the two chemicals and only the nonlinear part $f$ in \eqref{equ:NDE1} reflects their reaction and relation. To that end, the objective of the next sections is to derive the matrices for a generic water distribution network.
	
	As a preparation for the next sections, we list the symbols for different quantities in Tab. \ref{tab:Notation}. For example, $c(t)$ represents chemical concentration; $q(t)$ depicts the flow rate for a specific component depending on the superscript it has; $q^\mathrm{D}(t)$ is the demand drawn from a node; $q^\mathrm{B}(t)$ is the booster flow rate and in this case, the subscript represents the node with the injected flow; $V$ is the volume of the tank. As for the network components, $\mathrm{P,M,V,J,TK,R}$, and $\mathrm{B}$ are super/sub-scripts representing pipes, pumps. valves, junctions, tanks, reservoirs, and booster stations respectively; while super/sub-scripts $\mathrm{L}$ and $\mathrm{N}$ represent links and nodes. Note that links represent pipes, pumps, and valves, and nodes include reservoirs, tanks, and junctions. 
	
	It is worthwhile to note that all hydraulic variables and parameters (e.g., flow rates, velocities, water levels, pump setting, etc.) are assumed to be predetermined. This assumption is widely accepted due to the difference between the water quality modeling time-step and the one for the hydraulic simulation \cite{wang2005adaptive, wang2021effective}. Hydraulic modeling time-step is acceptable to be within an hourly scale to be aligned with the assumption of steady system simulation and to reflect the change in demand, while the range is between minutes and seconds for water quality modeling to allow a stable numerical simulation that captures chemicals reaction and evolution~\cite{seyoum2017integration,shang2008epanet}. 
	
	\begin{table}[h!]
		\centering
		\caption{Paper variables and parameters notation}~\label{tab:Notation}
		{\small\begin{tabular}{c|W|c}
				\hline
				Symbol & Description & Dimensions  \\
				\hline
				
				$c^\mathrm{N} := \{c^\mathrm{R}, c^\mathrm{J}, c^\mathrm{TK}\}$ & Concentrations at nodes (reservoirs, junctions, and tanks) & $\mathbb{R}^{n_\mathrm{N}}$  \\ 
				\hline
				$c^\mathrm{L} := \{c^\mathrm{M}, c^\mathrm{V}, c^\mathrm{P}\}$ & Concentrations at links (pumps, valves, and pipes) & $\mathbb{R}^{n_\mathrm{L}}$ \\
				\hline 
				$	x := \{ c^\mathrm{N}, c^\mathrm{L} \}$ & System states (i.e., chemical concentrations) & $\mathbb{R}^{n_x}$ \\
				\hline 
				$	u := \{ c^{\mathrm{B}_\mathrm{J}}, c^{\mathrm{B}_\mathrm{TK}} \}$ & System inputs (i.e., chemical injections at nodes) & $\mathbb{R}^{n_u}$  \\
				\hline 
				$q^\mathrm{M}, q^\mathrm{V}, q^\mathrm{P}$ & Flow rates in pumps, valves, and pipes & $\mathbb{R}^{n_\mathrm{M}}, \mathbb{R}^{n_\mathrm{V}}, \mathbb{R}^{n_\mathrm{P}}$ \\
				\hline
				$q^{\mathrm{D}_\mathrm{J}}$ & Demands from junctions & $\mathbb{R}^{n_\mathrm{J}}$ \\
				\hline
				$q_{\mathrm{in}}, q_{\mathrm{out}}$ & Inflows and outflows at nodes & --- \\
				\hline
				$q^{\mathrm{B}_\mathrm{J}}$ & Flow injected to junctions by booster stations & $\mathbb{R}^{n_{\mathrm{J}}}$ \\
				\hline
				$V^\mathrm{TK}$ & Tank volume & $\mathbb{R}^{n_\mathrm{TK}}$ \\
				\hline
				$V^{\mathrm{B}_\mathrm{TK}}$ & Volume injected to tank by booster station & $\mathbb{R}^{n_\mathrm{TK}}$ \\
				\hline
				$v$ & Flow velocity in pipes & $\mathbb{R}^{n_{\mathrm{P}}}$ \\ 
				\hline
				$L,r_\mathrm{P},s$ & Pipes length, radius, and number of segments to be divided into & $\mathbb{R}^{n_{\mathrm{P}}}$ \\
				\hline
				$\tilde{\lambda}$ & Courant number & $\mathbb{R}^{n_{\mathrm{P}_s}}$ \\
				\hline 
				$R_\mathrm{MS}, R_\mathrm{D}, R_\mathrm{M}$ & Multi-species, decay, and mutual reaction expressions & --- \\
				\hline
				$k_b, k_w, k_f, k_r$ & Chlorine bulk and wall reaction rate constants, mass transfer coefficient between bulk flow and pipe wall, and mutual reaction rate constant between two chemicals & --- \\
				\hline
				$k^\mathrm{P}, k^\mathrm{TK}$ & Chlorine decay-reaction rates for pipes and tanks & $\mathbb{R}^{n_{\mathrm{P}_s}}, \mathbb{R}^{n_\mathrm{TK}}$ \\
				\hline
				$T_s$ & Simulation period & --- \\						
				\hline
				\hline
		\end{tabular}}
	\end{table}
	
	\section{Transport and Reaction in Pipes: Discretization Techniques }\label{sec:TransPipes}
	Conservation of mass during transport and reaction in pipes is modeled by the one-dimension advection-reaction (1-D AR) equation. That is, we assume that the dispersion is neglected during our simulation (i.e., medium to high velocity ranges, limited number of dead-end nodes) \cite{tzatchkov2002advection}. To that end, in any Pipe $i$ of the network, the 1-D AR equation is expressed by a PDE as:
	\begin{equation}\label{equ:PDE}
		\frac{\partial c^\mathrm{P}_i}{\partial t}=-v_i(t) \frac{\partial c^\mathrm{P}_i}{\partial x} + R^\mathrm{P}_{\mathrm{MS}}(c^\mathrm{P}_i(x,t)),
	\end{equation}
	where $c^\mathrm{P}_i(x,t)$ is the concentration in Pipe $i$ at location $x$ and time $t$; $v_i(t)$ is the mean flow velocity which equals $\frac{4q^\mathrm{P}_i(t)}{\pi r_{\mathrm{P}_i}^2}$; $q^\mathrm{P}_i(t)$ is flow; $r_{\mathrm{P}_i}$ is the pipe radius; and $R^\mathrm{P}_{\mathrm{MS}}(c^\mathrm{P}_i(x,t))$ is the multiple-species reaction rate expression, which is explained briefly in Section \ref{sec:reaction}.
	
	The PDE \eqref{equ:PDE} can be solved using different numerical schemes that are Eulerian, Lagrangian, and mixed Eulerian-Lagrangian schemes \cite{rossman1996numerical,rossman1993discrete,liou1987modeling,basha2007eulerian}. The approach of the Eulerian Finite-Difference (EFD) schemes relies on defining a fixed numerical grid that discretizes the pipes in time and space resulting in approximate algebraic equation over the numerical grid. In Equation \eqref{equ:Segment}, Pipe $i$ with length $L_{i}$ is split into a number of segments $s_i$ of length $\Delta x_i$.
	
	\begin{equation}~\label{equ:Segment}
		s_i=\Bigg\lfloor \frac{L_i}{v_i(t) \Delta t} \Bigg\rfloor, \;\ \Delta x = \frac{L_i}{s_i}
	\end{equation}
	
\begin{figure}[h] 
		\centering
		\includegraphics[width=1\linewidth]{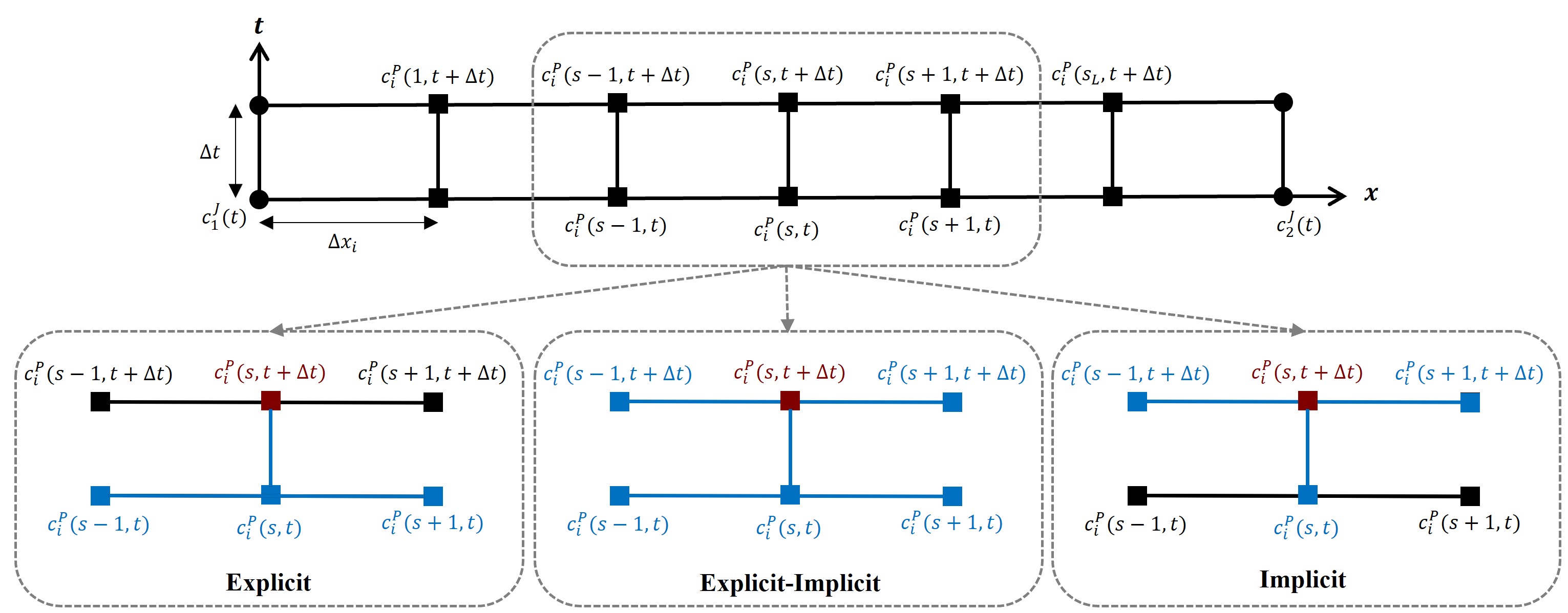}
		\caption{Eulerian Finite-Difference discretization schemes for Pipe $i$ connecting Junctions $1$ and $2$. Schemes calculate concentration  $c_i^\mathrm{P}(s,t+\Delta t)$ at segment $s$ (colored in maroon) depending on concentrations at the other segments/nodes colored in blue} 
		\label{fig:DiscPDE}
	\end{figure}

	As such, the EFD schemes calculates the concentration at any segment $s$ depending on the concentrations of the upstream and downstream nodes/segments. The scheme is considered \textit{implicit} when the upstream and downstream nodes/segments concentrations are taken at the current time-step; it is considered \textit{explicit} when they are taken at the previous time-step; \textit{explicit-implicit} when averaging the current and previous time-steps concentrations; see Fig. \ref{fig:DiscPDE}. 
	
	However, EPANET and its extension EPANET-MSX use the Lagrangian time-driven method (LTD) \cite{liou1987modeling, rossman2020epanet}. We give a full description of EPANET's algorithm in Appendix \ref{App:EPALag}. The main difference between EFD and LTD schemes is that LTD divides the pipe into non-overlapping changing-sized segments by allowing the most upstream segment to change its size at the expense of the most downstream segment. The change in the segment's size depends on the mass and flow entering the pipe every time step. However, EFD schemes work within a fixed grid that produces a state-space description of finite-dimension. This description allows us to formulate a control problem and solve an optimization problem to identify optimum solutions.
	
	In this paper, we use a Lagrangian-based method that uses an explicit method of characteristics (MOCs) to project concentrations forward in time which is achieved through integration and interpolation~\cite{tzatchkov2002advection,basha2007eulerian,abokifa2018developing}. This method allows for the state-space formulation of the concentrations profile with higher complexity than EFD schemes, more details on that are explained in Section \ref{sec:MoCs}.  
	
	We also investigate three EFD schemes: \textit{explicit, explicit-implicit,}  and \textit{implicit} schemes in addition to a Lagrangian-based method. When applying these schemes, the pipes are divided into a number of segments, thus, the size of vector $c^\mathrm{P}(t+\Delta t)$ is the summation of segments count for all pipes in the network. Hence, the size of the vector is $n_{\mathrm{P}_s} = \sum_{i=1}^{n_\mathrm{P}} s_i$ with $n_\mathrm{P}$ pipes in the network. For these techniques, the concentration at any segment $c^\mathrm{P}_i(s,t+\Delta t)$ depends on the concentrations at the previous time-step $c^\mathrm{P}_i(s,t)$ and the adjacent segments/nodes at the previous and/or current time-steps (i.e., $c^\mathrm{P}_i(s-1,t), c^\mathrm{P}_i(s+1,t), c^\mathrm{P}_i(s-1,t+\Delta t), c^\mathrm{P}_i(s+1,t+\Delta t)$) depending on the applied technique; see Fig. \ref{fig:DiscPDE} and Fig. \ref{fig:MoCs}. This dependency is formulated using the Courant number $\tilde{\lambda}_i(t)= {v_{i}(t) \frac{\Delta t}{\Delta x_{i}}}$, where $v_i(t)$ is the mean flow velocity in Pipe $i$; $\Delta t$ is the water quality modeling time-step; and $\Delta x_{i}$ is the chosen segment length. In the following sections, a brief description of each technique is provided.
	
	\subsection{Technique 1: Explicit Eulerian scheme --- Lax-Wendroff scheme}~\label{sec:LW}
	One of the EFD schemes is the Lax-Wendroff (L-W) explicit scheme \cite{rossman1996numerical,lax1964difference} which is widely used for solving hyperbolic differential equations with second-order accuracy \cite{smith1978numerical,tshehla2017state,wang2019state}. In water quality modeling research, a water quality state-space representation of a single-species model is previously developed for the first time in \cite{wang2021effective} using the Lax-Wendroff scheme to discretize the 1-D AR PDE. In this scheme, the chemical concentration at a segment $s$ and time $t + \Delta t$ depends on the concentrations at time $t$ of segment $s$ itself, upstream segment $s-1$, and downstream segment $s+1$.
	Hence, by applying the Lax-Wendroff scheme for segment $s$ of Pipe $i$, except for the first and last segments, the concentration is calculated as
	\begin{equation}\label{equ:LW_Si}
		c^\mathrm{P}_i(s,t+ \Delta t)= \underline{\lambda}_i(t) c^\mathrm{P}_i(s-1,t)+\lambda_i(t) c^\mathrm{P}_i(s,t)+\overline{\lambda}_i(t) c^\mathrm{P}_i(s+1,t)+R^\mathrm{P}_{\mathrm{MS}}(c^\mathrm{P}_i(s,t)) \Delta t,
	\end{equation}
	where $\underline{\lambda}_i(t), \lambda_i(t),$ and $\overline{\lambda}_i(t)$ are the weighting coefficients calculated as follow
	
	\begin{equation}~\label{equ:LW_coeff}
		\begin{aligned}
			\underline{\lambda}_i(t) &= 0.5\tilde{\lambda}_i(t)\left(1+{\tilde{\lambda}}_i(t)\right), \\
			\lambda_i(t) &=  1-{\tilde{\lambda}}_i^2(t),  \\
			\overline{\lambda}_i(t) &= -0.5\tilde{\lambda}_i(t)\left(1-\tilde{\lambda}_i(t)\right).
		\end{aligned}
	\end{equation}

	Note that the L-W scheme is conditionally stable by applying the Courant-Friedrichs-Lewy condition (CFL) \cite{lax1964difference}. The CFL condition puts the Courant number in the range of  $0<\tilde{\lambda}_i(t) \leq 1$. Thus, the choice of the number of segments and the water quality time-step must be in accordance to satisfy the CFL condition. The velocities are predetermined during the hydraulic simulation, subsequently, the first step for water quality simulation, when using this technique, is choosing the time-step or number of segments of all pipes in the network to ensure fulfilling the stability condition.
	
	\begin{myrem} \label{rem:PipeSeg}
		For the first segment $s=1$ in a pipe, there is no upstream segment and the concentration depends on the concentration of its upstream node. Likewise, the concentration of the last segment $s=s_\mathrm{L}$ depends on the downstream node's concentration. The concentrations in other network components connected to a pipe are affected by the properties and parameters of the pipe's first or last segment according to the flow direction.
	\end{myrem}
	
	According to Remark~\ref{rem:PipeSeg}, the concentrations in the first and last segment are expressed in Equation \eqref{equ:node2pipe} assuming that the connecting nodes are two junctions that are Junctions $j$ and $k$ --- a full description of how to calculate concentrations at junctions in the next section is given.
	\begin{subequations}~\label{equ:node2pipe}
		\begin{align}
			\begin{split} 	c^\mathrm{P}_i(1,t+ \Delta t) =& \underline{\lambda}_i(t) c^\mathrm{J}(t)+\lambda_{i}(t) c^\mathrm{P}_i(1,t)+\overline{\lambda}_i(t) c^\mathrm{P}_i(2,t)+R^\mathrm{P}_{\mathrm{MS}}(c^\mathrm{P}_i(1,t)) \Delta t, \label{equ:node2pipeA} 
			\end{split} \\
			\begin{split} 
				c^\mathrm{P}_i(s_{\mathrm{L}},t+ \Delta t)=& \underline{\lambda}_i(t) c^\mathrm{P}_i(s_{\mathrm{L}}-1,t)+\lambda_{i}(t) c^\mathrm{P}_i(s_{\mathrm{L}},t)+\overline{\lambda}_i(t) c^\mathrm{J}(t)+R^\mathrm{P}_{\mathrm{MS}}(c^\mathrm{P}_i(s_{L_i},t)) \Delta t. \label{equ:node2pipeB}
			\end{split}
		\end{align}
	\end{subequations}

	\subsection{Technique 2: Implicit Eulerian scheme --- Backward Euler scheme}~\label{sec:BackEuler} 
	Backward Euler scheme follows an implicit EFD approach \cite{mackenzie2007analysis}. The approach is for chemical concentration at segment $s$ of Pipe $i$ is expressed as:
	\begin{equation}\label{equ:BE_Si1}
		0.5\tilde{\lambda}_i(t+\Delta t) c^\mathrm{P}_i(s+1,t+ \Delta t)+c^\mathrm{P}_i(s,t+ \Delta t)- 0.5\tilde{\lambda}_i(t+\Delta t)c^\mathrm{P}_i(s-1,t+ \Delta t) \\
		= c^\mathrm{P}_i(s,t)+R^\mathrm{P}_{\mathrm{MS}}(c^\mathrm{P}_i(s,t)) \Delta t.
	\end{equation}
	
	\begin{asmp}~\label{asmp:lambda}
		Water quality time-step is much smaller than the hydraulic simulation time-step. That is, $\tilde{\lambda_i}$ is constant within the same hydraulic simulation time-step. 
	\end{asmp}
	
	By applying Assumption \ref{asmp:lambda} and Remark \ref{rem:PipeSeg}, Equation \eqref{equ:BE_Si1} is updated. To that end, for any, first, and last segment (assuming connected to Junctions $j$ and $k$) of the pipe, concentrations are expressed as Equations \eqref{equ:BE_Si2a}, \eqref{equ:BE_Si2b}, and \eqref{equ:BE_Si2c}.
	\begin{subequations}\label{equ:BE_Si2}
		\begin{align}
			& 0.5\tilde{\lambda}_i(t) c^\mathrm{P}_i(2,t+ \Delta t)+c^\mathrm{P}_i(1,t+ \Delta t)- 0.5\tilde{\lambda}_i(t)c^\mathrm{J}_i(t+ \Delta t)= c^\mathrm{P}_i(1,t)+R^\mathrm{P}_{\mathrm{MS}}(c^\mathrm{P}_i(1,t)) \Delta t,~\label{equ:BE_Si2a}\\
			& 0.5\tilde{\lambda}_i(t) c^\mathrm{P}_i(s+1,t+ \Delta t)+c^\mathrm{P}_i(s,t+ \Delta t)- 0.5\tilde{\lambda}_i(t)c^\mathrm{P}_i(s-1,t+ \Delta t)= c^\mathrm{P}_i(s,t)+R^\mathrm{P}_{\mathrm{MS}}(c^\mathrm{P}_i(s,t)) \Delta t,~\label{equ:BE_Si2b}\\
			& 0.5\tilde{\lambda}_i(t) c^\mathrm{J}_i(t+ \Delta t)+c^\mathrm{P}_i(s_\mathrm{L},t+ \Delta t)- 0.5\tilde{\lambda}_i(t)c^\mathrm{P}_i(s_\mathrm{L}-1,t+ \Delta t)= c^\mathrm{P}_i(s_\mathrm{L},t)+R^\mathrm{P}_{\mathrm{MS}}(c^\mathrm{P}_i(s_\mathrm{L},t)) \Delta t. ~\label{equ:BE_Si2c}
		\end{align}
	\end{subequations}
	
	Note that the Backward Euler scheme is different from the Lax-Wendroff scheme, and the stability check is not needed as it is unconditionally stable. Therefore, the choice of the number of segments does not depend on fulfilling any stability condition and can be chosen as an arbitrary number that preserves the desired accuracy.

	\subsection{Technique 3: Explicit-Implicit Eulerian scheme --- Crank-Nicolson scheme}~\label{sec:CN}
	Crank-Nicolson scheme \cite{crank1947practical} is as a hybrid method between the two previous ones and takes an average of the explicit and implicit schemes. And by considering Assumption \ref{asmp:lambda}and Remark \ref{rem:PipeSeg}, concentrations at any segment $s$, the first segment and the last segment are expressed through Equations \eqref{equ:CN_Sia}, \eqref{equ:CN_Sib}, and \eqref{equ:CN_Sic}. 
	\begin{subequations}\label{equ:CN_Si}
		\begin{align}
			\begin{split}
				0.25\tilde{\lambda}_i(t) c^\mathrm{P}_i(2,t+ \Delta t)&+c^\mathrm{P}_i(1,t+ \Delta t)  
				- 0.25\tilde{\lambda}_i(t)c^\mathrm{J}_i(t+ \Delta t)
				\\ & = 0.25\tilde{\lambda}_i(t) c^\mathrm{J}_i(t)+ c^\mathrm{P}_i(1,t)-0.25\tilde{\lambda}_i(t) c^\mathrm{P}_i(2,t)+R^\mathrm{P}_{\mathrm{MS}}(c^\mathrm{P}_i(1,t)) \Delta t,
			\end{split}\label{equ:CN_Sia} \\
			\begin{split}
				0.25\tilde{\lambda}_i(t) c^\mathrm{P}_i(s+1,t+ \Delta t)&+c^\mathrm{P}_i(s,t+ \Delta t)  
				- 0.25\tilde{\lambda}_i(t)c^\mathrm{P}_i(s-1,t+ \Delta t)
				\\ & = 0.25\tilde{\lambda}_i(t) c^\mathrm{P}_i(s-1,t)+ c^\mathrm{P}_i(s,t)-0.25\tilde{\lambda}_i(t) c^\mathrm{P}_i(s+1,t)+R^\mathrm{P}_{\mathrm{MS}}(c^\mathrm{P}_i(s,t)) \Delta t,
			\end{split} \label{equ:CN_Sib}\\
			\begin{split}
				0.25\tilde{\lambda}_i(t) c^\mathrm{J}_i(t+ \Delta t)&+c^\mathrm{P}_i(s_\mathrm{L},t+ \Delta t)  
				- 0.25\tilde{\lambda}_i(t)c^\mathrm{P}_i(s_\mathrm{L}-1,t+ \Delta t)
				\\ & = 0.25\tilde{\lambda}_i(t) c^\mathrm{P}_i(s_\mathrm{L}-1,t)+ c^\mathrm{P}_i(s_\mathrm{L},t)-0.25\tilde{\lambda}_i(t) c^\mathrm{J}_i(t)+R^\mathrm{P}_{\mathrm{MS}}(c^\mathrm{P}_i(s_\mathrm{L},t)) \Delta t. 
			\end{split} \label{equ:CN_Sic}
		\end{align}
	\end{subequations}
	
	Moreover, this method is unconditionally stable as the Backward Euler scheme where the choice of the number of segments does not depend on fulfilling the CFL condition. Yet, temporal numerical dispersion and oscillation may have a non-neglectable effect on the results \cite{chapra2008surface}.
	
	\subsection{Technique 4 --- Method of Characteristics}~\label{sec:MoCs}
\begin{figure}[t] 
		\centering
		\includegraphics[width=1\linewidth]{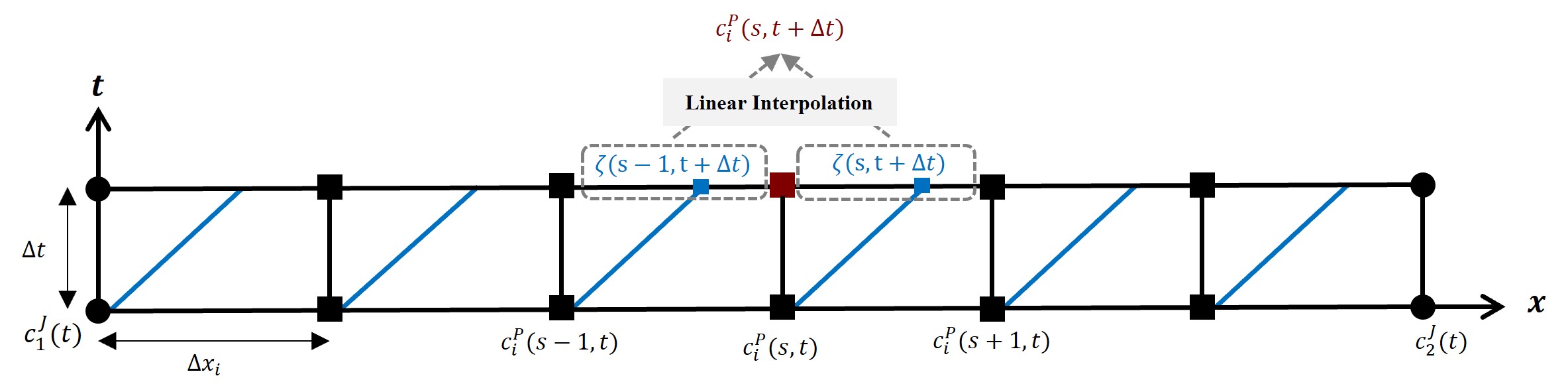}
		\caption{Lagrangian-MoCs for Pipe $i$ connecting Junctions $1$ and $2$.}
		\label{fig:MoCs}
	\end{figure}

	Method of Characteristics (MoCs) reduces Equation \eqref{equ:PDE} to an ODE along the advection characteristics line whose slope is $\dfrac{dt}{dx_i}$; see the blue line in Fig. \ref{fig:MoCs}. By the definition of advection, we find that $\frac{dx_i}{dt}=v_i(t)$ for that line. However, to guarantee the line falls within the same segment in the next time-step the number of segments and segment size are calculated by \eqref{equ:Segment}.
	
	The forward projection of concentrations from time $t$ through this line specifies new locations at $t+\Delta t$, shifted by a $v_i(t) \Delta t$ distance. The new location for concentrations are donated by $\zeta$ as illustrated in Fig. \ref{fig:MoCs} where the concentration $c^\mathrm{P}(s,t)$ is projected to $\zeta(s, t+\Delta t)$. However, to calculate concentrations profile for different $\zeta$ at time-step $t+ \Delta t$ according to the new reduced ODE, the reaction term should be added to the advection profile. That is, the derivation of the concentration expression is included in Section \ref{sec:reaction} after introducing the reaction and decay models. Henceforward, segments' concentrations profile at time-step $t + \Delta t$ is calculated by applying linear interpolation from the $\zeta$ profile at same time-step \cite{salih2016method,abokifa2018developing,basha2007eulerian}.
	
	\section{Mass balance at Different Network's Components}~\label{sec:MassBal}
	For components other than pipes (i.e., reservoirs, pumps, valves, junctions, and tanks), the conservation of mass is applied to obtain concentrations at different time steps. Different descriptions are provided for components depending on their characteristics and connected nodes/links.    
	
	\subsection{Mass balance at reservoirs} 
	Reservoirs are assumed to have constant concentrations where they can present a continuous source of a specific chemical. In vector $c^\mathrm{R}(t +\Delta t) \in \mathbb{R}^{n_\mathrm{R}}$, for each Reservoir $i$ concentration is expressed as
	\begin{equation} \label{equ:reservoir}
		c_i^\mathrm{R}(t+\Delta t)=c_i^\mathrm{R}(t).
	\end{equation}
	
	\subsection{Mass balance at pumps and valves} 
	Links other than pipes which are represented in pumps and valves are assumed to have no defined length or storage. Subsequently, the model deals with them as transmission links with concentration equals to the concentration of the node upstream while heir functions are reflected in the hydraulic model. That being said, the concentrations for Pump $i$ or Valve $j$, in vectors $c^\mathrm{M}(t +\Delta t) \in \mathbb{R}^{n_\mathrm{M}}$ and $c^\mathrm{V}(t +\Delta t) \in \mathbb{R}^{n_\mathrm{V}}$, installed after Reservoir $k$ (as an example), are expressed as
	\begin{align}
		c_i^{\mathrm{M}}(t+\Delta t) &= c_k^{\mathrm{R}}(t+\Delta t), \label{equ:pump} \\
		c_j^{\mathrm{V}}(t+\Delta t) &= c_k^{\mathrm{R}}(t+\Delta t). \label{equ:valve}
	\end{align}
	
	\subsection{Mass balance at junctions}~\label{sec:mbjunction} 
	Chemicals are assumed to have a complete and instantaneous mixing in the junctions with no storage volume \cite{shang2008epanet,boulos2006comprehensive,rossman1996numerical}. Thus, the concentration at each Junction $i$ in vector $c^\mathrm{J}(t) \in \mathbb{R}^{n_\mathrm{J}}$ is expressed as
	\begin{equation}~\label{equ:mb-junc} 
		c_i^\mathrm{J}(t)= \frac{\sum_{j \in L_{\mathrm{in}}} q_{\mathrm{in}}^{j}(t) c_\mathrm{in}^j(t)+q^\mathrm{B_\mathrm{J}}_i(t) c^\mathrm{B_\mathrm{J}}_i(t)}{q^{\mathrm{D}_\mathrm{J}}_i(t)+\sum_{k \in L_{\mathrm{out}}} q_{\mathrm{out}}^{k}(t)},
	\end{equation}
	where $j$ and $k$ are the counters for total $L_{\mathrm{in}}$ links flowing into the junction and $L_{\mathrm{out}}$ links extracting flow from the junction; $q_{\mathrm{in}}^{j}(t)$ and $q_{\mathrm{out}}^{k}(t)$ are the inflows and outflows from these links connected to the junction; $c_\mathrm{in}^j(t)$ is the concentration in the inflow solute;  $q^\mathrm{B_\mathrm{J}}_i(t)$ is the flow injected to the junction with concentration $c^\mathrm{B_\mathrm{J}}_i(t)$ by booster station if located; and $q^{\mathrm{D}_\mathrm{J}}_i(t)$ is demand.
	
	Equation \eqref{equ:mb-junc} can be applied at any arbitrary modeling time $t + \Delta t$ and according to Remark \ref{rem:PipeSeg}, with inflow entering the junction from a pipe, $c_\mathrm{in}^j(t)$ is taken as the concentration of the last segment of the pipe calculated in Equation \eqref{equ:node2pipeB}.  
	
	\subsection{Mass balance at tanks}~\label{sec:MB_Tank} Tanks are used as storage facilities being filled during low demands to cover the high demands. Mass conservation in tanks assumes complete instantaneous mixing of all inflows, outflows, and stored water following the continuously stirred tank reactor (CSTR) model \cite{shang2008epanet,butcher2016numerical}.
	\begin{myrem} \label{rem:Tank}
		The CSTR model assumes that the stored water gets affected by any inflow or injected disinfection instantaneously. It also assumes that all outflows and stored volume have the same solute concentrations.
	\end{myrem}
	The change in solute concentration at tanks is expressed by an ODE \cite{schmidt1998engineering} as
	\begin{equation} \label{equ:TankODE}
		\frac{\mathrm d (c^\mathrm{TK}V^\mathrm{TK})}{\mathrm d t} =\sum_{j \in L_{\mathrm{in}}} q^j_\mathrm{in}(t) c^j_\mathrm{in}(t)
		- \sum_{k \in L_{\mathrm{out}}} q^k_\mathrm{out}(t) c^\mathrm{TK}(t) +R^\mathrm{TK}_{\mathrm{MS}}(c^\mathrm{TK}(t)) V^\mathrm{TK}(t).
	\end{equation}
	where $V^\mathrm{TK}$ is the stored water volume at time $t$; $j$ and $k$ are the counters for total $L_{\mathrm{in}}$ links flowing into the tank and $L_{\mathrm{out}}$ links extracting flow from the tank, these links are either part of the network or not; $q^j_\mathrm{in}(t)$ is inflow either from links or injected; $c^j_\mathrm{in}(t)$ is substance concentration in the inflow solute; $q^k_\mathrm{out}(t)$ is outflow by links connected to the tank; $c^\mathrm{TK}(t)$ is concentration at tank and any flow drawn from it according to Remark \ref{rem:Tank}; and $R^\mathrm{TK}_{\mathrm{MS}}(c^\mathrm{TK}_i)$ is the multi-species reaction rate expression explained briefly in Section \ref{sec:reaction}.
	
	Applying Forward Euler discretization method~\cite{norsett1993solving} in Equation~\eqref{equ:TankODE}, the concentration at each Tank $i$  in $c^\mathrm{TK}(t +\Delta t) \in \mathbb{R}^{n_\mathrm{TK}}$ is expressed as 
	
	\begin{multline}\label{equ:tank1}
		V_i^\mathrm{TK}(t + \Delta t) c_i^\mathrm{TK}(t+ \Delta t)= V_i^\mathrm{TK}(t) c_i^\mathrm{TK}(t) + \sum_{j \in L_{\mathrm{in}}} q^j_\mathrm{in}(t) c^j_\mathrm{in}(t) \Delta t
		-\sum_{k \in L_{\mathrm{out}}} q^k_\mathrm{out}(t) c_i^\mathrm{TK}(t) \Delta t  +R^\mathrm{TK}_{\mathrm{MS}}(c_i^\mathrm{TK}(t)) V_i^\mathrm{TK}(t) \Delta t.
	\end{multline}
	Yet, according to Remark \ref{rem:Tank}, the effect of the injections from are instantaneous. resulting in an immediate effect \cite{wang2019state}. Updating Equation \eqref{equ:tank1} to be
	\begin{multline}\label{equ:tank2}
		V_i^\mathrm{TK}(t + \Delta t) c_i^\mathrm{TK}(t+ \Delta t)= V_i^\mathrm{TK}(t) c_i^\mathrm{TK}(t) +\sum_{j \in L_{\mathrm{in}}} q^j_\mathrm{in}(t)c^j_\mathrm{in}(t) \Delta t
		+V^\mathrm{B_\mathrm{TK}}_i(t+\Delta t)c^\mathrm{B_\mathrm{TK}}_i(t+\Delta t)\\
		- \sum_{k \in L_{\mathrm{out}}} q^k_\mathrm{out}(t)c_i^\mathrm{TK}(t) \Delta t
		+R^\mathrm{TK}_{\mathrm{MS}}(c_i^\mathrm{TK}(t)) V_i^\mathrm{TK}(t) \Delta t,
	\end{multline}
	where $V^\mathrm{B_\mathrm{TK}}_i(t+\Delta t)$ is the volume injected to the tank with concentration $c^\mathrm{B_\mathrm{TK}}_i(t+\Delta t)$ by booster station if located.
	
	Similarly to junctions, $c_\mathrm{in}^j(t)$ is taken as the concentration of the last segment of the pipe calculated in Equation \eqref{equ:node2pipeB} for inflows entering the tank.
	
	\section{Multi-species Reaction and Decay Model}~\label{sec:reaction}
	Having modeled mass balance in various components as well as discretization of pipes, herein we showcase the chlorine decay and reaction models. Chlorine decay and reaction models have been broadly studied as stated in Section \ref{sec:Into-Lit}. In Tab. \ref{tab:CLModels}, we provide a list of different, widely used and studied, bulk decay and reaction models \cite{fisher2011suitability, helbling2009modeling}. 
	
	\begin{table}[h!]
		\centering
		\caption{Chlorine bulk decay and reaction models}~\label{tab:CLModels}
		{\small\begin{tabular}{R c L c}
				\hline
				Model  & Model formulation & Model description & Citation\\
				\hline
				First-order & $\frac{dc}{dt}=-kc(t)$ & the simplest model with a first-order decay rate & \cite{jonkergouw2009variable, rossman2020epanet, haas1984kinetics} \\
				\hline
				First-order with stable component & $\frac{dc}{dt}=-k(c(t)-c_\mathrm{L})$ & this model is built on the first-order one and adding a stable component that represents the limiting concentration for chlorine $c_\mathrm{L}$ & \cite{haas1984kinetics, powell2000performance} \\
				\hline
				Parallel first-order & 
				$\begin{aligned}
					& \frac{dc_1}{dt} \Big\vert_{\text{fast}}=-k_{\text{fast}}c_1(t) \\
					& \frac{dc_2}{dt} \Big\vert_{\text{slow}}=-k_{\text{slow}}c_2(t) \\
					& c_t(t) = c_1(t) + c_2(t)
				\end{aligned}$
				& this model divides chlorine decay into two stages, and each stage decays independently, according to a first-order reaction with its own individual decay rate. The two stages are a slow and a fast decay dynamics & \cite{powell2000performance, haas1984kinetics} \\
				\hline
				Parallel second-order & 
				$\begin{aligned}
					& \frac{dc_\mathrm{F}}{dt} \Big\vert_{\text{fast}}=-k_{\text{fast}}c(t)c_\mathrm{F}(t) \\
					& \frac{dc_\mathrm{S}}{dt} \Big\vert_{\text{slow}}=-k_{\text{slow}}c(t)c_\mathrm{S}(t) \\
					& \frac{dc}{dt} = \frac{dc_\mathrm{F}}{dt} + \frac{dc_\mathrm{S}}{dt}
				\end{aligned}$
				& In comparison to the previous model, the decay rate of chlorine is divided into fast and slow second-order dynamics where $c_\mathrm{F}$ and $c_\mathrm{s}$ are representing the concentrations of fast and slow reducing agents.  & \cite{fisher2012suitable,fisher2017comprehensive,wang2019quantifying} \\
				\hline
				n\textsuperscript{th}-order & $\frac{dc}{dt}=-kc^n(t)$ & the rate of reaction can generally be described as a power function of concentration  & \cite{powell2000performance, haas1984kinetics} \\
				\hline
				n\textsuperscript{th}-order with stable component & $\frac{dc}{dt}=-k(c(t)-c_\mathrm{L}) c^{(n-1)}$ & $c_\mathrm{L}$ puts a cut-off for the regular n\textsuperscript{th}-order model & \cite{johnson1975measurement, haas1984kinetics, feben1951studies} \\
				\hline
				Second-order with fictitious component & $\begin{aligned}
					& \frac{dc}{dt}=-k c(t) \tilde{c}(t) \\
					& \frac{d \tilde{c}}{dt}=-kc(t) \tilde{c}(t)
				\end{aligned}$ & a fictitious reactant with concentration $\tilde{c}(t)$ is introduced in the system and accounts for all components reacting with chlorine & \cite{boccelli2003reactive, clark2002predicting, clark1998chlorine} \\
				\hline
				Second-order with multiple components & $\begin{aligned}
					& \frac{dc_i}{dt}=-k_i c(t) \tilde{c}_i(t) \\
					& \frac{d \tilde{c}_i}{dt}=-k_i c(t) \tilde{c}_i(t)\\
					& \frac{dc}{dt}= \sum_{i} \frac{dc_i}{dt}
				\end{aligned}$ & separate rate is considered for every component reacting with chlorine, where $\tilde c_i$ is the concentration of the i\textsuperscript{th} reacting substance & \cite{jadas1992chlorine, jonkergouw2009variable}  \\
				\hline
				\hline
		\end{tabular}}
	\end{table}
	
	In this paper, we consider a hybrid model that accounts for both chlorine decay with a constant rate and a reaction dynamic with a fictitious constituent. This model description has been used to detect different contamination events and their effect on chlorine residual in \cite{umberg2008performance}. Dividing the model into decay and mutual reaction dynamics allows it to consider a substance with relatively different reaction rates than the decay rate, and allows the model to be less sensitive to the other reactants' concentrations. Hence, the model formulation is
	\begin{equation}\label{equ:decayandmutual}
		R_{\mathrm{MS}}(c(t))=R_{\mathrm{D}}(c(t))+R_{\mathrm{M}}(c(t)),
	\end{equation}
	where $R_{\mathrm{D}}(c(t))$ is the decay reaction expression; and $R_{\mathrm{M}}(c(t))$ is the mutual reaction expression.
	
	Note that the decay and reaction model~\eqref{equ:decayandmutual} is assumed to occur in pipes and tanks only. We assume that there is no storage nor enough contact time in the other network components (e.g., pumps, valves, junctions). Likewise, the reservoir is considered a fixed source of chemicals with no decay or reactions \cite{rossman2020epanet, shang2008epanet}. 
	
	\subsection{Decay reaction model} \label{sec:decay}
	This decay model is a first-order model that depends only on chlorine concentration and constant decay rate. In addition to chlorine bulk decay models listed in Tab. \ref{tab:CLModels}, chlorine decay also takes place at the pipe's wall. While disinfectant bulk fluid reactions occur in pipes and tanks, wall reactions only considered to take place in pipes. The bulk decay is a result of the reaction with natural organic matter, while pipe wall decay is due to the reaction with the materials released from its boundary layer \cite{rossman1996numerical,basha2007eulerian,shang2008epanet}. Different wall reaction models have been developed/used varying between the zero-order and first-order models available in EPANET and its multi-species extension \cite{rossman2000epanet} and the developed and validated EXPBIO model by \cite{fisher2017new}. The wall decay zero-order model assumes constant decay while in the first-order model the decay rate reduces with the disinfactant's concentration, on the other hand, they have negative relation in the EXPBIO model. In this paper the focus is on the different bulk reaction models and we consider the first-order wall decay model in our simulation and when comparing our results with EPANET-MSX.

	Hence, the chlorine decay reaction rates \cite{vasconcelos1996characterization} for Pipe $i$ and Tank $j$ are 
	\begin{align}~\label{equ:decayreactionA}
		k_i^\mathrm{P} = k_{b}+\frac{2k_{w}k_{f}}{r_{\mathrm{P}_i}(k_{w}+k_{f})},\,\,\,\,\, k_j^\mathrm{TK} = k_{b},
	\end{align}
	where $ k_{b}$ is the bulk reaction rate constant; $k_{w}$ is the wall reaction rate constant; $k_{f}$ is the mass transfer coefficient between the bulk flow and the pipe wall; $r_{\mathrm{P}_i}$ is the pipe radius.
	
	Eventually, the chlorine decay reaction expressions for segment $s$ of Pipe $i$ and Tank $j$ are
	\begin{equation} ~\label{equ:decay_PTK}
		R^{\mathrm{P}}_{\mathrm{D}}(c^\mathrm{P}_i(s,t))=-k_i^\mathrm{P} c^\mathrm{P}_i(s,t), \,\,\,\,\, R^{\mathrm{TK}}_{\mathrm{D}}(c^\mathrm{TK}_j(t))=-k_j^\mathrm{TK} c^\mathrm{TK}_j(t).
	\end{equation}
	\subsection{Mutual reaction model} \label{sec:MutualReact}
	The reaction model is introduced in a form two-species reaction dynamics between chlorine and a fictitious reactant. The model is covered by the general reaction model of any two substances\cite{jadas1992chlorine, helbling2009modeling, uber2007evaluating}). Therefore, the mutual reaction model is expressed by a second-order nonlinear ODEs as 
	\begin{equation} \label{equ:mrNL}
		\begin{aligned}
			\frac{\mathrm d c}{\mathrm d t} = -k_r c(t) \tilde{c}(t), \;\;\; 
			\frac{\mathrm d \tilde{c}}{\mathrm d t} = -k_r c(t) \tilde{c}(t),
		\end{aligned}
	\end{equation}
	where  $c(t), \tilde{c}(t)$ are the concentrations for chlorine and fictitious reactant respectively; and $k_r$ is the mutual reaction rate between them.
	
	The different techniques described in Section \ref{sec:TransPipes} to solve the advection-reaction PDE \eqref{equ:PDE} for pipes need different approaches to include Equation \eqref{equ:mrNL}. First, when adopting one of the EFD discritization techniques (see Sections \ref{sec:LW}, \ref{sec:BackEuler}, and \ref{sec:CN}), the mutual reaction model in Equation \eqref{equ:mrNL} (i.e., ODEs) is discretized applying the forward Euler discrtezation method resulting in
	
	\begin{equation} \label{equ:mrL}
		\begin{aligned}
			c(t+\Delta t)-c(t) &=-k_r \Delta t(c(t) \tilde{c}(t)),\\
			\tilde{c}(t+\Delta t)-\tilde{c}(t) &=-k_r \Delta t(c(t)\tilde{c}(t)).
		\end{aligned}
	\end{equation}
	
	Similarly, discretized reaction model in Equation \eqref{equ:mrL} is used to formulate the reaction expression in tank Equation \eqref{equ:tank2} along with Equation \eqref{equ:decay_PTK}. Note that, for the techniques considered for pipes and for tanks they are discretized through the Eulerian discretization schemes in Sections \ref{sec:TransPipes} and \ref{sec:MB_Tank}. Thus, to avoid double discretizing, the left-hand-side of Equation \eqref{equ:mrL} is not reconsidered, and we have
	\begin{subequations} ~\label{equ:MR_PTK}
		\begin{align}
			R^{\mathrm{P}}_{\mathrm{M}}(c^\mathrm{P}_i(s,t))=-k_r c^\mathrm{P}_i(s,t) \tilde{c}^\mathrm{P}_i(s,t), \,\,\,\,\, R^{\mathrm{TK}}_{\mathrm{M}}(c^\mathrm{TK}_j(t))=-k_r \tilde{c}^\mathrm{TK}_j(t) c^\mathrm{TK}_j(t),\\
			R^{\mathrm{P}}_{\mathrm{M}}(\tilde{c}^\mathrm{P}_i(s,t))=-k_r c^\mathrm{P}_i(s,t) \tilde{c}^\mathrm{P}_i(s,t), \,\,\,\,\, R^{\mathrm{TK}}_{\mathrm{M}}(\tilde{c}^\mathrm{TK}_j(t))=-k_r \tilde{c}^\mathrm{TK}_j(t) c^\mathrm{TK}_j(t).
		\end{align}
	\end{subequations}
	
	Second, for pipes when applying the Lagrangian-MoCs technique explained in Section \ref{sec:MoCs} the following approach is followed. We start by showing that when the MoCs method is applied on a first-order decay model (Equation \eqref{equ:decay_PTK}) only, we get
	\begin{equation}~\label{MoC_FO}
		\zeta (s,t +\Delta t) = c^\mathrm{P}_i (s,t)  \exp(-k_i^\mathrm{P} \Delta t).
	\end{equation}
	However, the system of equations in \eqref{equ:mrNL} can be turned into a pseudo-first-order reaction model by assuming that the other chemical's concentration in the previous time-step is not variable. The limitation of this approach is to have one chemical's initial concentration much greater than the other's concentration. Yet, the desired difference between initial concentrations depends directly on the mutual reaction rate \cite{chapra2008surface, fisher2011suitability}.  To that end, similar to Equation \eqref{MoC_FO} we obtain
	\begin{equation}~\label{MoC_PFO}
		\begin{gathered}
			\zeta (s,t +\Delta t) = c^\mathrm{P}_i (s,t) \exp(-(k_i^\mathrm{P}+k_r \tilde{c}^\mathrm{P}_i (s,t)) \Delta t),\\
			\tilde{\zeta} (s,t +\Delta t) = \tilde{c}^\mathrm{P}_i (s,t)  \exp(-k_r c^\mathrm{P}_i (s,t)) \Delta t).
		\end{gathered}
	\end{equation}
	By applying linear interpolation at time-step $t+\Delta t$; see Fig. \ref{fig:MoCs}, concentrations at segment $s$ of Pipe $i$ at time-step $t+\Delta t$ are expressed as
	\begin{equation}~\label{Seg_MoC1}
		\begin{gathered}
			c^\mathrm{P}_i (s,t+\Delta t) = \frac{v_i(t) \Delta t}{\Delta x} \zeta (s-1,t +\Delta t) + (1-\frac{v_i(t) \Delta t}{\Delta x}) \zeta (s,t +\Delta t),\\
			\tilde{c}^\mathrm{P}_i (s,t+\Delta t) = \frac{v_i(t) \Delta t}{\Delta x} \tilde{\zeta} (s-1,t +\Delta t) + (1-\frac{v_i(t) \Delta t}{\Delta x}) \tilde{\zeta} (s,t +\Delta t).
		\end{gathered}
	\end{equation}
	Since Courant number $\tilde{\lambda}_i(t)$ equals ${v_{i}(t) \frac{\Delta t}{\Delta x_{i}}}$, and after substituting Equation \eqref{MoC_PFO} to Equation \eqref{Seg_MoC1}, we have
	\begin{equation}~\label{Seg_MoC2}
		\begin{gathered}
			c^\mathrm{P}_i (s,t+\Delta t) = \tilde{\lambda_i}(t) c^\mathrm{P}_i (s-1,t)  \exp(-(k_i^\mathrm{P}+k_r \tilde{c}^\mathrm{P}_i (s-1,t)) \Delta t) + (1-\tilde{\lambda_i}(t)) c^\mathrm{P}_i (s,t)  \exp(-(k_i^\mathrm{P}+k_r \tilde{c}^\mathrm{P}_i (s,t)) \Delta t),\\
			\tilde{c}^\mathrm{P}_i (s,t+\Delta t) = \tilde{\lambda_i}(t) \tilde{c}^\mathrm{P}_i (s-1,t) \exp(-k_r c^\mathrm{P}_i (s-1,t)) \Delta t) + (1-\tilde{\lambda_i}(t)) \tilde{c}^\mathrm{P}_i (s,t)  \exp(-k_r c^\mathrm{P}_i (s,t)) \Delta t)).
		\end{gathered}
	\end{equation}
	Note that in the Method of Characteristics, the first segment's concentration is taken as equal to the concentration of the upstream node. While the last segment's concentration is calculated depending on a virtual forward projection of its concentration at the previous time-step and the projection of the previous segment.
	
	\section{State-space Formulation}~\label{sec:SSForm}
	The multi-species water quality models have been analyzed in Sections \ref{sec:TransPipes}, \ref{sec:MassBal}, and \ref{sec:MutualReact} and the equations for WDN components have been derived. In this section, the detailed derivation of a state-space representation of the system based on these equations is shown in this section. We note that each technique listed in Section \ref{sec:TransPipes} for pipes leads to a different formulation. That is, the general formulation is expressed in Equation \eqref{equ:NDE} with $E_{11}(t)=E_{22}(t)=E(t)$. Moreover, $E(t)$ matrix is a non-singular matrix and invertible  allowing $x(t + \Delta t)$ to be calculated for every time-step according to~Equation~\eqref{equ:NDE_inv}. As such, Tab. \ref{tab:SSTechs} lists the formulation for each of the techniques which applies to any of the two chemicals. Differences in the techniques' approaches are reflected in some of the representation matrices --- highlighted in a different color for each technique.

	\begin{equation}~\label{equ:NDE_inv}
		\resizebox{.92\hsize}{!}{$	\begin{bmatrix}
				x_1(t+\Delta t) \\ x_2(t+\Delta t)
			\end{bmatrix} = 
			\begin{bmatrix}
				E'(t)A_{11}(t) & 0 \\ 0 & E'(t)A_{22}(t)
			\end{bmatrix}
			\begin{bmatrix}
				x_1(t) \\ x_2(t)
			\end{bmatrix} +
			\begin{bmatrix}
				E'(t)B_{11}(t) & 0 \\ 0 & E'(t)B_{22}(t)
			\end{bmatrix} 
			\begin{bmatrix}
				u_1(t) \\ u_2(t)
			\end{bmatrix}+ \begin{bmatrix}
				E'(t) & 0 \\ 0 & E'(t)
			\end{bmatrix} f(x_1,x_2,t).$}
	\end{equation}

	In these formulations, each of the chemicals $x_i(t)$ collects the concentrations in nodes and links of the network:
	\begin{eqnarray*}
		x_i(t) := \{ c_i^\mathrm{N}(t), c_i^\mathrm{L}(t) \}  =  \{ c_i^\mathrm{R}(t),   c_i^\mathrm{J}(t),   c_i^\mathrm{TK}(t),  c_i^\mathrm{M}(t),   c_i^\mathrm{P}(1,t), \dots , c_i^\mathrm{P}(s_\mathrm{L},t),  c_i^\mathrm{V}(t) \}, ~\label{equ:xstate}
	\end{eqnarray*}
	
	where $x(t) \in \mathbb{R}^{n_x}$; $n_x$ is the total number of nodes and links including pipes segments: $n_x=n_\mathrm{R}+n_\mathrm{J}+n_\mathrm{TK}+n_\mathrm{M}+n_{\mathrm{P}_\mathrm{s}}+n_\mathrm{V}$. 
	
	This state-space formulation is updated every water quality time-step within a hydraulic time-step to reflect parameters and variables change. However, the system dimensions depend on how many segments each pipe is divided into, the number of segments depends on the water quality time-step and velocity (refer to Equation \eqref{equ:Segment}) and the water quality time-step is fixed throughout the whole simulation. On the other hand, velocities change from hydraulic time-step to another for a system in a dynamic state (i.e., changing demands and flows). Consequently, in a dynamic system, with every hydraulic time-step applying \eqref{equ:Segment} gives a different number of segments for each pipe which leads to different system dimensions. 
	
	\begin{myrem}~\label{rm:Segments}
		For a specific network, model dimensions change with changing the hydraulic parameters and variables (e.g., demands, levels, diameters, etc.) leading to different concentrations evolution throughout the network.
	\end{myrem}
	
	Therefore, to be able to apply our model for an extended hydraulic period, we fix the number of segments for each pipe $i$ according to the following approach: 
	\begin{subequations}~\label{equ:ChooseSegment}
		\begin{align}
			& s_i=\Bigg\lfloor \frac{L_i}{\max_{t \in [0,T_s]}(v_i(t)) \Delta t} \Bigg\rfloor, \;\ \Delta x = \frac{L_i}{s_i},\\
			\intertext{for  $v_i(t) \leq \max_{t \in [0,T_s]}(v_i(t)),$} & 0 \leq \tilde{\lambda}_i(t)= {v_{i}(t) \frac{\Delta t}{\Delta x_{i}}} \leq 1.~\label{equ:ChooseSegment2}  
		\end{align}
	\end{subequations}
	
	To that end, the state-space formulation process for multi-species water quality dynamics in WDN is described in Algorithm \ref{alg:FormNDE}. WDN topology and hydraulics throughout the simulation period are the algorithm inputs with an initialization step that constructs the numerical grid in time and space. Afterwards, within every hydraulic time-step parameters are updated and the state-space representation is formulated for each water quality time-step according to Tab. \ref{tab:SSTechs} and the governing equations in Sections \ref{sec:TransPipes}, \ref{sec:MassBal}, and \ref{sec:reaction}.

	\begin{algorithm}[h]
		\small	\DontPrintSemicolon
		\caption{State-space formulation for multi-species water quality dynamics in WDN \label{alg:FormNDE}}
		\KwIn{WDN topology, components' charachtristics, and hydraulics parameters (velocities, flow rates, and flow directions)}
		\KwOut{State-space representation for the WDN for Techniques 1, 2, 3, and 4} 
		\Init{}{ Choose $\Delta t$ \;
			For each pipe $i$ in set $\mathcal{P}$, calculate $s_i$ and $\Delta x_i$ by applying \eqref{equ:ChooseSegment} \;
			Obtain initial concentrations for both chemicals and reaction rates
		}
		\ForEach{Hydraulic time-step ($t_\mathrm{H}$)}{
			Run hydraulic simulation and update velocities and flow rates  \;
			Update flow directions for each pipe and order segments accordingly \;
			Update Courant number $\tilde{\lambda}_i(t)$ \eqref{equ:ChooseSegment2} for set $\mathcal{P}$ \;
			\ForEach{$t \leq t_\mathrm{H}$}{ 
				\For{both chemicals}{
					Construct $A^\mathrm{R}_\mathrm{M}, A^\mathrm{J}_\mathrm{M}$ by applying \eqref{equ:pump} for set $\mathcal{M}$ \;
					Construct $A^\mathrm{J}_\mathrm{V}, A^\mathrm{TK}_\mathrm{V} $ by applying \eqref{equ:valve} for set $\mathcal{V}$ \;
					Construct $A^\mathrm{J}_\mathrm{J}, A^\mathrm{M}_\mathrm{J}, A^\mathrm{V}_\mathrm{J}, A^\mathrm{P}_\mathrm{J}$ and $B_\mathrm{J}$ by applying \eqref{equ:mb-junc} for set $\mathcal{J}$ \;
					Obtain reaction trems reflecting decay $R^{\mathrm{TK}}_{\mathrm{D}}$ and $R^{\mathrm{P}}_{\mathrm{D}}$ via \eqref{equ:decayreactionA} for chlorine and no decay for the other reactant \;
					Construct $A^\mathrm{TK}_\mathrm{TK}, A^\mathrm{P}_\mathrm{TK}$, and $B_\mathrm{TK}$ by applying \eqref{equ:tank2} for set $\mathcal{T}$ \;
					Construct $f(c^\mathrm{TK}(t),\tilde{c}^\mathrm{TK}(t))$ via \eqref{equ:MR_PTK} for set $\mathcal{T}$ \;
					\uIf{Applying Lax-Wendroff scheme (Tech. 1)
					}{Construct $A^\mathrm{R}_\mathrm{P}, A^\mathrm{J}_\mathrm{P}, A^\mathrm{TK}_\mathrm{P},$ and  $A^\mathrm{P}_\mathrm{P}$ by applying \eqref{equ:LW_Si} and \eqref{equ:node2pipe} for set $\mathcal{P}$ \; 
						Construct $f(c^\mathrm{P}(t),\tilde{c}^\mathrm{P}(t))$ via \eqref{equ:MR_PTK} for set $\mathcal{P}$}
					\uElseIf{Applying Backward Euler scheme (Tech. 2)}{Construct $E^\mathrm{R}_\mathrm{P}, E^\mathrm{J}_\mathrm{P}, E^\mathrm{TK}_\mathrm{P}, E^\mathrm{P}_\mathrm{P}$ and  $A^\mathrm{P}_\mathrm{P}$ by applying \eqref{equ:BE_Si2} for set $\mathcal{P}$ \; 
						Construct $f(c^\mathrm{P}(t),\tilde{c}^\mathrm{P}(t))$ via \eqref{equ:MR_PTK} for set $\mathcal{P}$}
					\uElseIf{Applying Crank-Nicolson scheme (Tech. 3)}{Construct $E^\mathrm{R}_\mathrm{P}, E^\mathrm{J}_\mathrm{P}, E^\mathrm{TK}_\mathrm{P}, E^\mathrm{P}_\mathrm{P}, A^\mathrm{R}_\mathrm{P}, A^\mathrm{J}_\mathrm{P}, A^\mathrm{TK}_\mathrm{P},$ and  $A^\mathrm{P}_\mathrm{P}$ by applying \eqref{equ:CN_Si} for set $\mathcal{P}$\; 
						Construct $f(c^\mathrm{P}(t),\tilde{c}^\mathrm{P}(t))$ via \eqref{equ:MR_PTK} for set $\mathcal{P}$}
					\Else{\textit{Applying Method of Characteristics (Tech. 4)}: \; 
						Construct $A^\mathrm{R}_\mathrm{P}, A^\mathrm{J}_\mathrm{P}, A^\mathrm{TK}_\mathrm{P},$ and  $A^\mathrm{P}_\mathrm{P}$ by applying \eqref{Seg_MoC2} for set $\mathcal{P}$
					}
				}
				Formulate the state-space representation according to Tab. \ref{tab:SSTechs}	\;
				Calculate concentrations for both chemicals by applying state-space representation \eqref{equ:NDE_inv} \;
				$t=t+\Delta t$
			} 
		}
	\end{algorithm}
	
	\begin{landscape}
		\begin{table}[h!]
			{\small\begin{threeparttable}
					\setlength{\tabcolsep}{0.4pt}
					\caption{State-space representation with different advection-reaction discretization techniques\tnote{a}}~\label{tab:SSTechs}
					\begin{tabular}{c|c c c c c c c c c c}
						\hline
						Tech. & $E(t)$ & $x(t +\Delta t)$ & $=$ & $A(t)$ & $x(t)$ & $+$ &  $B(t)$ & $u(t)$ & $+$ &   $f(x_1,x_2,t)$ \\
						\hline
						& & & & & & & & & & \\
						1 & $I_{n_x}$ & $x(t +\Delta t)$ & $=$ & $\begin{bmatrix} \begin{array}{c c c | c c c}
								I_{n_\mathrm{R}} & &  \\
								& A^\mathrm{J}_\mathrm{J} & & A^\mathrm{M}_\mathrm{J} & A^\mathrm{V}_\mathrm{J} & A^\mathrm{P}_\mathrm{J} \\
								& & A^\mathrm{TK}_\mathrm{TK} & & & A^\mathrm{P}_\mathrm{TK} \\
								\hline
								A^\mathrm{R}_\mathrm{M} & A^\mathrm{J}_\mathrm{M} & & & & \\
								& A^\mathrm{J}_\mathrm{V} & A^\mathrm{TK}_\mathrm{V} \\
								\tikzmarknode{Apr}{\highlight{red}{$A^\mathrm{R}_\mathrm{P}$}} & \tikzmarknode{Apj}{\highlight{red}{$A^\mathrm{J}_\mathrm{P}$}} & 
								\tikzmarknode{Aptk}{\highlight{red}{$A^\mathrm{TK}_\mathrm{P}$}}
								& & &
								\tikzmarknode{App}{\highlight{red}{$A^\mathrm{P}_\mathrm{P}$}}   
							\end{array}
						\end{bmatrix} $ & 
						$\begin{bmatrix}
							\begin{array}{c}
								c^\mathrm{R} \\
								c^\mathrm{J} \\
								c^\mathrm{TK} \\
								\hline
								c^\mathrm{M} \\
								c^\mathrm{V} \\
								c^\mathrm{P}
							\end{array}
						\end{bmatrix} $ & $+$ &  
						$\begin{bmatrix}
							\begin{array}{c c c}
								O_{n_\mathrm{R}} & & \\
								& B_\mathrm{J} & \\
								& & B_\mathrm{TK} \\
								\hline
								\mathrm{} \\
								\multicolumn{3}{c}{O_{n_\mathrm{L} \times n_\mathrm{N}}}\\
								\mathrm{}
							\end{array}
						\end{bmatrix} $ &
						$\begin{bmatrix}
							c^{\mathrm{B}_\mathrm{R}} \\
							c^{\mathrm{B}_\mathrm{J}} \\
							c^{\mathrm{B}_\mathrm{TK}}
						\end{bmatrix}$ & $+$ & 
						$\begin{bmatrix}
							\begin{array}{c}
								O_{n_\mathrm{R}} \\
								O_{n_\mathrm{J}} \\
								f(c^\mathrm{TK}(t),\tilde{c}^\mathrm{TK}(t))\\
								\hline
								O_{n_\mathrm{M}} \\
								O_{n_\mathrm{V}} \\
								\tikzmarknode{fb}{\highlight{red}{$f(c^\mathrm{P}(t),\tilde{c}^\mathrm{P}(t))$}}
							\end{array}
						\end{bmatrix}$ \\ 
						& & & & & & & & & & \\ 
						\hline
						& & & & & & & & & & \\
						2 & $\begin{bmatrix} \begin{array}{c c c|c c c}
								& & &  \\
								\multicolumn{3}{c|} {I_{n_\mathrm{L} \times n_\mathrm{L}}} & \multicolumn{3}{c} {O_{n_\mathrm{L} \times n_\mathrm{N}}} \\
								& & & & & \\
								\hline
								& & & I_{n_\mathrm{M}} & & \\
								& & & & I_{n_\mathrm{V}} & \\
								\tikzmarknode{Epj}{\highlight{blue}{$E^\mathrm{R}_\mathrm{P}$}} & \tikzmarknode{Epj}{\highlight{blue}{$E^\mathrm{J}_\mathrm{P}$}} & \tikzmarknode{Eptk}{\highlight{blue}{$E^\mathrm{TK}_\mathrm{P}$}} & & & \tikzmarknode{Epp}{\highlight{blue}{$E^\mathrm{P}_\mathrm{P}$}}   
							\end{array}
						\end{bmatrix}$ & $x(t +\Delta t)$ & $=$ & $\begin{bmatrix} \begin{array}{c c c | c c c}
								I_{n_\mathrm{R}} & &  \\
								& A^\mathrm{J}_\mathrm{J} & & A^\mathrm{M}_\mathrm{J} & A^\mathrm{V}_\mathrm{J} & A^\mathrm{P}_\mathrm{J} \\
								& & A^\mathrm{TK}_\mathrm{TK} & & & A^\mathrm{P}_\mathrm{TK} \\
								\hline
								A^\mathrm{R}_\mathrm{M} & A^\mathrm{J}_\mathrm{M} & & & & \\
								& A^\mathrm{J}_\mathrm{V} & A^\mathrm{TK}_\mathrm{V} \\
								&  &  & & & \tikzmarknode{App}{\highlight{blue}{$A^\mathrm{P}_\mathrm{P}$}}    
							\end{array}
						\end{bmatrix} $ & 
						$\begin{bmatrix}
							\begin{array}{c}
								c^\mathrm{R} \\
								c^\mathrm{J} \\
								c^\mathrm{TK} \\
								\hline
								c^\mathrm{M} \\
								c^\mathrm{V} \\
								c^\mathrm{P}
							\end{array}
						\end{bmatrix} $ & $+$ &  
						$\begin{bmatrix}
							\begin{array}{c c c}
								O_{n_\mathrm{R}} & & \\
								& B_\mathrm{J} & \\
								& & B_\mathrm{TK} \\
								\hline
								\mathrm{} \\
								\multicolumn{3}{c}{O_{n_\mathrm{L} \times n_\mathrm{N}}} \\
								\mathrm{}
							\end{array}
						\end{bmatrix} $ &
						$\begin{bmatrix}
							c^{\mathrm{B}_\mathrm{R}} \\
							c^{\mathrm{B}_\mathrm{J}} \\
							c^{\mathrm{B}_\mathrm{TK}}
						\end{bmatrix}$ & $+$ & 
						$\begin{bmatrix}
							\begin{array}{c}
								O_{n_\mathrm{R}} \\
								O_{n_\mathrm{J}} \\
								f(c^\mathrm{TK}(t),\tilde{c}^\mathrm{TK}(t))\\
								\hline
								O_{n_\mathrm{M}} \\
								O_{n_\mathrm{V}} \\
								\tikzmarknode{fb}{\highlight{blue}{$f(c^\mathrm{P}(t),\tilde{c}^\mathrm{P}(t))$}}
							\end{array}
						\end{bmatrix}$ \\
						& & & & & & & & & & \\
						\hline
						& & & & & & & & & & \\
						3 & $\begin{bmatrix} \begin{array}{c c c|c c c}
								& & &  \\
								\multicolumn{3}{c|} {I_{n_\mathrm{L} \times n_\mathrm{L}}} & \multicolumn{3}{c} {O_{n_\mathrm{L} \times n_\mathrm{N}}} \\
								& & & & & \\
								\hline
								& & & I_{n_\mathrm{M}} & & \\
								& & & & I_{n_\mathrm{V}} & \\
								\tikzmarknode{Epj}{\highlight{green}{$E^\mathrm{R}_\mathrm{P}$}} & \tikzmarknode{Epj}{\highlight{green}{$E^\mathrm{J}_\mathrm{P}$}} & \tikzmarknode{Eptk}{\highlight{green}{$E^\mathrm{TK}_\mathrm{P}$}} & & & \tikzmarknode{Epp}{\highlight{green}{$E^\mathrm{P}_\mathrm{P}$}}
							\end{array}
						\end{bmatrix}$ & $x(t +\Delta t)$ & $=$ & $\begin{bmatrix} \begin{array}{c c c | c c c}
								I_{n_\mathrm{R}} & &  \\
								& A^\mathrm{J}_\mathrm{J} & & A^\mathrm{M}_\mathrm{J} & A^\mathrm{V}_\mathrm{J} & A^\mathrm{P}_\mathrm{J} \\
								& & A^\mathrm{TK}_\mathrm{TK} & & & A^\mathrm{P}_\mathrm{TK} \\
								\hline
								A^\mathrm{R}_\mathrm{M} & A^\mathrm{J}_\mathrm{M} & & & & \\
								& A^\mathrm{J}_\mathrm{V} & A^\mathrm{TK}_\mathrm{V} \\
								\tikzmarknode{Apr}{\highlight{green}{$A^\mathrm{R}_\mathrm{P}$}} & \tikzmarknode{Apj}{\highlight{green}{$A^\mathrm{J}_\mathrm{P}$}} & 
								\tikzmarknode{Aptk}{\highlight{green}{$A^\mathrm{TK}_\mathrm{P}$}}
								& & &
								\tikzmarknode{App}{\highlight{green}{$A^\mathrm{P}_\mathrm{P}$}}  
							\end{array}
						\end{bmatrix} $ & 
						$\begin{bmatrix}
							\begin{array}{c}
								c^\mathrm{R} \\
								c^\mathrm{J} \\
								c^\mathrm{TK} \\
								\hline
								c^\mathrm{M} \\
								c^\mathrm{V} \\
								c^\mathrm{P}
							\end{array}
						\end{bmatrix} $ & $+$ &  
						$\begin{bmatrix}
							\begin{array}{c c c}
								O_{n_\mathrm{R}} & & \\
								& B_\mathrm{J} & \\
								& & B_\mathrm{TK} \\
								\hline
								\mathrm{} \\
								\multicolumn{3}{c}{O_{n_\mathrm{L} \times n_\mathrm{N}}} \\
								\mathrm{}
							\end{array}
						\end{bmatrix} $ &
						$\begin{bmatrix}
							c^{\mathrm{B}_\mathrm{R}} \\
							c^{\mathrm{B}_\mathrm{J}} \\
							c^{\mathrm{B}_\mathrm{TK}}
						\end{bmatrix}$ & $+$ & 
						$\begin{bmatrix}
							\begin{array}{c}
								O_{n_\mathrm{R}} \\
								O_{n_\mathrm{J}} \\
								f(c^\mathrm{TK}(t),\tilde{c}^\mathrm{TK}(t))\\
								\hline
								O_{n_\mathrm{M}} \\
								O_{n_\mathrm{V}} \\
								\tikzmarknode{fb}{\highlight{green}{$f(c^\mathrm{P}(t),\tilde{c}^\mathrm{P}(t))$}}
							\end{array}
						\end{bmatrix}$ \\
						& & & & & & & & & & \\
						\hline
						& & & & & & & & & & \\
						4 & $I_{n_x}$ & $x(t +\Delta t)$ & $=$ & $\begin{bmatrix} \begin{array}{c c c | c c c}
								I_{n_\mathrm{R}} & &  \\
								& A^\mathrm{J}_\mathrm{J} & & A^\mathrm{M}_\mathrm{J} & A^\mathrm{V}_\mathrm{J} & A^\mathrm{P}_\mathrm{J} \\
								& & A^\mathrm{TK}_\mathrm{TK} & & & A^\mathrm{P}_\mathrm{TK} \\
								\hline
								A^\mathrm{R}_\mathrm{M} & A^\mathrm{J}_\mathrm{M} & & & & \\
								& A^\mathrm{J}_\mathrm{V} & A^\mathrm{TK}_\mathrm{V} \\
								\tikzmarknode{Apr}{\highlight{yellow}{$A^\mathrm{R}_\mathrm{P}$}} & \tikzmarknode{Apj}{\highlight{yellow}{$A^\mathrm{J}_\mathrm{P}$}} & 
								\tikzmarknode{Aptk}{\highlight{yellow}{$A^\mathrm{TK}_\mathrm{P}$}}
								& & &
								\tikzmarknode{App}{\highlight{yellow}{$A^\mathrm{P}_\mathrm{P}$}}   
							\end{array}
						\end{bmatrix} $ & 
						$\begin{bmatrix}
							\begin{array}{c}
								c^\mathrm{R} \\
								c^\mathrm{J} \\
								c^\mathrm{TK} \\
								\hline
								c^\mathrm{M} \\
								c^\mathrm{V} \\
								c^\mathrm{P}
							\end{array}
						\end{bmatrix} $ & $+$ &  
						$\begin{bmatrix}
							\begin{array}{c c c}
								O_{n_\mathrm{R}} & & \\
								& B_\mathrm{J} & \\
								& & B_\mathrm{TK} \\
								\hline
								\mathrm{} \\
								\multicolumn{3}{c}{O_{n_\mathrm{L} \times n_\mathrm{N}}} \\
								\mathrm{}
							\end{array}
						\end{bmatrix} $ &
						$\begin{bmatrix}
							c^{\mathrm{B}_\mathrm{R}} \\
							c^{\mathrm{B}_\mathrm{J}} \\
							c^{\mathrm{B}_\mathrm{TK}}
						\end{bmatrix}$ & $+$ & 
						$\begin{bmatrix}
							\begin{array}{c}
								O_{n_\mathrm{R}} \\
								O_{n_\mathrm{J}} \\
								f(c^\mathrm{TK}(t),\tilde{c}^\mathrm{TK}(t))\\
								\hline
								O_{n_\mathrm{M}} \\
								O_{n_\mathrm{V}} \\
								\tikzmarknode{Op}{\highlight{yellow}{$O_{n_\mathrm{P}}$}} 	
							\end{array}
						\end{bmatrix}$ \\
						& & & & & & & & & & \\
						\hline
						\hline
					\end{tabular}
					\begin{tablenotes}
						\item[a] Tech. (1) Lax-Wendroff scheme, (2) Backward Euler scheme, (3) Crank-Nicolson scheme, and (4) Method of Characteristics.
					\end{tablenotes}
			\end{threeparttable}}
		\end{table}
	\end{landscape}

	\section{Case Studies}~\label{sec:Case-Studies}
\begin{figure}[t] 
		\centering
		\includegraphics[width=0.9\linewidth]{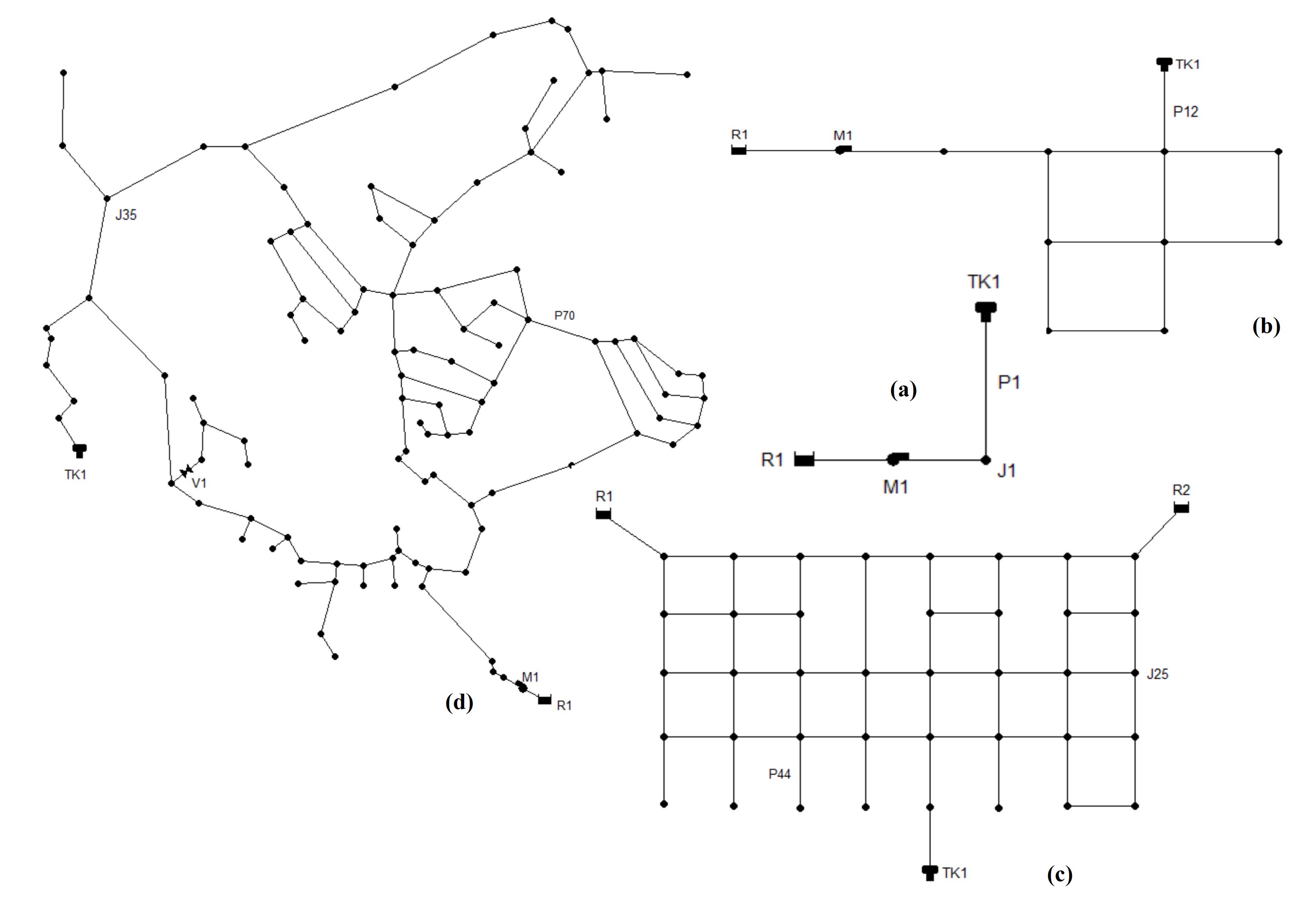}
		\caption{Case studies' layouts: (a) Three-node network, (b) Net1, (c) Net2, and (d) FFCL-1.}
		\label{fig:Networks}
	\end{figure}

	In this section,  several networks are used to validate the accuracy of the proposed two-species water quality dynamics and compare the results of different discretization schemes and their applicability. Four networks are considered, three-node, Net1, Net2, and FFCL-1 \cite{rossman2020epanet}. Note that the FFCL-1 network is modified accordingly based on the topology from EPANET to include a reservoir, a pump, and a valve. The topologies of tested networks are illustrated in Fig. \ref{fig:Networks} and the details of their components are listed in Tab. \ref{tab:NetComp}. The Three-node network is used as an example of how to build and apply our state-space formulation for all the techniques due to the simple and clear topology it has. Net1 and Net2 are used to show how the inclusion of a fictitious reactant affects chlorine concentrations and compare results from applying all techniques, and the scalability of the proposed method is tested on the FFCL-1 network.
	
	\begin{table}[h!]
		\centering
		\caption{Test networks components}~\label{tab:NetComp}
		{\small\begin{tabular}{c|c|c|c|c|c|c}
				\hline
				\multirow{2}{*}{Network} & \multicolumn{6}{c}{Components Count} \\
				\cline{2-7}
				& Reservoir & Tank & Junction & Pipe & Pump & Valve \\
				\hline
				Three-node & 1 & 1 & 1 & 1 & 1 & 0 \\
				\hline 
				Net1 & 1 & 1 & 9 & 12 & 1 & 0 \\
				\hline 
				Net2 & 2 & 1 & 39 & 60 & 0 & 0 \\
				\hline 
				FFCL-1 & 1 & 1 & 108 & 121 & 1 & 1 \\ 
				\hline
				\hline
		\end{tabular}}
	\end{table}

	\subsection{State-space formulation illustration using Three-node network} \label{sec:SS-Ex}
	
	In this section, the details of how the state-space formulation is derived are given with the Three-node network that has a simple topology consisting of  four components (i.e., a reservoir, a pump, a junction, a pipe, and a tank) as listed in Tab. \ref{tab:NetComp}. We formulate the state-space representation using the Lax-Wendroff scheme (Tech. \#1) explained in Section \ref{sec:LW} and Appendix \ref{App:SS-Ex} covers the formulation for the rest of the techniques. We follow the approach of Algorithm \ref{alg:FormNDE} assuming that the initialization step has been carried out.  
	
	We assume that Pipe $\mathrm{P}1$ of Three-node network shown in Figure \ref{fig:Networks} is divided into three segments $s_1=3$ to help readers understand our method; flow $q^\mathrm{M}_1(t)$ is pumped from Reservoir $\mathrm{R}1$ to Junction $\mathrm{J}1$ by Pump $\mathrm{M}1$; flow in $\mathrm{P}1$ is $q^\mathrm{P}_1(t)$; and demand at $\mathrm{J}1$ is $q^{\mathrm{D}_\mathrm{J}}_1(t)$. One chlorine booster station is located at $\mathrm{TK}1$ injecting volume of $V^\mathrm{B}(t+\Delta t)$ with concentration $c^\mathrm{B_\mathrm{TK}}(t + \Delta t)$. In this water quality model, we consider $u_1(t) = c^{\mathrm{B}_\mathrm{TK}}(t + \Delta t)$. This can be explained physically as the amount of chlorine to be injected in the time step we are calculating at. 
	
	Tab. \ref{tab:WQ-3n} summarizes the governing equations for two species that are chlorine and fictitious reactant. That is, the following state-space representation is formulated for $x_1$ as the vector of chlorine concentrations and $x_2$ as the vector of fictitious component concentrations.
	
	\input{Example.tex}

	\begin{table}[h!]
		\centering
		{\small\begin{threeparttable}
				\caption{Water quality model equations for the three-node example\tnote{a}}
				\begin{tabular}{c|c|c}
					\hline
					Network Component     &  Chlorine & Fictitious Reactive Component \\
					\hline
					$\mathrm{R}1$    & \multicolumn{2}{c}{$c_1^\mathrm{R}(t+\Delta t)=c_1^\mathrm{R}(t)$}\\
					\hline
					$\mathrm{M}1$    & \multicolumn{2}{c}{$c_1^{\mathrm{M}}(t+\Delta t) = c_1^{\mathrm{R}}(t+\Delta t)=c_1^\mathrm{R}(t)$}\\
					\hline
					$\mathrm{J}1$    & \multicolumn{2}{c}{$c_1^\mathrm{J}(t+ \Delta t)= \frac{q^\mathrm{M}_{1}(t+ \Delta t) c^\mathrm{M}_{1}(t+ \Delta t)}{q^{\mathrm{D}_\mathrm{J}}_1(t+ \Delta t)+q^\mathrm{P}_{1}(t+ \Delta t)}=\frac{q^\mathrm{M}_{1}(t+ \Delta t) c^\mathrm{R}_{1}(t)}{q^{\mathrm{D}_\mathrm{J}}_1(t+ \Delta t)+q^\mathrm{P}_{1}(t+ \Delta t)}$}\\
					\hline
					\multirow{3}{*}{$\mathrm{P}1$\tnote{b}}    & \multicolumn{2}{c}{$c^\mathrm{P}_1(1,t+ \Delta t) = \underline{\lambda}_1(t) c^\mathrm{J}_1(t)+\lambda_{1}(t) c^\mathrm{P}_1(1,t)+\overline{\lambda}_1(t) c^\mathrm{P}_1(2,t)+R^\mathrm{P}_{\mathrm{MS}}(c^\mathrm{P}_i(1,t))\Delta t$}\\
					& \multicolumn{2}{c}{$c^\mathrm{P}_1(2,t+ \Delta t)= \underline{\lambda}_1(t) c^\mathrm{P}_1(1,t)+\lambda_1(t) c^\mathrm{P}_1(2,t)+\overline{\lambda}_1(t) c^\mathrm{P}_1(3,t)+R^\mathrm{P}_{\mathrm{MS}}(c^\mathrm{P}_1(2,t))\Delta t$}\\
					& \multicolumn{2}{c}{$c^\mathrm{P}_1(3,t+ \Delta t)= \underline{\lambda}_1(t) c^\mathrm{P}_1(2,t)+\lambda_{1}(t) c^\mathrm{P}_1(3,t)+\overline{\lambda}_1(t) c^\mathrm{TK}_1(t)+R^\mathrm{P}_{\mathrm{MS}}(c^\mathrm{P}_1(3,t))\Delta t$}\\
					\hline
					$\mathrm{TK}1$\tnote{b}   & $\begin{aligned}
						& V_1^\mathrm{TK}(t + \Delta t) c_1^\mathrm{TK}(t+ \Delta t)= V_1^\mathrm{TK}(t) c_1^\mathrm{TK}(t) \\  & +q^\mathrm{P}_1(t)c^\mathrm{P}_1(3,t) \Delta t +V^\mathrm{B}(t+\Delta t)c^{\mathrm{B}_\mathrm{TK}}(t+\Delta t) \\
						& 					+R^\mathrm{TK}_{\mathrm{MS}}(c_1^\mathrm{TK}(t)) V_1^\mathrm{TK}(t) \Delta t
					\end{aligned}$
					& $\begin{aligned}
						& V_1^\mathrm{TK}(t + \Delta t) \tilde{c}_1^\mathrm{TK}(t+ \Delta t)= V_1^\mathrm{TK}(t) \tilde{c}_1^\mathrm{TK}(t) \\  & +q^\mathrm{P}_1(t)\tilde{c}^\mathrm{P}_1(3,t) \Delta t +R^\mathrm{TK}_{\mathrm{MS}}(\tilde{c}_1^\mathrm{TK}(t)) V_1^\mathrm{TK}(t) \Delta t 
					\end{aligned}$\\
					\hline
					\hline
				\end{tabular}
				\label{tab:WQ-3n}
				\begin{tablenotes}
					\item[a] For equations applicable for both chemicals, $c$ is the concentration for any of the two species. Otherwise, $c$ is chlorine concentration while $\tilde{c}$ is fictitious reactant concentration.
					\item[b] Reaction expressions are according to Equations \eqref{equ:decay_PTK} and \eqref{equ:MR_PTK}.
				\end{tablenotes}
		\end{threeparttable}}
	\end{table}
	
	\newpage
	\subsection{NDE model implementation with different discretization techniques, validity, and complexity}~\label{sec:validity}
	In this section, we test the state-space representation formulated using the considered discretization techniques on different case studies. Then, we compare the results with the ones obtained from EPANET and its multi-species simulation extension EPANET-MSX. All parameters are matched for our NDE and EPANET to have a reliable comparison. Codes are run on MATLAB and EPANET results are obtained using its toolkit. Bare in mind, EPANET-MSX is not yet inherent in EPANET software. One way to run it is on the Command Prompt in Windows, which results in a form of \text{.rpt}-extended file. This file extension can be opened by any text editor. Yet, to do any further analysis/modeling, the procedure to read the results needs to follow a certain complex format. Another way to run it is through the toolkit on MATLAB. Both approaches take seconds to run the model. However, using the toolkit to get these results takes a longer time. To that end, the run time results listed in the following subsections are the actual total time needed to run and extract the results of the simulation. On the other hand, according to Remark \ref{rm:Segments} the dimension of the model depends on the hydraulics within the system for the same network. Thus, we list the number of states in the model for each network simulation. 
	
	Additionally, pipes are divided into a number of segments with a fixed length per pipe, thus, an average concentration over pipe segments represents the simulation result for each pipe. The same approach is followed in EPANET-MSX but with variable-sized segments per pipe.
	
	\subsubsection{Multi-species model vs. single-species reaction and transport models}
	
	\begin{figure}[t!]
		\centering
		\subfloat[\label{fig:Net1LW_a}]{\includegraphics[keepaspectratio=true,scale=0.6]{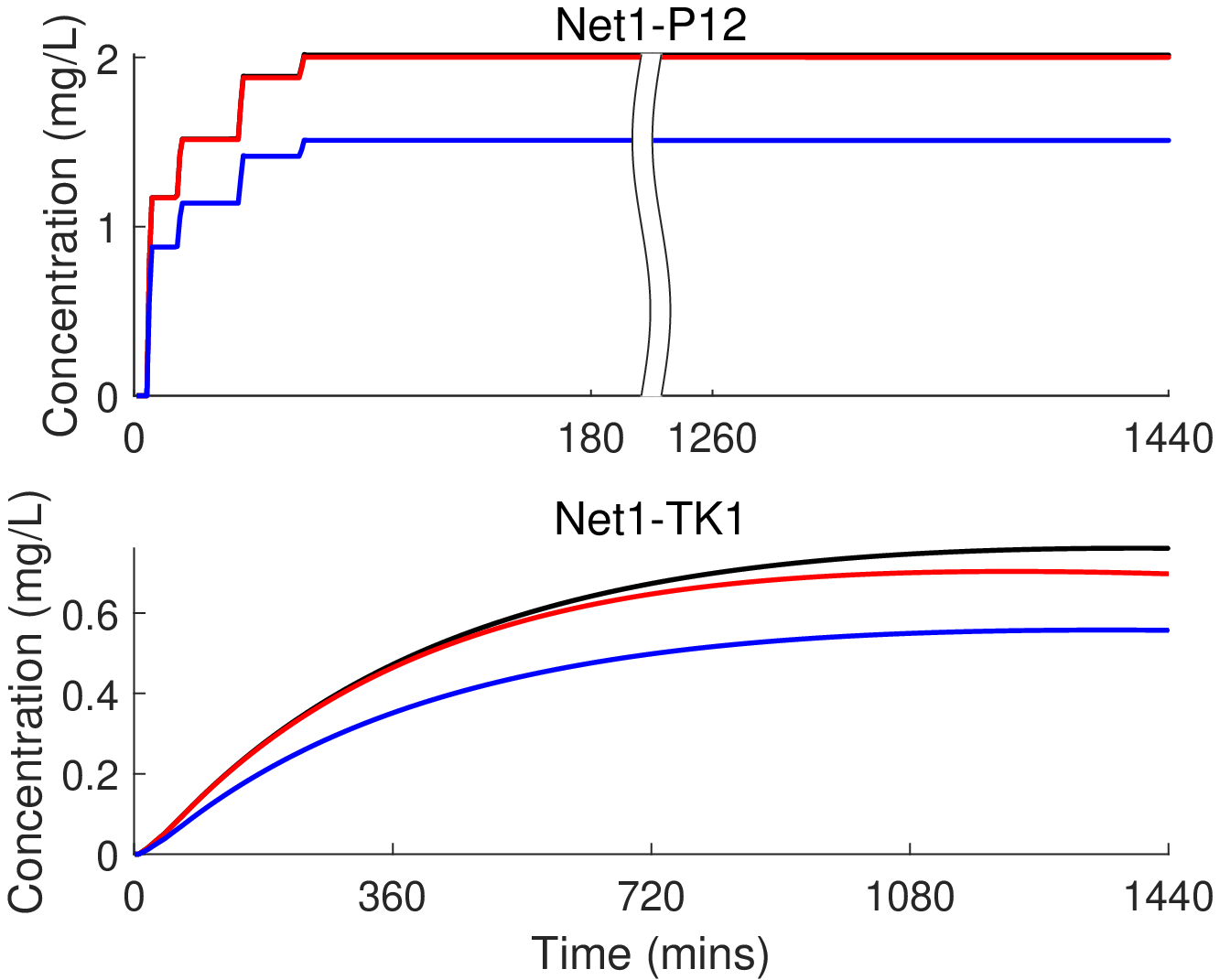}}{}\vspace{-0.1cm} \hspace{0.1cm}
		\subfloat[\label{fig:Net1LW_b}]{\includegraphics[keepaspectratio=true,scale=0.6]{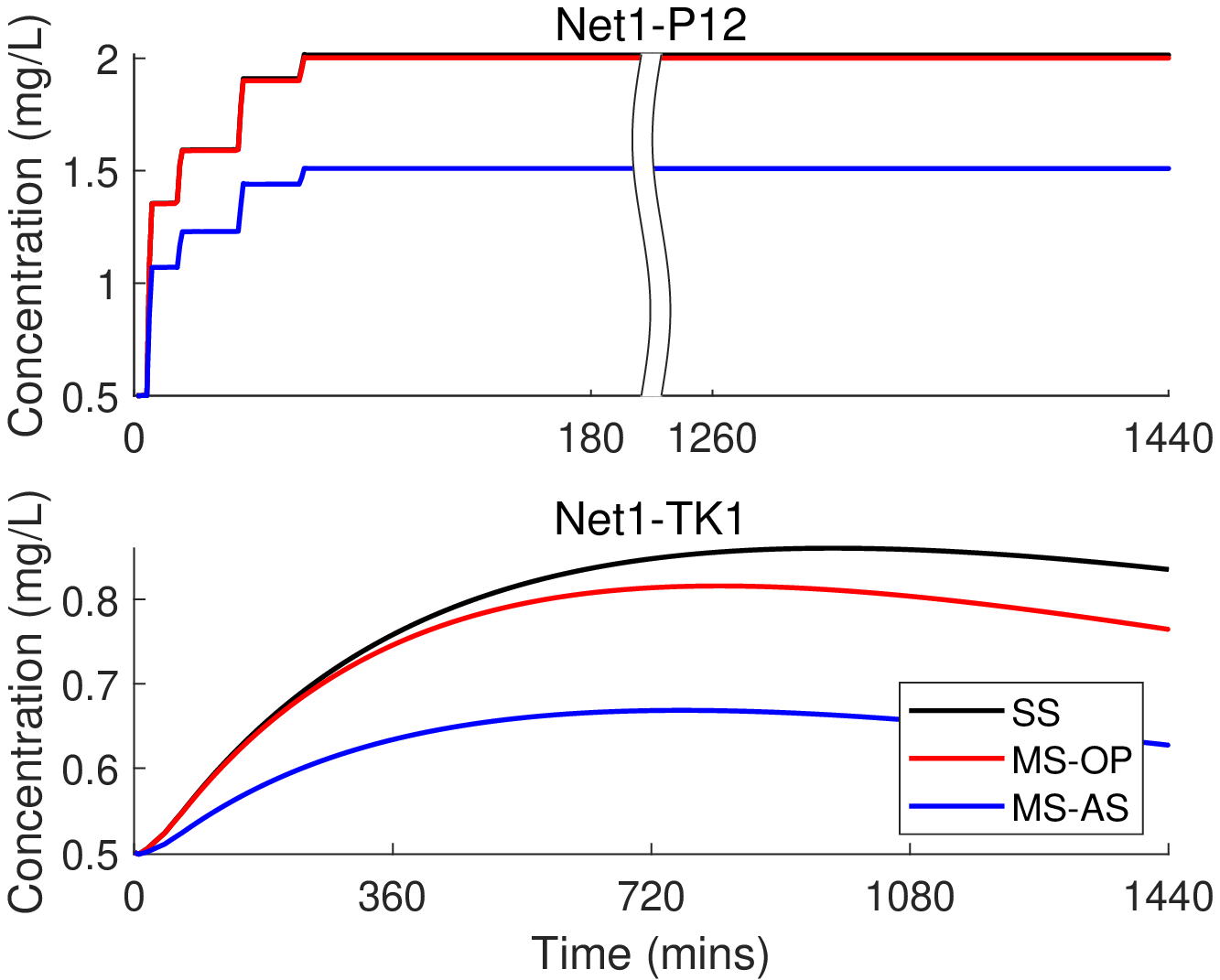}}{}\vspace{-0.1cm}\hspace{-0.1cm}
		\caption{Chlorine concentrations at Tank TK1 and Pipe P12 in Net1 in case of (a) zero initial concentrations for both chemicals (b) 0.5 mg/L chlorine concentration, in all components except for Reservoir R1 with source concentrations.\\
			{\scriptsize{*SS: Single-species model, MS-OP: Multi-species model with a fictitious reactant with slow reaction rate (e.g., pro-oxonic organo-phosphate), MS-AS: Multi-species model with a fictitious reactant with relatively rapid reaction rate (e.g., sodium arsenite). Multi-species model is simulated by applying Tech. \#1 - Lax-Wendroff scheme}}}
		\label{fig:Net1LW}
	\end{figure}
	
	Firstly, we showcase the impact of the inclusion of the fictitious reactant by simulating three cases using Lax-Wendroff scheme (Tech. \#1): no fictitious reactant included, a fictitious reactant with a slow reaction rate (e.g., pro-oxonic organo-phosphate), and a fictitious reactant with relatively rapid reaction rate (e.g., sodium arsenite) \cite{umberg2008performance}. Simulations are performed on Net1 a constant source of both chemicals at Reservoir R1 of 2 mg/L and 0.3 mg/L with chlorine and fictitious reactant, respectively. The water quality time-step is considered to be $\Delta t=1$ sec and pipes are divided into a different number of segments satisfying the stability condition. Moreover, we consider two scenarios for initial chlorine concentrations in networks components other than Reservoir R1, zeros (Fig. \ref{fig:Net1LW_a}) and 0.5 mg/L (Fig. \ref{fig:Net1LW_b}). Results from both scenarios confirm how chlorine concentrations can be affected by the presence of another substance. Furthermore, depending on the substance and its reaction rate with chlorine, it can consume chlorine slowly or rapidly.
	
	\subsubsection{Discretization schemes/techniques performance}
\begin{figure}[t]
		\centering
		\subfloat[\label{fig:Net1Rest_a}]{\includegraphics[keepaspectratio=true,scale=0.42]{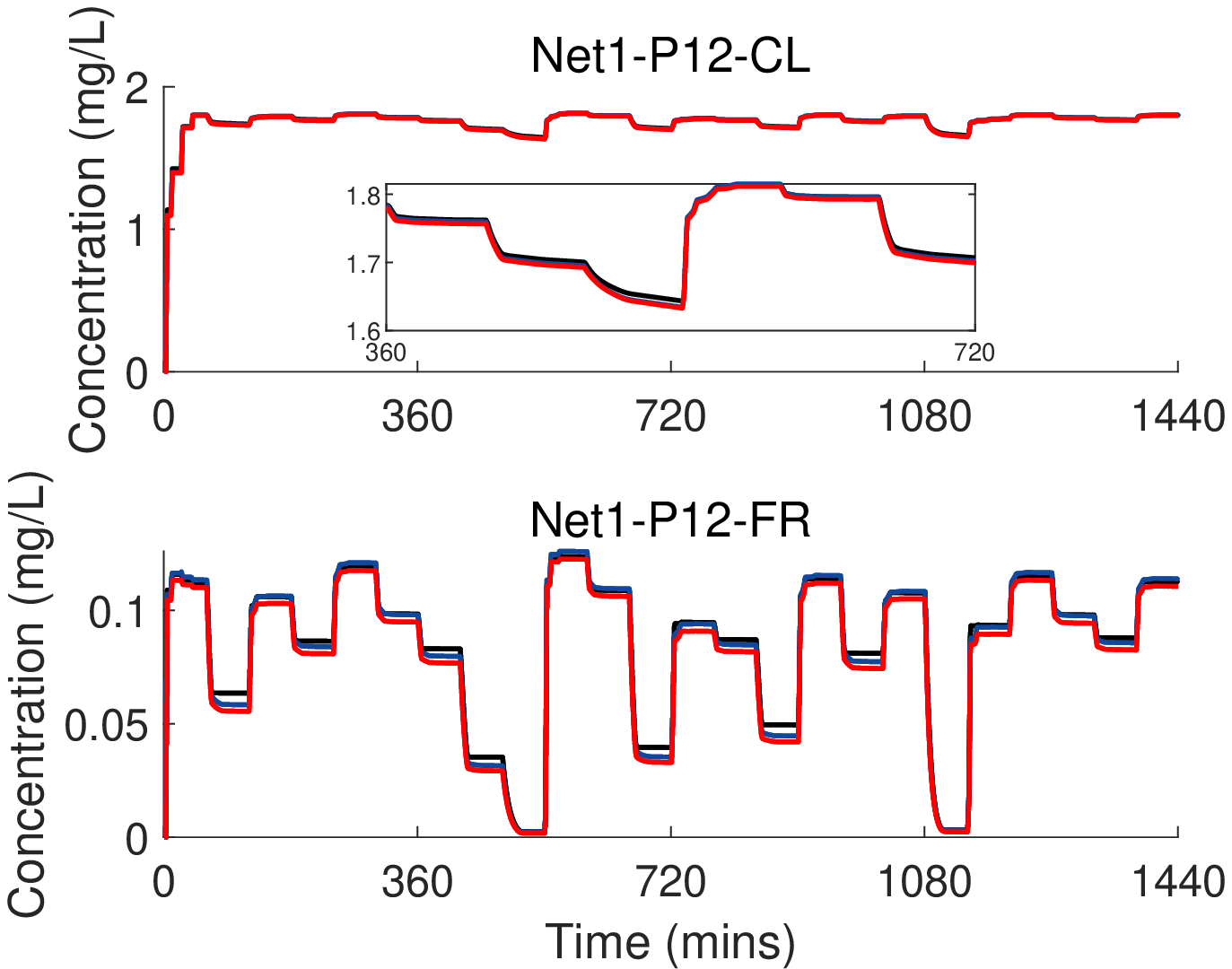}}{}\vspace{0cm} \hspace{0cm}
		\subfloat[\label{fig:Net1Rest_b}]{\includegraphics[keepaspectratio=true,scale=0.42]{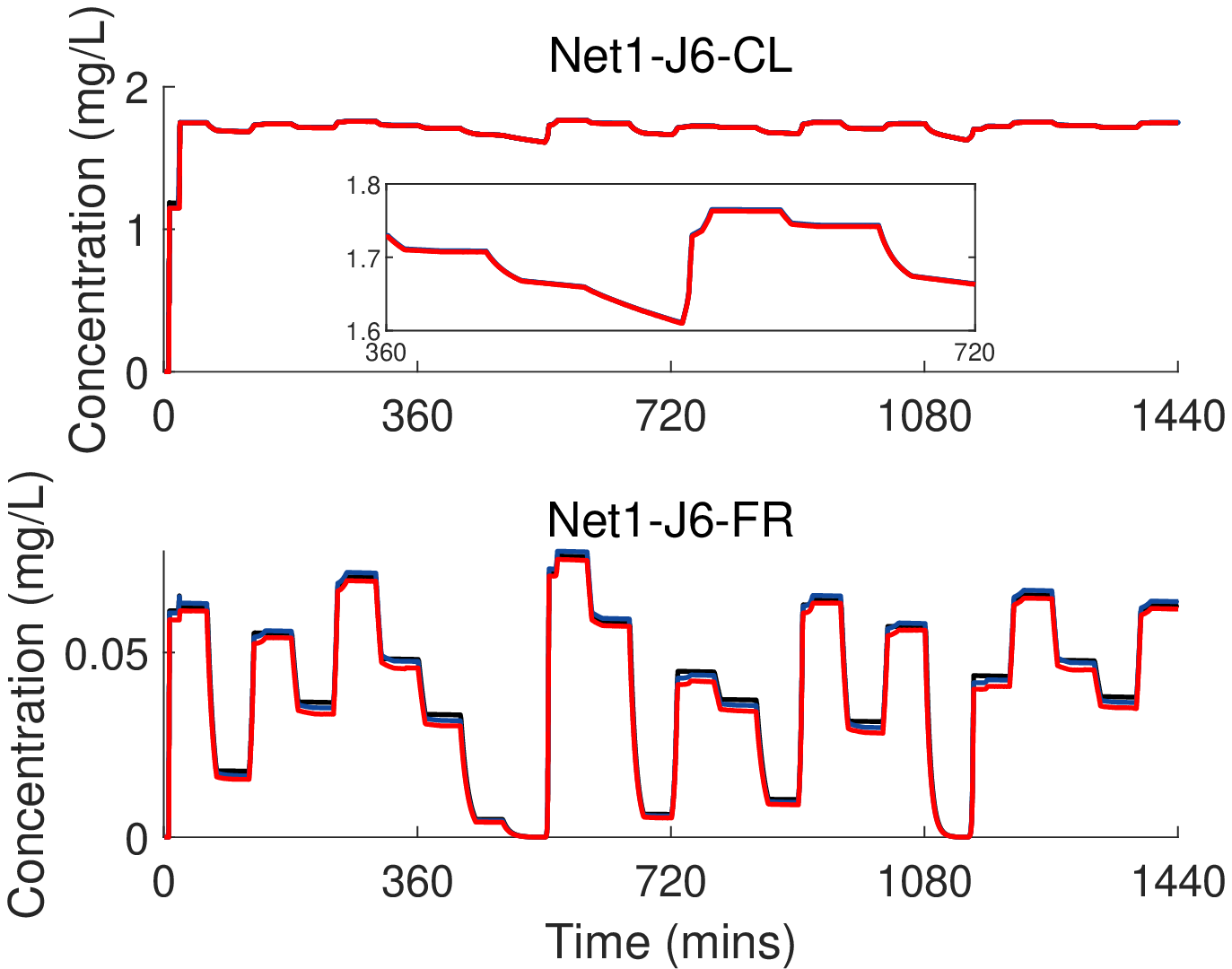}}{}\vspace{0cm}\hspace{-0cm}
		\subfloat[\label{fig:Net1Rest_c}]{\includegraphics[keepaspectratio=true,scale=0.42]{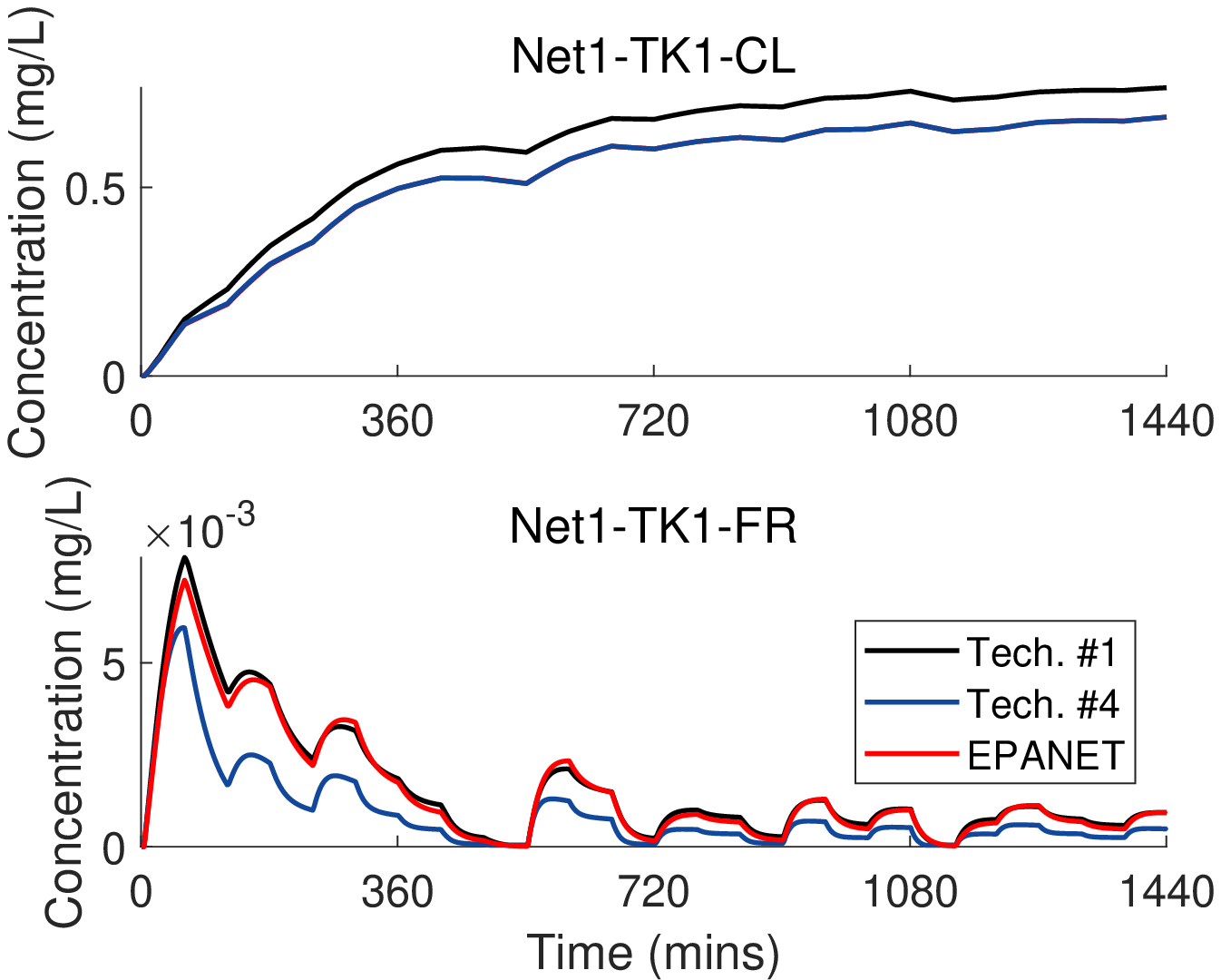}}{}\vspace{0cm}\hspace{-0cm}
		\caption{Simulation Results at (a) Pipe P12 and (b) Junction J6 (c) Tank TK1 in Net1 with reservoir source concentrations of 2 and 0.3 mg/L, and zero initial concentrations at the rest of the components. Simulation is performed over 24hrs period with demand pattern changing every 1hr by applying Tech. \#1 - Lax-Wendroff scheme and Tech. \# 4 - Method of Characteristics.}
		\label{fig:Net1Rest}
	\end{figure}

	Each of the four discretization schemes follows a different approach as explained. Henceforward, we investigate the behavior of each and compare the results with EPANET-MSX. We apply each of the schemes on Net1 with a fixed chlorine concentration of 2 mg/L and 0.3 mg/L for the fictitious reactive substance at Reservoir R1. A dynamic hydraulic model is considered that resulted in a number of segments of pipes ranging between 44 to 301 segments. Fig. \ref{fig:Net1Rest} depicts the results of the L-W scheme (Tech. \#1), MoCs (Tech. \#4), and EPANET with patterned demand and zero initial concentrations for chlorine in the first row and the fictitious reactant in the second one. MoCs gives almost exact chlorine concentrations for all components as EPANET-MSX but relatively different for fictitious reactant with cases of low concentrations (e.g., TK1). On the contrary, results of the L-W scheme are closer in case of low concentrations of the fictitious reactant and chemicals concentrations in pipes and junctions, while maintaining a maximum relative difference of 9.8\% for chlorine in TK1. In general, we can see that both methods give a good representation of both chemicals' evolution, follow similar behavior and give close results to EPANET-MSX. 
	
	\begin{figure}[h!]
		\centering
		\subfloat[\label{fig:Net1BE_b}]{\includegraphics[keepaspectratio=true,scale=0.4]{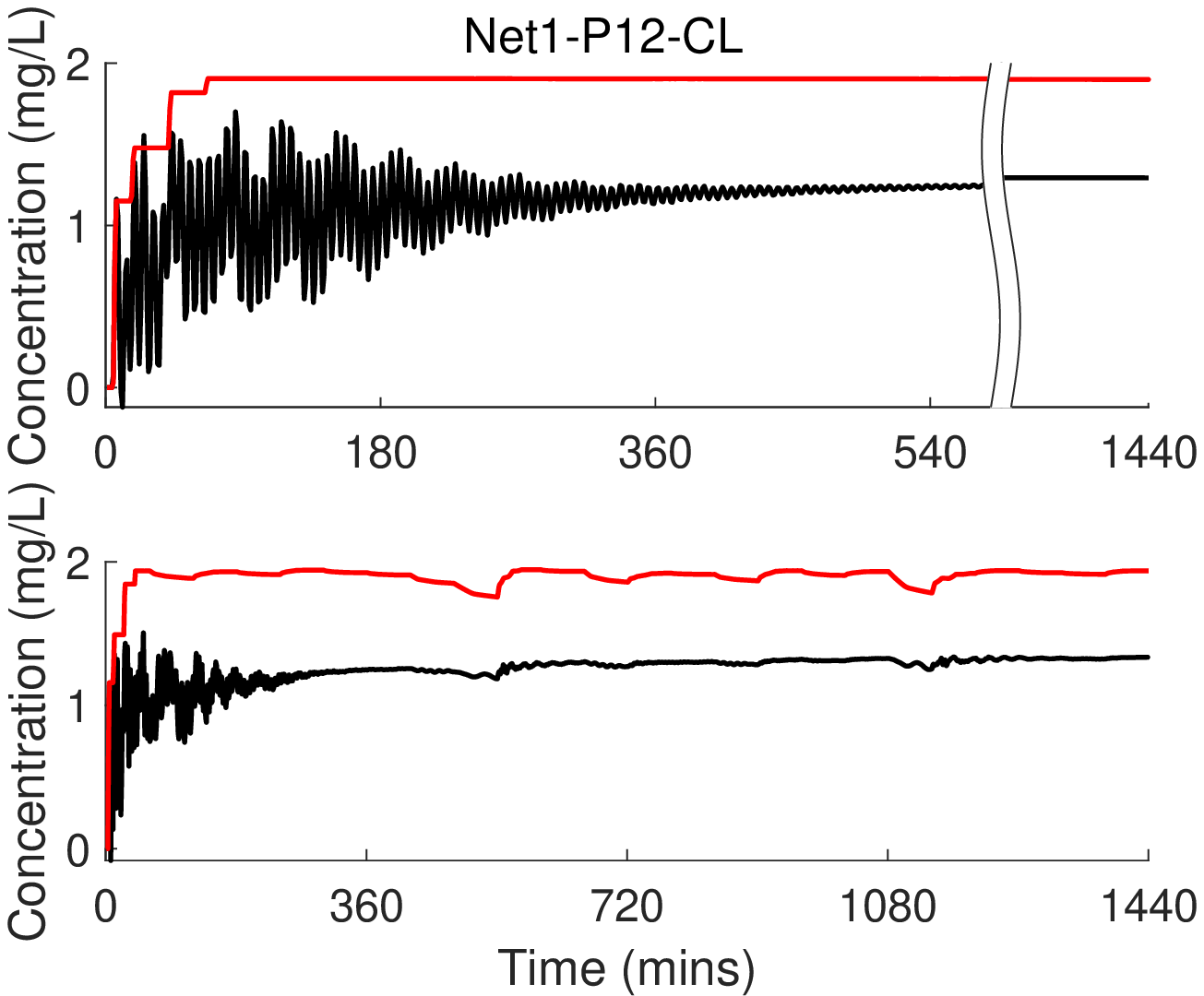}}{}\vspace{0cm}\hspace{0.05cm}
		\subfloat[\label{fig:Net1BE_c}]{\includegraphics[keepaspectratio=true,scale=0.4]{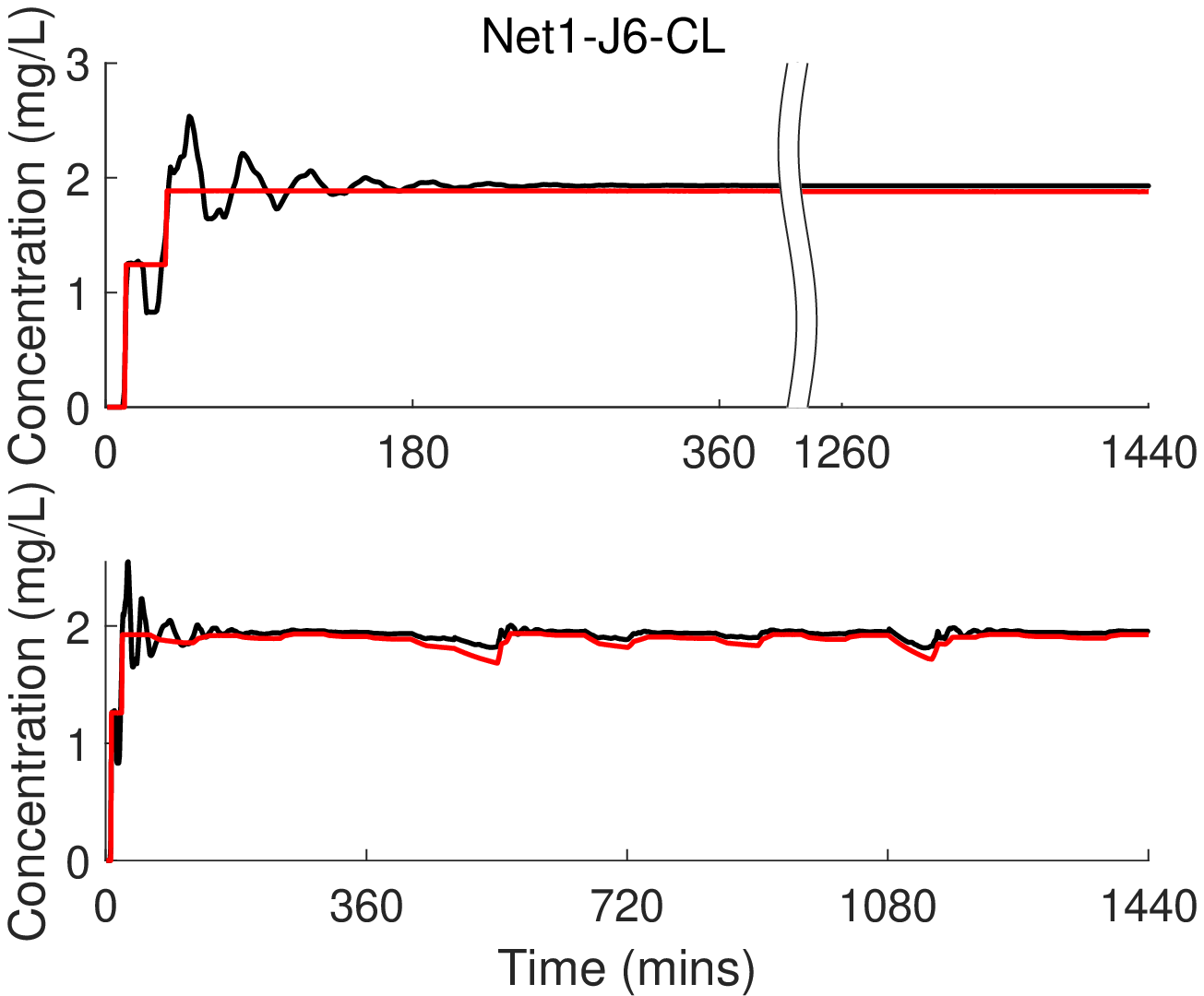}}{}\vspace{0cm} \hspace{0.05cm}
		\subfloat[\label{fig:Net1BE_a}]{\includegraphics[keepaspectratio=true,scale=0.4]{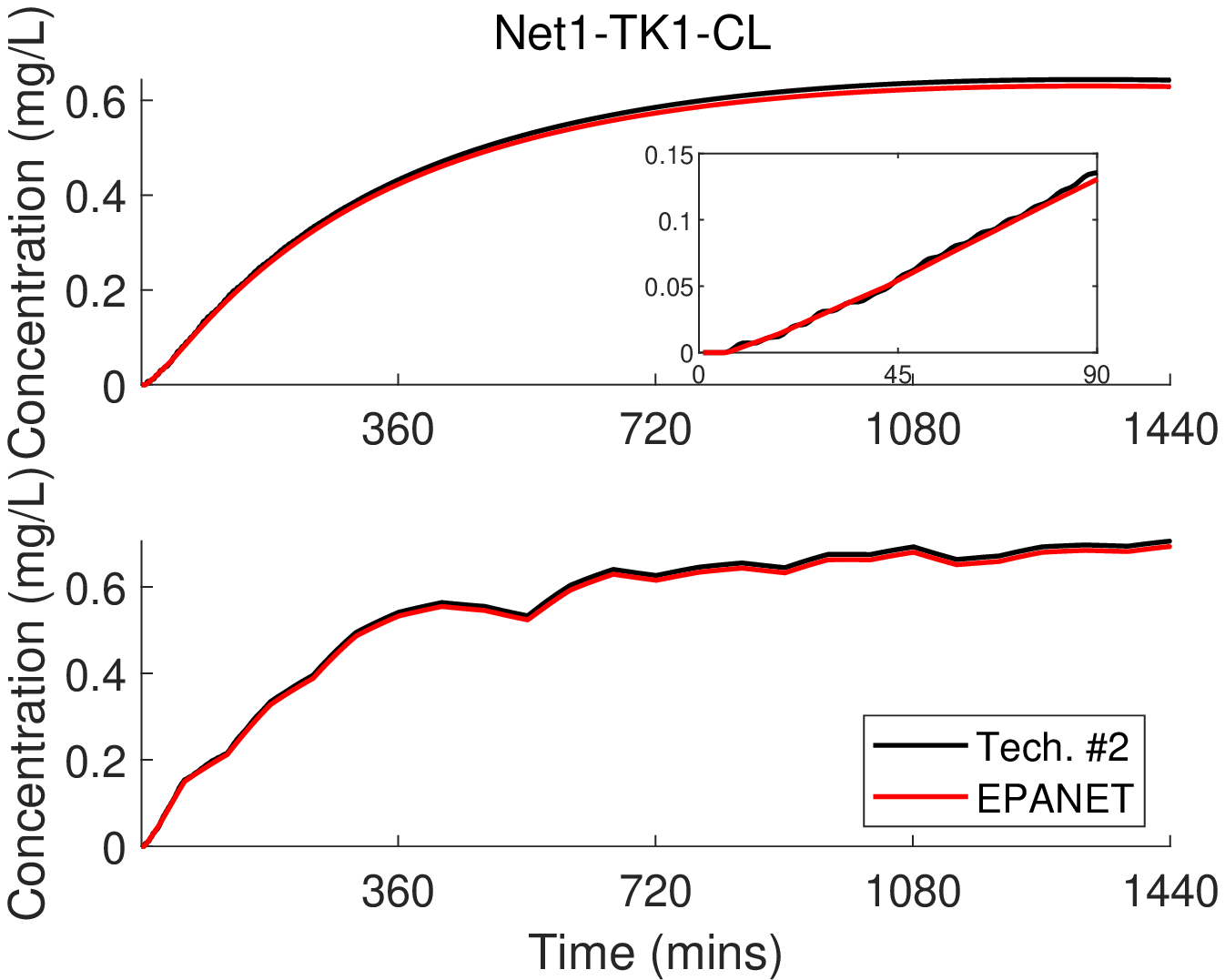}}{}\vspace{0cm} \hspace{0.05cm}
		\caption{Chlorine concentrations at (a) Pipe P12, (b) Junction J6, and (c) Tank TK1 in Net1 under static- and dynamic-states by applying Tech. \#2 - Backward Euler scheme.}
		\label{fig:Net1BE}
	\end{figure} 
	
	On the other hand, Fig. \ref{fig:Net1BE} shows results from simulating chlorine using the Backward Euler scheme (Tech \#2) with the same setting on Net1 within static-state in the first row and dynamic hydraulics in the second one for Pipe P12, Junction J6, and Tank TK1. It is obvious that it generates numerical dispersion. Although Tech. \#2 is unconditionally stable, a relatively small time step is required to achieve accurate solutions. Otherwise, it generates time- step-dependent numerical dispersion. The numerical dispersion drops out as the static condition is reached. Moreover, with dynamic hydraulics it starts to show at the transition phase between one hydraulic step to the next one. Furthermore, this dispersion affects both chemicals simultaneously and each of them directly affects the other one implicitly through their mutual reaction which effect appears in Pipe P12. On the other hand, Tank TK1 is affected by the last segment in Pipe P12 which does not participate in its concentration/volume leading to small dispersion (see Fig. \ref{fig:Net1BE_c}). However, This dispersion leads to misleading unrealistic overestimated/underestimated results that controllers act depends on. Whilst, to avoid dispersion a significantly smaller time-step compared to other methods is needed which increases the run time exponentially.
	
\begin{figure}[h]
		\centering
		\subfloat[\label{fig:Net1CN2a}]{\includegraphics[keepaspectratio=true,scale=0.4]{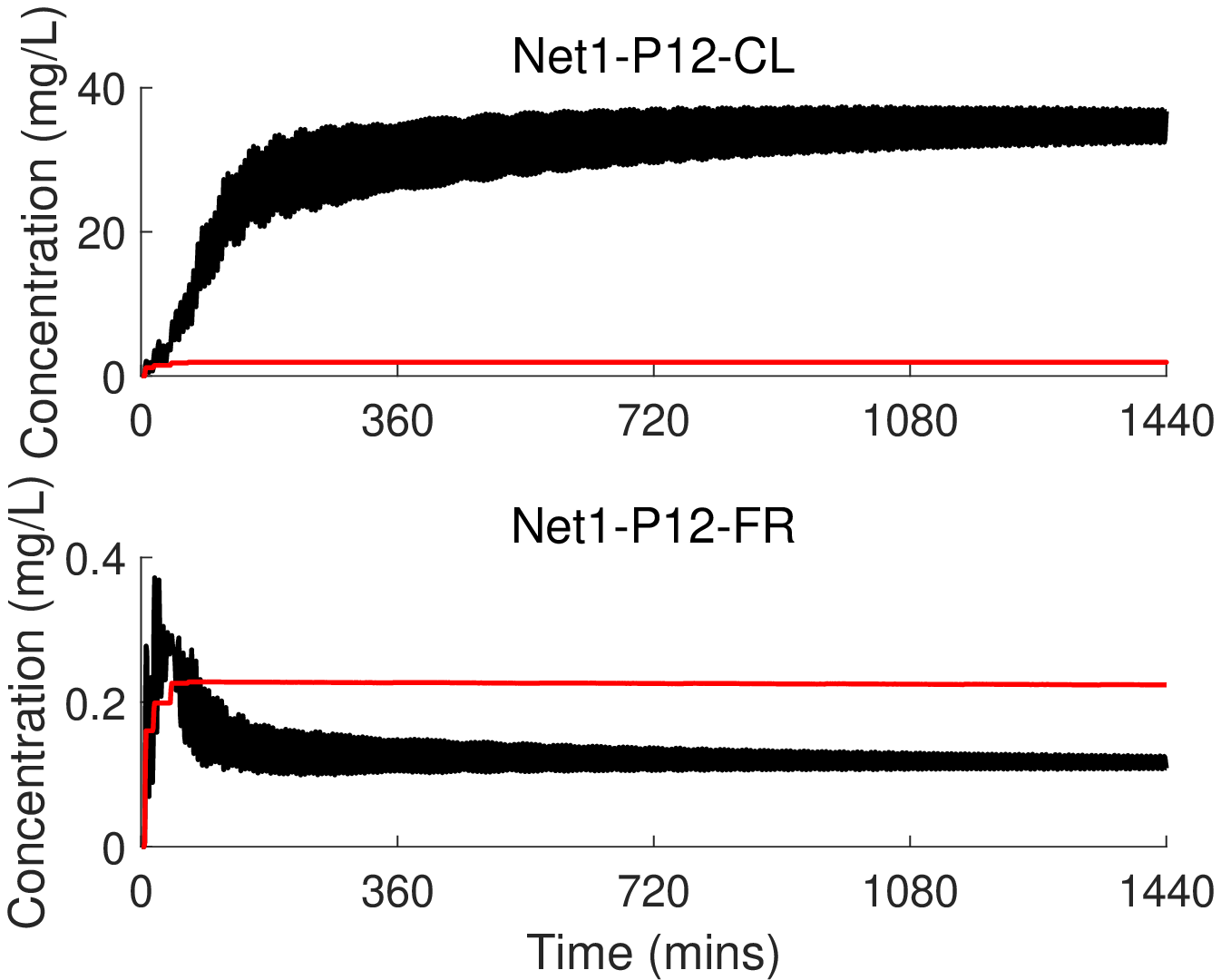}}{}\vspace{-0.05cm} \hspace{-0.1cm}
		\subfloat[\label{fig:Net1CN2b}]{\includegraphics[keepaspectratio=true,scale=0.4]{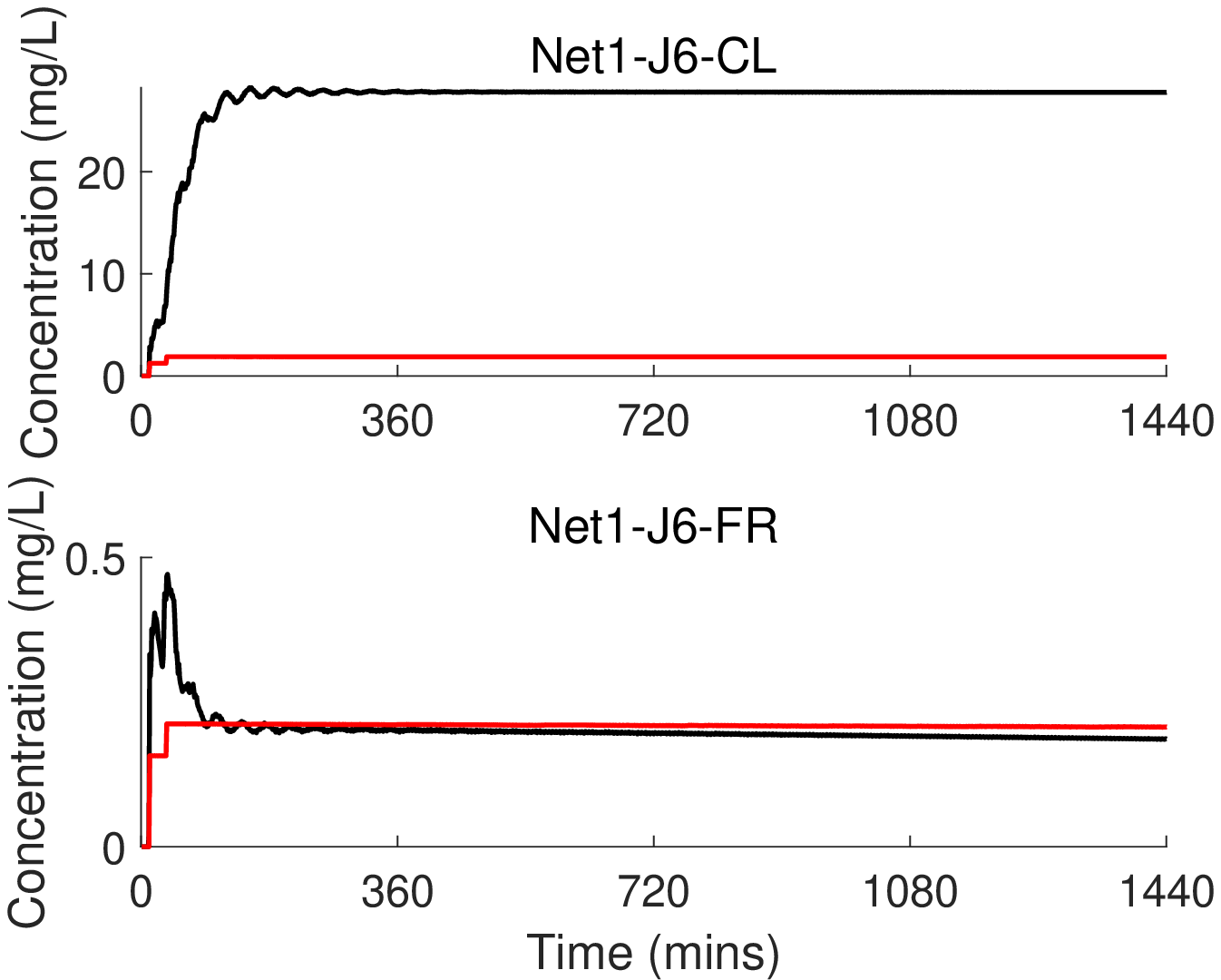}}{}\vspace{-0.05cm} \hspace{-0.1cm}
		\subfloat[\label{fig:Net1CN2c}]{\includegraphics[keepaspectratio=true,scale=0.4]{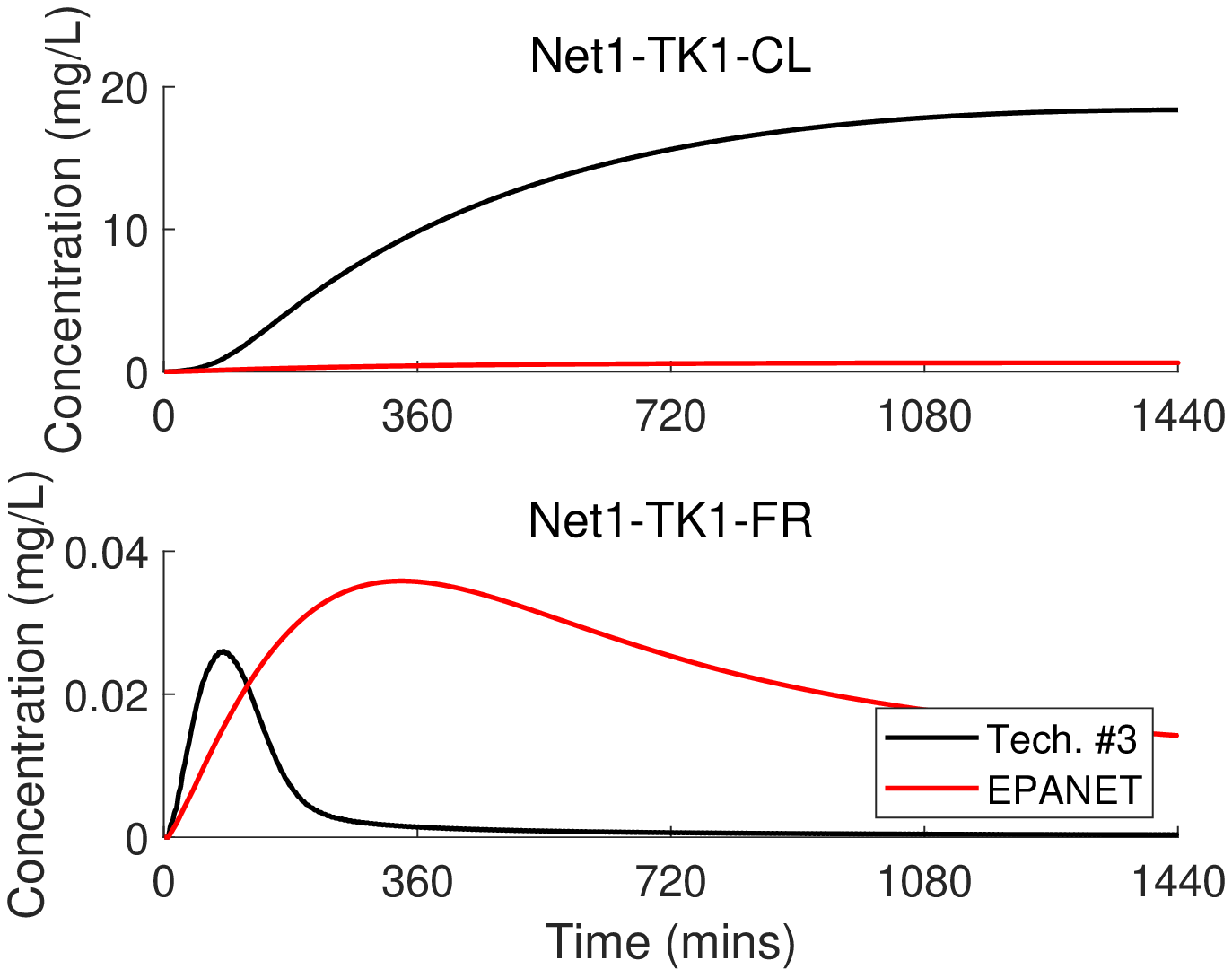}}{}\vspace{-0.05cm} \hspace{-0.1cm}
		\caption{Simulation results for chlorine and the fictitious reactant on Net1 by applying Tech. \#3 - Crank-Nicolson scheme at (a) Pipe P12, (b) Junction J6, and (c) Tank TK1.}
		\label{fig:Net2CNkr01}
	\end{figure}

\begin{figure}[h]
		\centering
		\subfloat[\label{fig:Net1CNa}]{\includegraphics[keepaspectratio=true,scale=0.42]{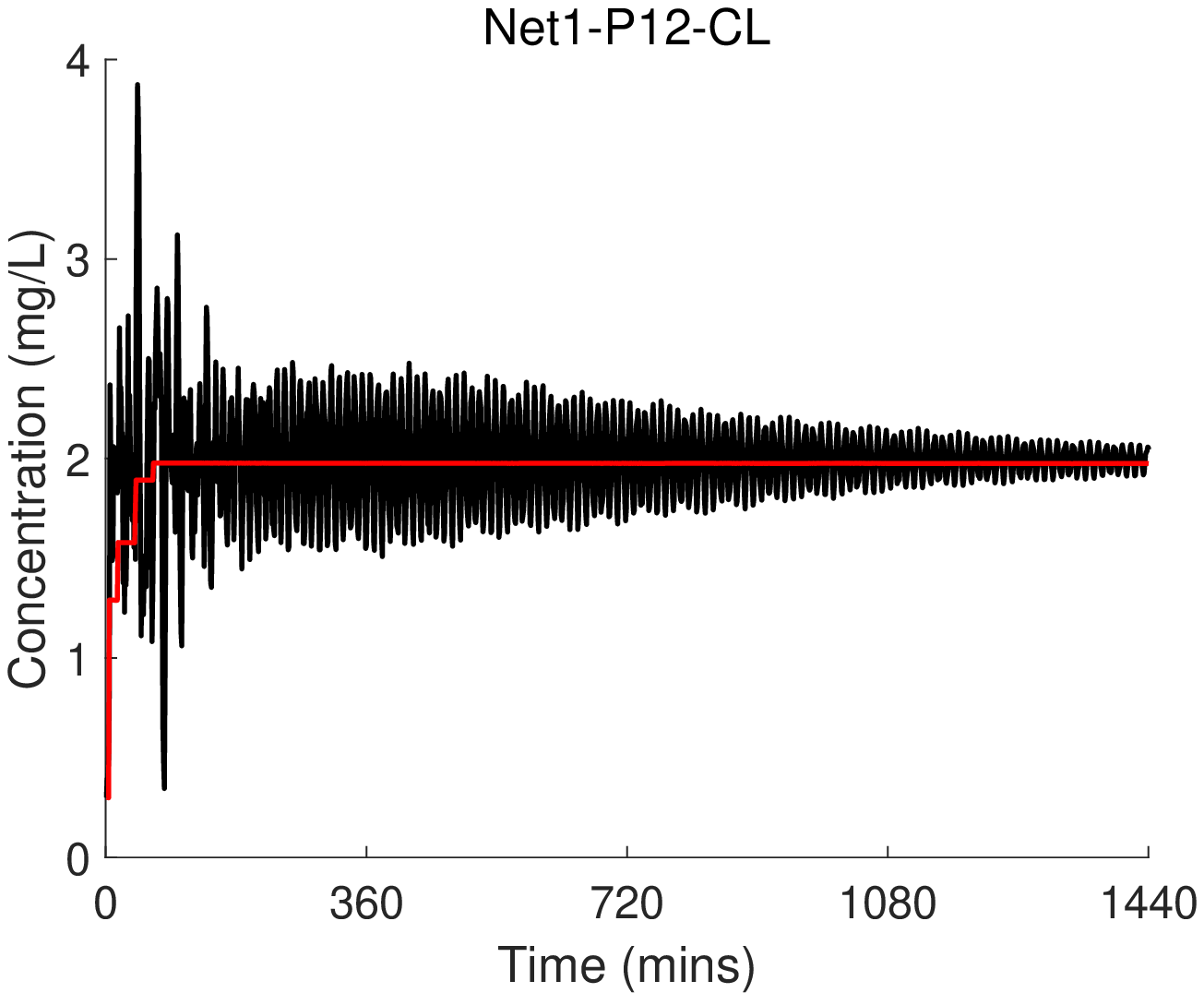}}{}\vspace{-0cm} \hspace{-0cm}
		\subfloat[\label{fig:Net1CNb}]{\includegraphics[keepaspectratio=true,scale=0.42]{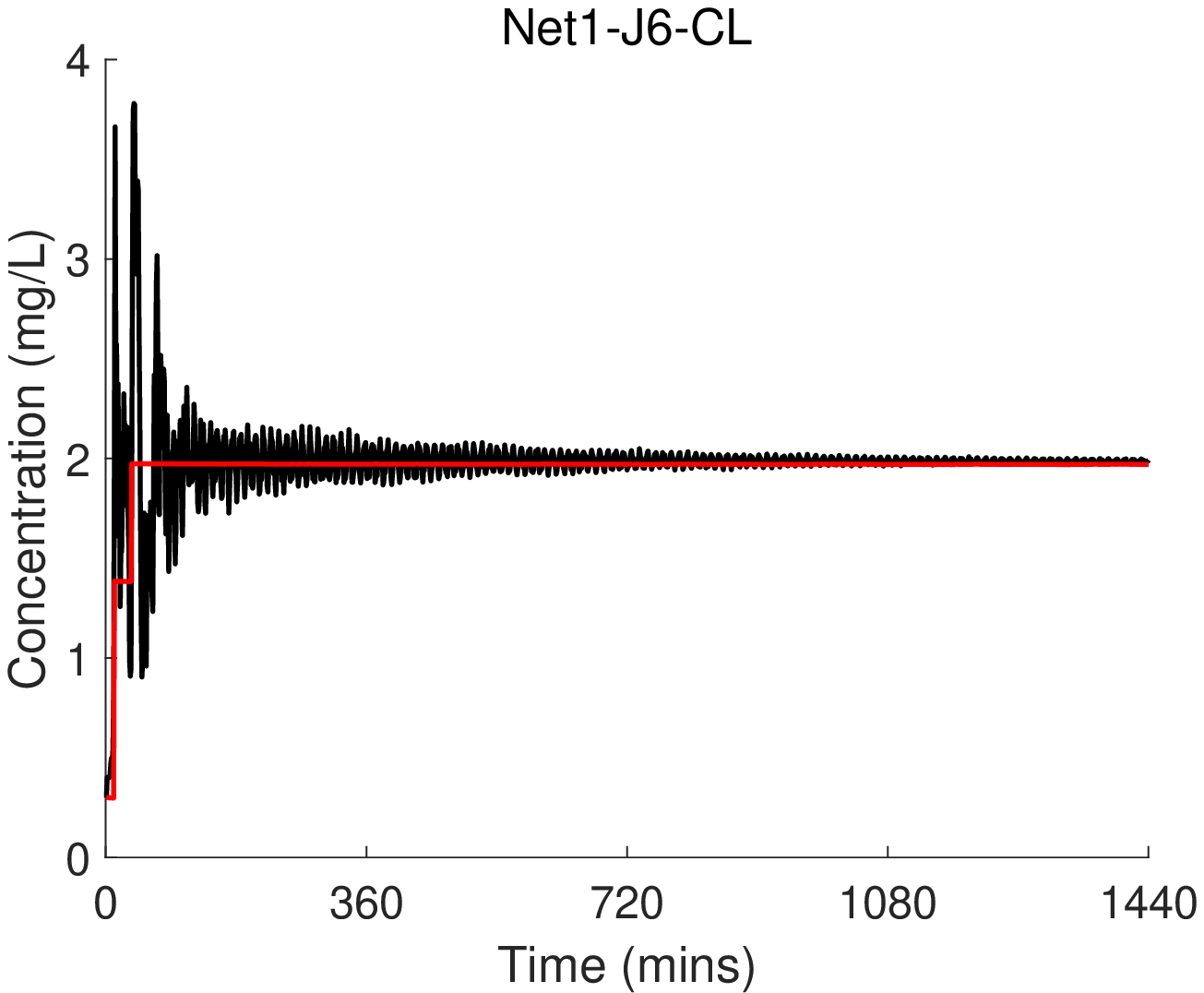}}{}\vspace{-0cm} \hspace{-0cm}
		\subfloat[\label{fig:Net1CNc}]{\includegraphics[keepaspectratio=true,scale=0.42]{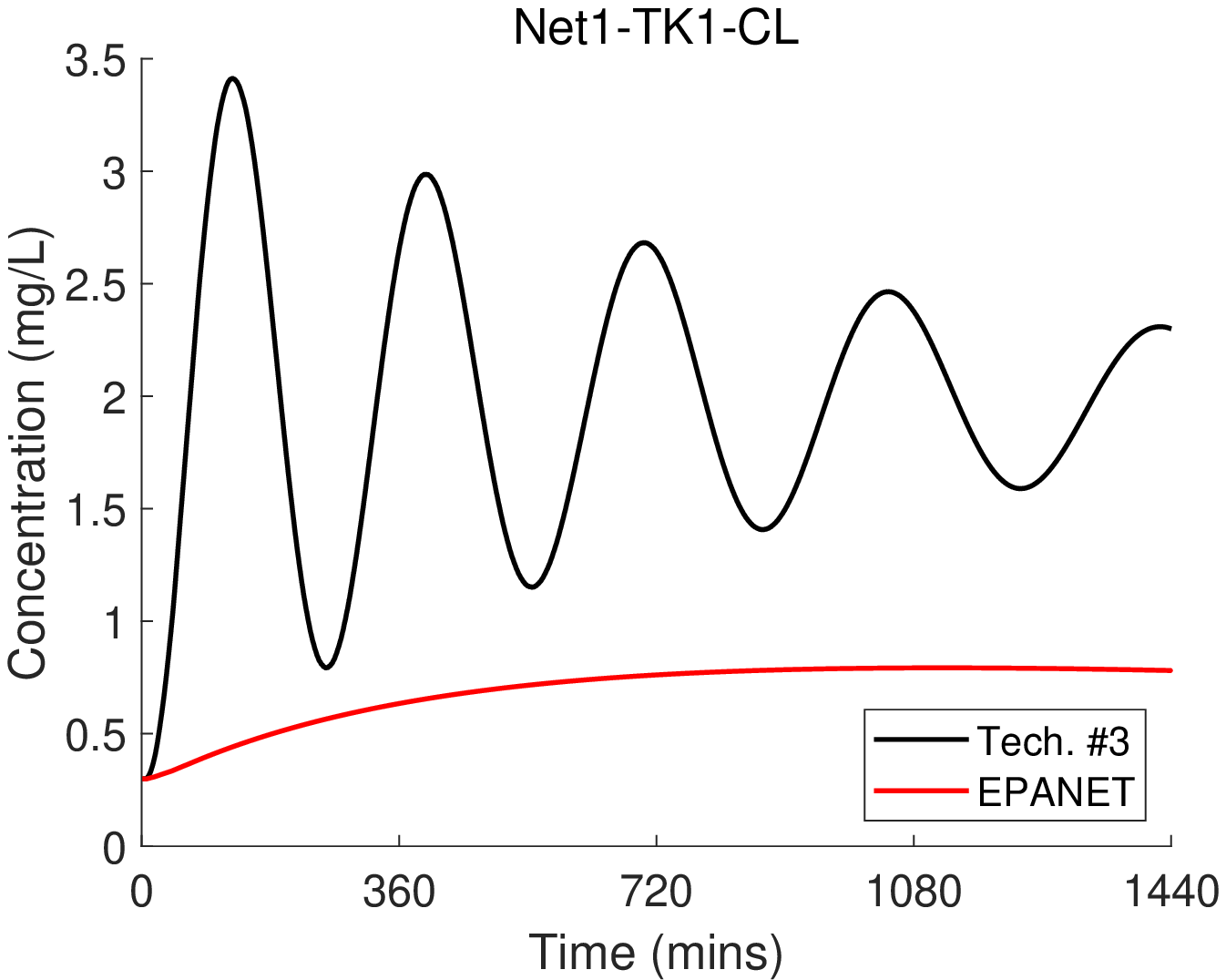}}{}\vspace{-0cm} \hspace{-0cm}
		\caption{Simulation results for chlorine on Net1 by applying Tech. \#3 - Crank-Nicolson scheme on single-species first-order model with initial chlorine concentration of 0.3 mg/L and source concentration of 2 mg/L at (a) Pipe P12, (b) Junction J6, and (c) Tank TK1.}
		\label{fig:Net2CNkr0In}
	\end{figure}

	As for Crank-Nicolson scheme (Tech. \#3), Fig. \ref{fig:Net2CNkr01} shows chemicals concentrations evolution in Net1 with the aforementioned fixed setting. Results exhibit high oscillation reaching illogical concentrations values for both chemicals. That is, this scheme which follows an explicit-implicit discretization approach is sensitive to sharp initial transients that cause an oscillatory response. This oscillatory affects both chemicals separately and they depend on each other which results in a dramatically large error. To test this sensitivity, the scheme is applied on Net1 with only single-species chlorine first-order decay dynamics and with initial conditions of 0.3 mg/L for all components other than the source. Results from this scenario are shown in Fig. \ref{fig:Net2CNkr0In}. We can see that the oscillation effect is reduced but not avoided completely. In fact, sharp transitions in concentrations take place repeatedly during dynamic hydraulic and water quality simulation with demands fluctuation and chemicals injections/intrusion. Comprehensively, it is concluded that the Backward Euler scheme (Tech. \#2) and Crank-Nicolson scheme (Tech. \#3) put stiff limitations to be used in water quality modeling including extremely small time-step to avoid numerical dispersion and no sharp transitions in chemical concentrations to eliminate the oscillation action.
	
\begin{figure}[h]
		\centering
		\subfloat[\label{fig:Net2a}]{\includegraphics[keepaspectratio=true,scale=0.42]{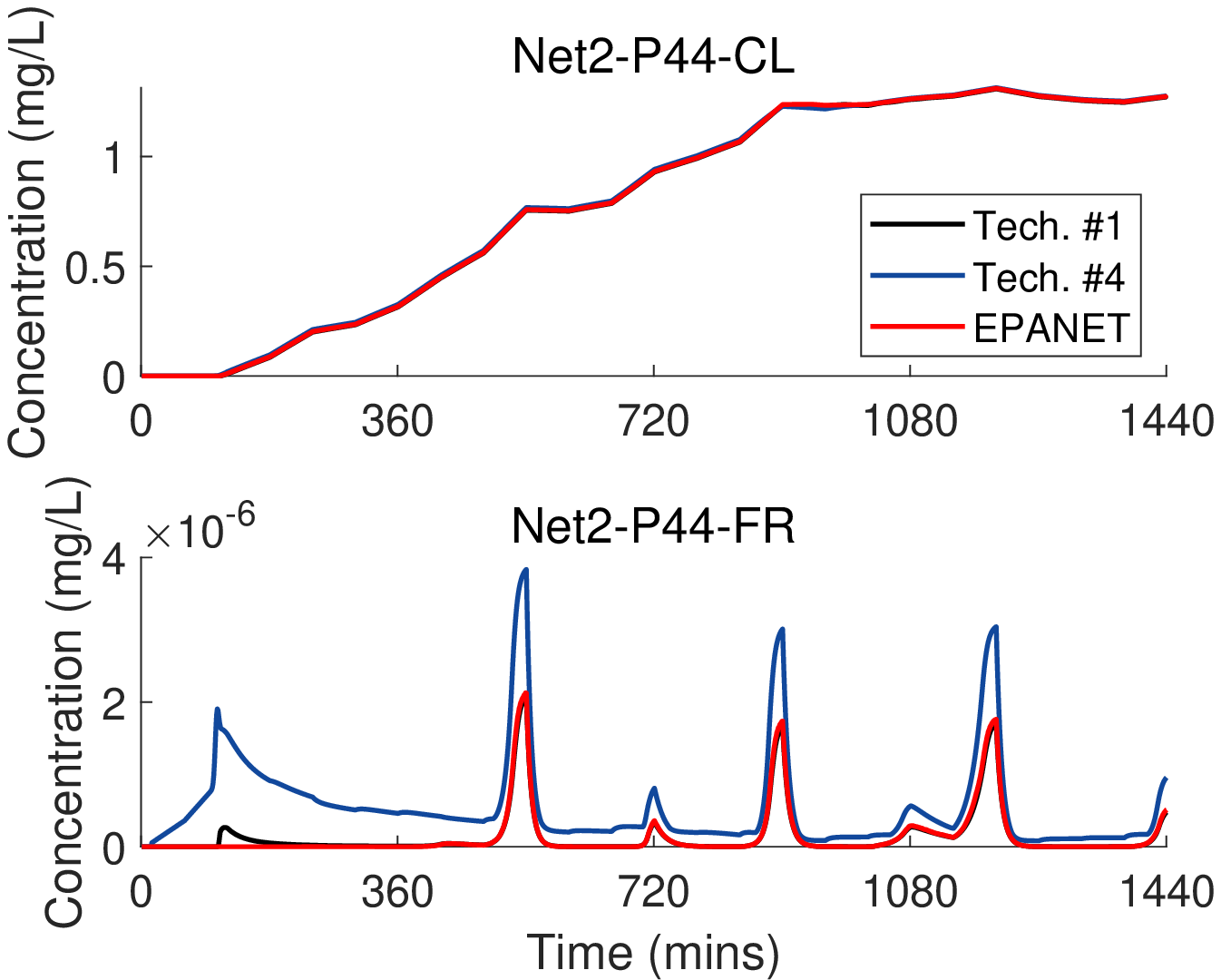}}{}\vspace{-0cm} \hspace{-0cm}
		\subfloat[\label{fig:Net2b}]{\includegraphics[keepaspectratio=true,scale=0.42]{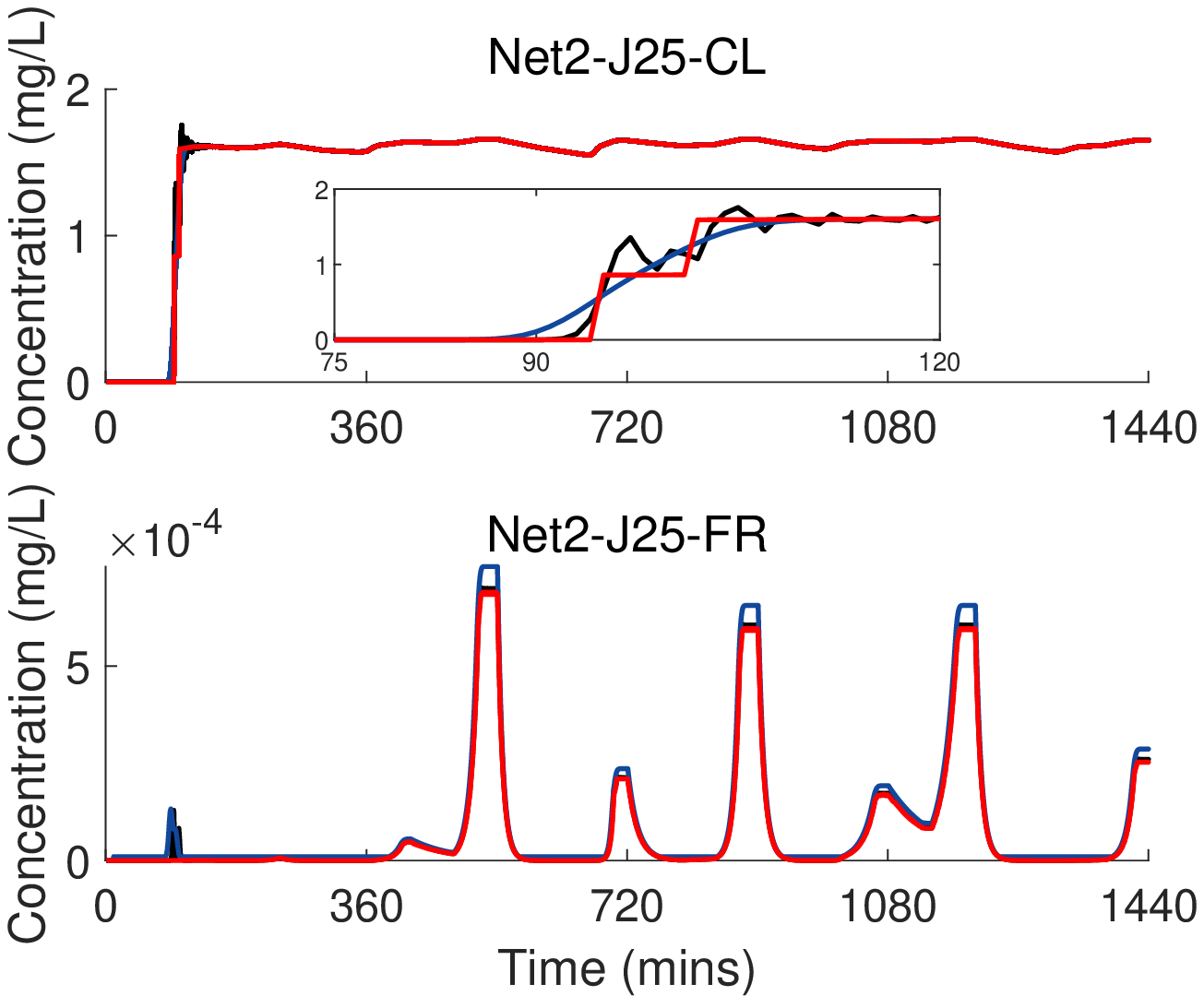}}{}\vspace{-0cm} \hspace{-0cm}
		\subfloat[\label{fig:Net2c}]{\includegraphics[keepaspectratio=true,scale=0.42]{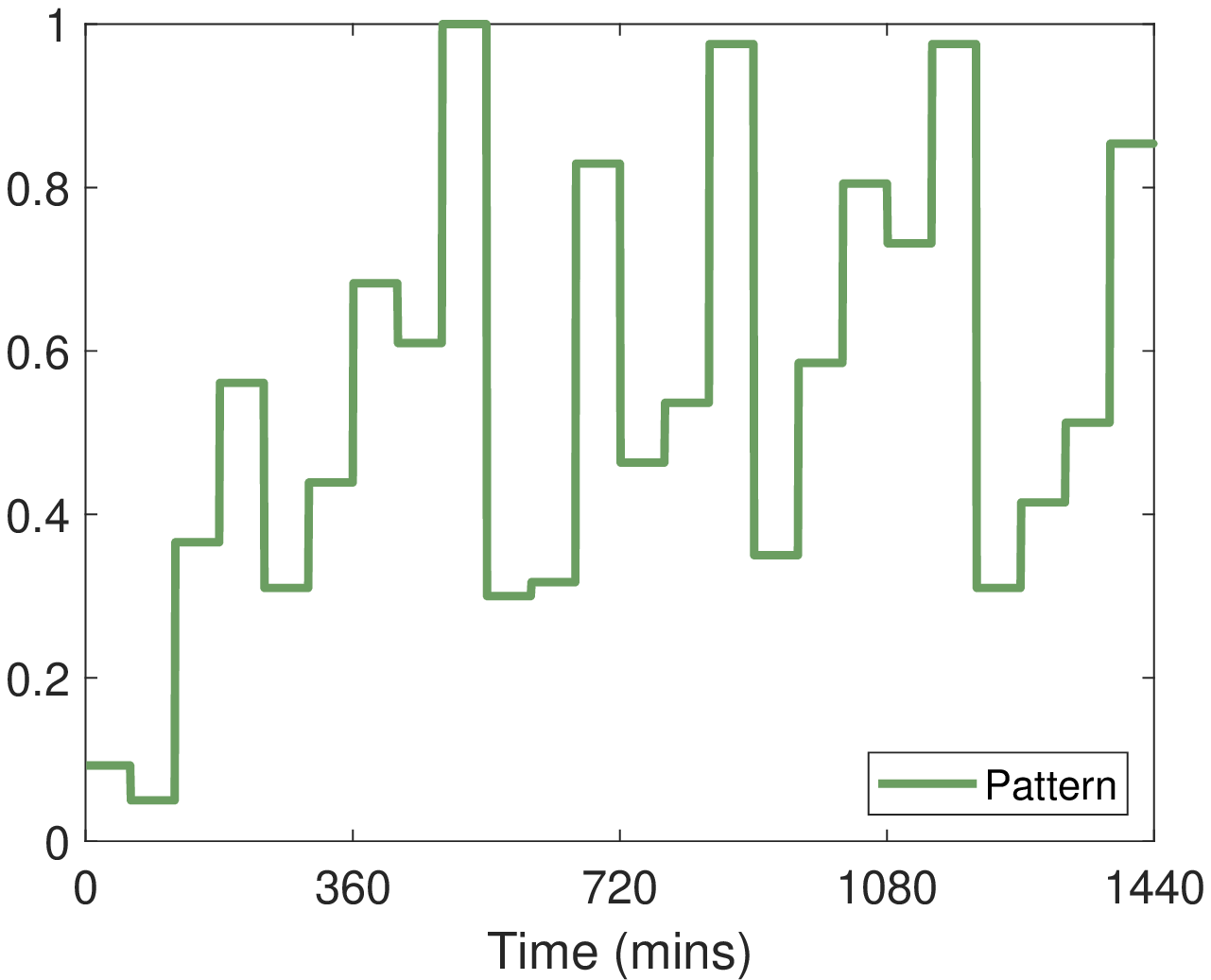}}{}\vspace{-0cm} \hspace{-0cm}
		\caption{Simulation results on Net2 by applying Tech. \#1 - L-W scheme and Tech. \#4 - MoCs at (a) Pipe P44 and (b) Junction J25 with demand pattern shown in (c).}
		\label{fig:Net2}
	\end{figure}

	L-W scheme and MoCs are applied on networks with different scales under various hydraulic states to test their performance. First, we apply both techniques on Net2 with chlorine and fictitious reactant concentrations of 2 and 0.3 mg/L at both Reservoirs R1 and R2. Results in Fig. \ref{fig:Net2} imply the reliable performance of both techniques. Yet, the L-W scheme results in numerical dispersion at sharp fronts in Junction J25 as shown in Fig. \ref{fig:Net2b}. This dispersion is likely to take place when the Courant number $\tilde{\lambda}$ is small close to zero. Such a scenario occurs when fixing the number of segments via Equation \eqref{equ:ChooseSegment} for a system with a high range of velocities resulted from extremely high or low hydraulic states (see Fig. \ref{fig:Net2c}). However, this dispersion can be eliminated by dividing the simulation period to separate these extreme states and maintain the Courant number as close as possible to 1. First, the system dimensions are defined by velocities in the first phase then the system is resized according to velocities in the following phase. Average concentration over pipe at the end of the first phase is used as the initial value for the next one.
	
\begin{figure}[h]
		\centering
		\subfloat[\label{fig:FFCLa}]{\includegraphics[keepaspectratio=true,scale=0.5]{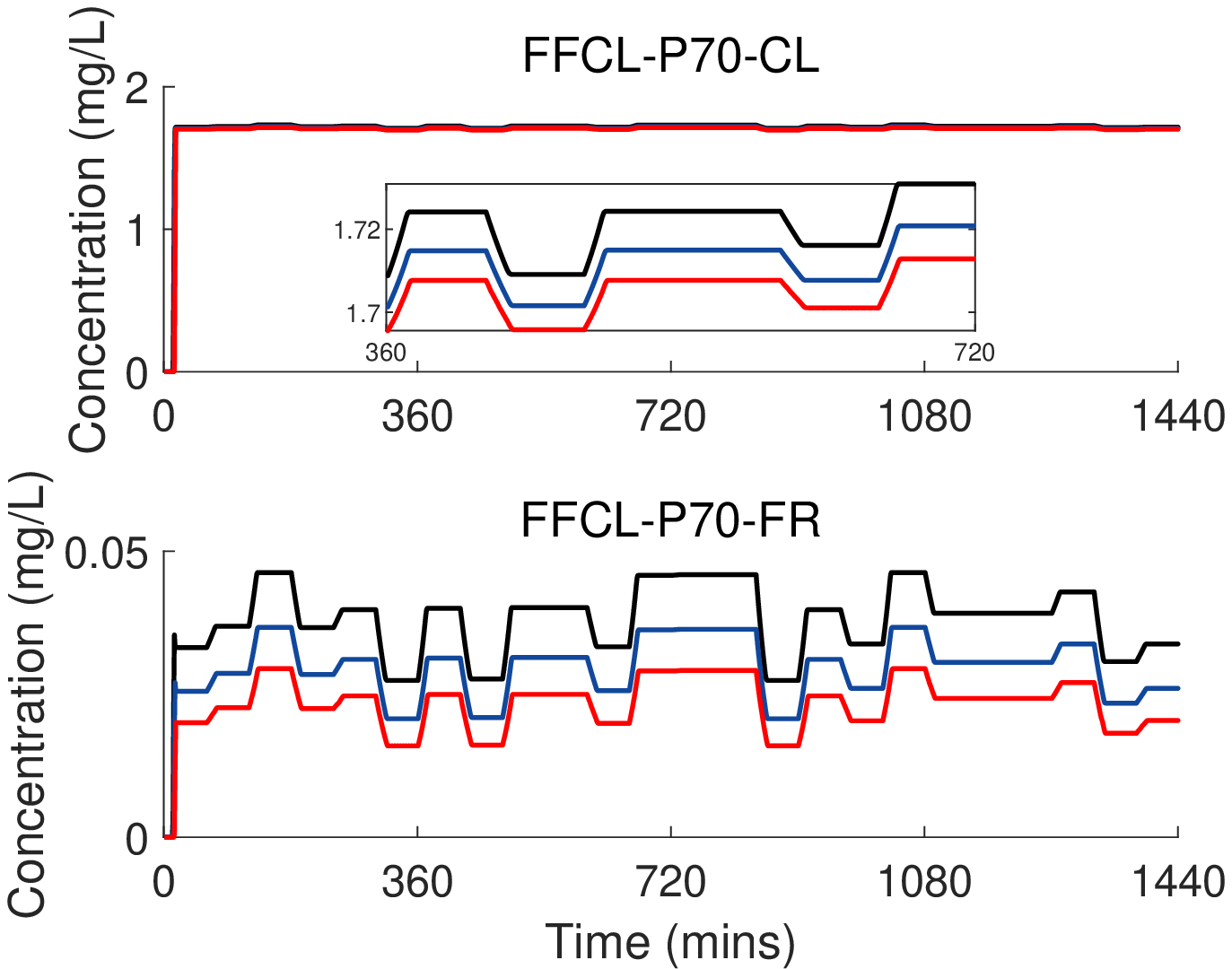}}{}\vspace{-0.05cm} \hspace{0.1cm}
		\subfloat[\label{fig:FFCLb}]{\includegraphics[keepaspectratio=true,scale=0.5]{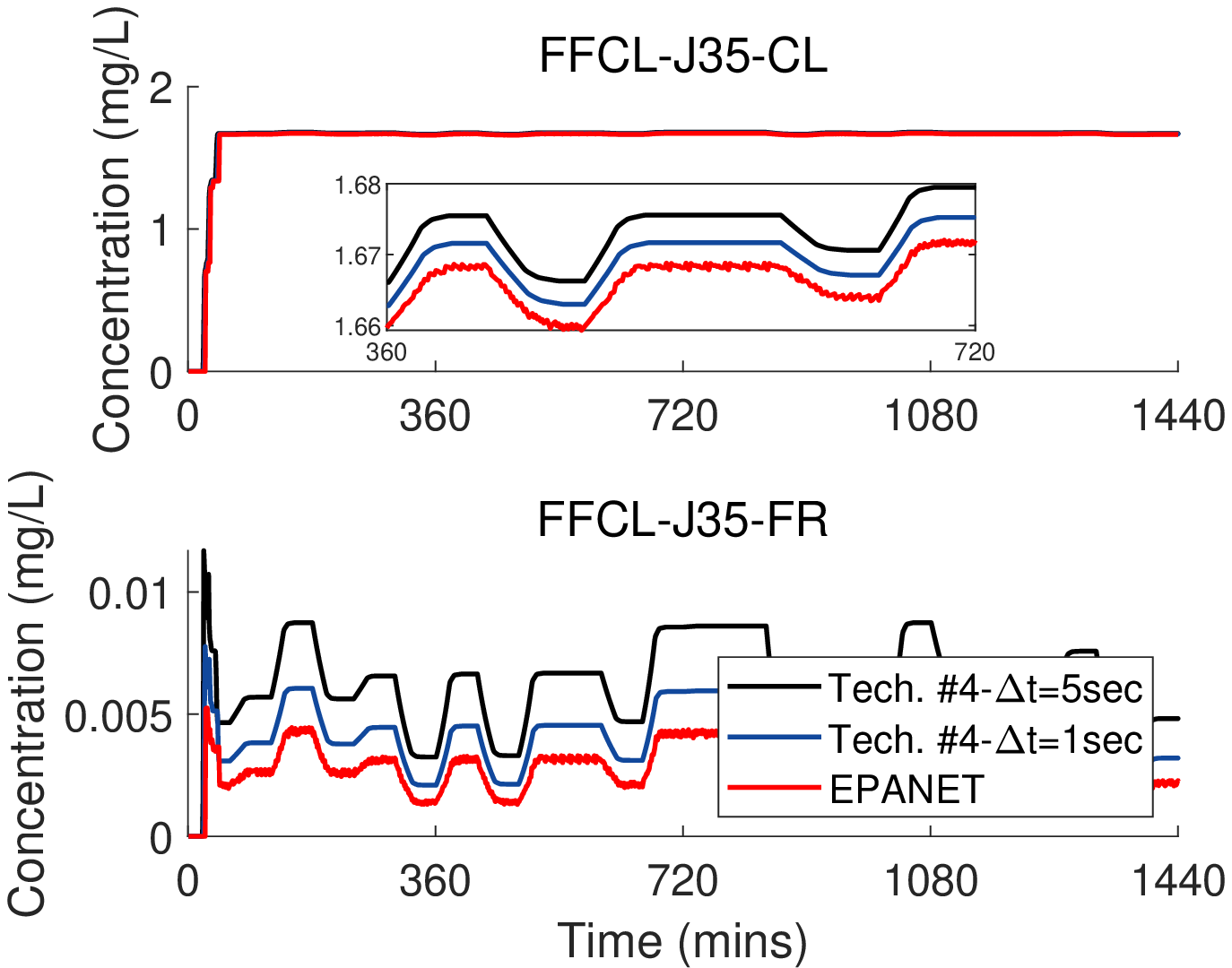}}{}\vspace{-0.05cm} \hspace{0.1cm}
		\caption{Simulation results for chlorine and the fictitious reactant on FFCL-1 by applying Tech. \#4 - Method of Characteristics at (a) Pipe P70, and (b) Junction J35 with water quality time-steps of 5 and 1 sec.}
		\label{fig:FFCL}
	\end{figure}

	Moreover, both techniques are applied on FFCL-1 with fixed concentrations of 2 and 0.3 mg/L at Reservoir R1 to test their scalability. To compare results, we calculate relative difference ratio between the applied technique and EPANET-MSX to be $$ \textit{Relative Difference } (RD)= \dfrac{c_\text{EPANET-MSX} - c_\text{model}}{c_\textit{EPANET-MSX}}.$$ The maximum relative difference ratio for the L-W scheme (Tech. \#1) for both chlorine and fictitious reactant within these case studies ranges between 5-23\%. However, the maximum relative difference ratio for the MoCs scheme (Tech. \#4) varies between 1-11\% for chlorine and between 7-33\% for fictitious reactants. This difference is caused by the assumption made to transform the second-order dynamics into pseudo-first-order which affects the substance with smaller concentrations. Nonetheless, the error decreases with decreasing the water quality time-step as shown in Fig. \ref{fig:FFCL}.   
	
	Moreover, the effect of hydraulic parameters variability and its direct impact on system dimensions as stated in Remark \ref{rm:Segments} is tested. For each network, demands values and patterns are changed leading to different velocities and different number of states. Simulation run-time for each case study is listed in Tab. \ref{tab:runtime}. Notice that the hydraulic simulation output almost does not affect EPANET AND EPANET-MSX run-time due to the fact that the time needed is for extracting the results. On the other hand, increasing number of states increases computational time, especially for Method of Characteristics (Tech. \#4).

	\begin{table}[t]
		\centering
		\caption{Simulations run-time (secs)}~\label{tab:runtime}
		{\small\begin{threeparttable}
				\caption{Water quality model equations for the three-node example\tnote{a}}
				\begin{tabular}{c||c|c|c||c}
					\hline
					Network & \# States\tnote{a} & Tech. \#1 - Lax-Wendroff scheme & Tech. \#4 - Method of Characteristics & EPANET-MSX \\
					\hline
					\multirow{2}{*}{Net1} & 4586 & 81.9  & 96.5 & 615  \\	
					\cline{2-5}
					& 6360 & 113  & 116 & 643  \\	
					\hline
					\multirow{2}{*}{Net2} & 23680 & 466  & 483 & 745  \\
					\cline{2-5}
					& 57186 & 1207.5  & 2180 & 763  \\
					\hline
					\multirow{2}{*}{FFCL-1} & 35370 & 611 & 1044 & 2092\\
					\cline{2-5}
					& 52454 & 1225 & 3929 & 2201\\
					\hline
					\hline
				\end{tabular}
				\begin{tablenotes}
					\item[a] States count is the summation of the number of states (i.e., $x_i(t)$) for both chemicals.
				\end{tablenotes}
		\end{threeparttable}}
	\end{table}
	
	\begin{table}[h!]
		\centering
		\caption{Techniques/methods applicability and limitations in multi-species water quality modeling}~\label{tab:TechComp}
		{\small\begin{tabular}{c|K|K}
				\hline
				Technique/method & Applicability & Limitation \\
				\hline
				EPANET and EPANET-MSX & Widely used and EPANET-MSX allows full control to specify the reaction model to be considered & High run-time to obtain results even for small networks. Complexity to couple a control model with it \\
				\hline
				Tech. \#1 - Lax-Wendroff Scheme & Gives a reliable results in a form of nonlinear state-space representation and traces the evolution of both chemicals  & Conditionally stable. Stability should be checked throughout whole simulation period. According to networks topology and hydraulics parameters, may results in high numbers of state-space variables, hence, high run-time.\\
				\hline
				Tech. \#2 - Backward Euler Scheme & Sensitive to simulation time-step causing dispersion that results in non-realistic results especially at the start of the simulation & To avoid numerical dispersion extremely small time-step is needed even for small- and mid-sized network leading to high run-time \\
				\hline
				Tech. \#3 - Crank-Nicolson Scheme & Results in high oscillatory response to transients & These transitions take place repetitively in water quality modeling and control \\
				\hline
				Tech. \#4 - Method of Characteristics & Depicts chemical evolution in WDN while overcoming the nonlinearity complexity & Presents the second-order reaction model in a form of pseudo-first order one resulting in approximation error that increases with the high difference between chemical concentrations, network's scale and hydraulic parameters change \\
				\hline
				\hline
		\end{tabular}}
	\end{table}

	\newpage
	\section{Conclusion, Paper's Limitations, and Recommendations for Future Work}~\label{sec:Conc-Rec}
	This paper presents a full description of the formulation of multi-species water quality state-space representation. This description is provided and tested over a fixed time and space grid discretized using distinguishably based techniques. Three Eulerian Finite Difference schemes, Lax-Wendroff, Backward Euler, and Crank-Nicolson schemes, and one Lagrangian time-driven method, Method of Characteristics are considered for implementation and testing. Case studies show that Backward Euler and Crank-Nicolson schemes are strictly limited to avoid numerical dispersion and oscillation. On the contrary, the Lax-Wendroff scheme and Method of Characteristics reliably trace chlorine contractions within different networks with various scales. Yet, the Lax-Wendroff scheme's performance worsens with highly variable hydraulic dynamics, however, it is proposed to divide the simulation period to separate extremely low and high hydraulic states as mitigation to this problem. Tab. \ref{tab:TechComp} lists techniques' applicability and limitations concluded from the research output.   
	
	An educational and theoretical value with a broader impact of this paper is providing a comprehensive framework and detailed example of how the state-space representation for water quality dynamics is formulated. The algorithm provided is easily editable to consider different scenarios and different chlorine reaction models.
	
	However, the presented research has its limitations and recommended extensions. First, this paper tests numerically-driven discretization methods, however, more methods varying from Eulerian-, Lagrangian-, hybrid-based are available and can lead to a more accurate/limitation-free simulation of the model. Second, the multi-species model is valid for substances actively reacting with chlorine only. Also, in this paper we consider the first order wall reaction model, while applying other models including the zero-order and the validated EXPBIO models in the closed-form state space representation is worth investigating. Additionally, discretization methods result in large model dimensions that vary with the considered hydraulic parameter even for mid-sized networks. Subsequently, it is proposed for the authors' future work to perform a model order reduction taking into consideration the nonlinearity associated with applying the Lax-Wendroff scheme for the multi-species dynamics. And to investigate and apply different control algorithms to maintain chlorine residuals within the desired standard range under multi-species water quality dynamics.
	
	\bibliographystyle{IEEEtran}
	\bibliography{IEEEabrv,MS_arXiv}

	\newpage
	\appendices
	
	\section{EPANET's AR Discretization Algorithm}~\label{App:EPALag}
	EPANET and its extension, EPANET-MSX, use the Lagrangian time-driven method to solve the advection-reaction PDE in pipes. Hereinafter, Algorithm \ref{alg:LAG_EPANET} gives a full description of how the method is applied for each hydraulic time-step within the simulation period $[0, T_s]$.
	
	\begin{algorithm}[h]
		\small	\DontPrintSemicolon
		\KwIn{Hydraulic parameters (e.g., velocities, flow rates), pipe's length, and adjustment tolerance ($\tau_\text{adj}$)}
		\KwOut{Chemical concentrations in each pipe}
		\eIf{It is the first hydraulic time-step}{Pipe is considered as one segment with initial concentration equal to the upstream node's concentration}{Segments are ordered in the direction of the flow}
		\While { $t \leq T_s$ }{
			Apply reaction to each segment \;
			Calculate volume entering each downstream node, which equal pipe's flow rate times time-step \;
			\eIf{volume is larger than last segment's volume}
			{Last segment is distroyed \;
				Segment before the last one contributes its volume
			}{Last segment's volume is decreased by the volume contributed}
			Update concentrations at nodes according to entering volumes and demands  \;
			\eIf{difference between concentrations of the upstream node and the first segment $\leq \tau_\text{adj}$ }{Update first segment's size by the inflowing volume}{Create new segment with the upstream node's concentration}
			Obtain the average of the segments' concentrations to represent the whole pipe's concentration at time $t$  \;
			$ t = t + \Delta t$
		}
		\caption{EPANET's Lagrangian time-driven method to solve advection-reaction PDE for each hydraulic time-step}
		\label{alg:LAG_EPANET}
	\end{algorithm}

	\section{State-space formulation example - Different Techniques}~\label{App:SS-Ex}
	
	In this appendix we apply Backward Euler, Crank-Nicolson schemes, and Method of Characteristics (Tech. 2,3, and 4) on the same example explained in Section \ref{sec:SS-Ex} by highlighting the differences from the Lax-Wendroff scheme (Tech. 1). As noticed in Tab. \ref{tab:SSTechs}, the main differences between the techniques are in $E(t), A(t),$ and $f(x_1,x_2,t)$. That is, we give a general form for them (Tab. \ref{tab:SSGnrlMt}) for the example then list the specific application for each of the techniques (Tab. \ref{tab:TechsEx}). We only list elements for the pipe's segments in Tab. \ref{tab:TechsEx}) as the remaining elements are the same as explained in Section \ref{sec:SS-Ex}. Notice that, the listed values are for chlorine. As for the fictitious reactant, $k^\mathrm{P}_1$ is considered equal to zero.
	
	\input{TechniquesExample}

\end{document}

%% file: preambleT.tex
\usepackage{epsfig,color,amsmath,cite}

\IEEEoverridecommandlockouts
\usepackage{wrapfig}
\usepackage{setspace}  
\usepackage{xcolor}
\usepackage{epstopdf}
\usepackage{amssymb}
\usepackage{url}
\usepackage{mathtools, nccmath}
\usepackage{enumitem}
\usepackage{multirow}
\usepackage{makecell}
\usepackage{amsmath}
\usepackage{stackrel}
\usepackage{booktabs}
\usepackage[flushleft]{threeparttable}
\usepackage{makecell}

\usepackage{subfig}
\usepackage{graphicx}
\usepackage{lscape}

\usepackage{comment}

\usepackage[linesnumbered,lined,boxed,commentsnumbered,ruled,longend]{algorithm2e}
\makeatletter

\SetKwProg{Init}{Initialization}{}{Proceed}
\makeatother

\setlist[itemize]{leftmargin=*}

\makeatother

\allowdisplaybreaks[4]
\usepackage[colorlinks = true,
linkcolor = blue,
urlcolor  = blue,
citecolor = blue,
anchorcolor = blue]{hyperref}

\usepackage{amsthm}

\newtheorem{myrem}{Remark}
\newtheorem{asmp}{Assumption}

\usepackage[usestackEOL]{stackengine}
\stackMath
\usepackage[framemethod=TikZ]{mdframed}
\usepackage{bm}
\mdfdefinestyle{MyFrame}{%
	linecolor=black,
	outerlinewidth=1.5pt,
	roundcorner=1.5pt,
	innerrightmargin=5pt,
	innerleftmargin=5pt,}

\usepackage{tikz}
\usetikzlibrary{matrix,positioning,decorations.pathreplacing}
\usetikzlibrary{shapes.geometric, arrows}
\usetikzlibrary{backgrounds}
\usetikzlibrary{shapes}
\usetikzlibrary{tikzmark}
\usetikzlibrary{calc}
\usetikzlibrary{arrows,shapes,positioning,shadows,trees,mindmap}
\usepackage[edges]{forest}
\usetikzlibrary{arrows.meta}
\colorlet{linecol}{black!75}
\usepackage{xkcdcolors} 

\tikzstyle{startstop} = [rectangle, rounded corners, minimum width=3cm, minimum height=1cm,text centered, draw=black, fill=red!10]
\tikzstyle{io} = [trapezium, trapezium left angle=70, trapezium right angle=110, minimum width=3cm, minimum height=1cm, text centered, text width=5cm, draw=black, fill=blue!10]
\tikzstyle{process} = [rectangle, minimum width=3cm, minimum height=1cm, text centered, draw=black, fill=orange!10]
\tikzstyle{decision} = [diamond, minimum width=2cm, minimum height=2cm, text centered, text width=3cm, draw=black, fill=green!10]
\tikzstyle{arrow} = [thick,->,>=stealth]

\newcommand{\highlight}[2]{\colorbox{#1!17}{$\displaystyle #2$}}

\renewcommand{\highlight}[2]{\colorbox{#1!17}{#2}}

\newcolumntype{L}{>{\arraybackslash}m{3in}}
\newcolumntype{R}{>{\arraybackslash}m{1.2in}}
\newcolumntype{W}{>{\centering\arraybackslash}m{4in}}
\newcolumntype{K}{>{\centering\arraybackslash}m{2.2in}}

%% file: Example.tex
\begin{figure}[htb]
    \vspace{2\baselineskip}
\begin{subequations}
	\begin{align*}
		x_1(t  + \Delta  t) &= \begin{bmatrix}
			\tikzmarknode{Ir}{\highlight{red}{$1$}} & 0 & 0 & 0 & 0 & 0 & 0 \\
			0 & 0 & 0 & \tikzmarknode{Amj}{\highlight{olive}{$a_{24}$}} & 0 & 0 & 0 \\ 
			0 & 0 & \tikzmarknode{Atktk}{\highlight{orange}{$a_{33}$}} & 0 & 0 & 0 & \tikzmarknode{Aptk}{\highlight{orange}{$a_{37}$}} \\ 
			\tikzmarknode{Arm}{\highlight{blue}{$1$}} & 0 & 0 & 0 & 0 & 0 & 0 \\
			0 & \tikzmarknode{Ajp}{\highlight{violet}{$a_{52}$}} & \tikzmarknode{Atkp11}{$0$} & 0 & \tikzmarknode{App11}{\highlight{violet}{$a_{55}$}} & \tikzmarknode{App12}{\highlight{violet}{$a_{56}$}} & \tikzmarknode{App13}{$0$} \\
			0 & \tikzmarknode{Ajp22}{$0$} & \tikzmarknode{Atkp22}{$0$} & 0 & \tikzmarknode{App21}{\highlight{violet}{$a_{65}$}} & \tikzmarknode{App22}{\highlight{violet}{$a_{66}$}} & \tikzmarknode{App23}{\highlight{violet}{$a_{67}$}} \\
			0 & \tikzmarknode{Ajp33}{$0$} & \tikzmarknode{Atkp}{\highlight{violet}{$a_{73}$}} & 0 & \tikzmarknode{App31}{$0$} & \tikzmarknode{App32}{\highlight{violet}{$a_{76}$}} & \tikzmarknode{App33}{\highlight{violet}{$a_{77}$}}
		\end{bmatrix}
		\begin{bmatrix}
			c^\mathrm{R}_1 (t) \\ c^\mathrm{J}_1 (t) \\ c^\mathrm{TK}_1 (t) \\ c^\mathrm{M}_1 (t) \\ c^\mathrm{P}_1 (1,t) \\ c^\mathrm{P}_1 (2,t) \\ c^\mathrm{P}_1 (3,t)
		\end{bmatrix}
		+ \begin{bmatrix}
			0 \\ 0 \\ \tikzmarknode{Btk}{\highlight{orange}{$b_{3}$}} \\ 0 \\ 0 \\ 0 \\ 0
		\end{bmatrix}
		\begin{bmatrix}
			c^{\mathrm{B}_\mathrm{TK}}(t + \Delta t)
		\end{bmatrix}
		 -k_r \begin{bmatrix}
			0 \\ 0 \\ \tikzmarknode{ftk}{\highlight{orange}{$\tilde{c}^\mathrm{TK}_1 (t) c^\mathrm{TK}_1 (t)$}} \\ 0 \\
			\tikzmarknode{fp1}{\highlight{violet}{$\tilde{c}^\mathrm{P}_1 (1,t) c^\mathrm{P}_1 (1,t)$}} \\  \tikzmarknode{fp2}{\highlight{violet}{$\tilde{c}^\mathrm{P}_1 (2,t) c^\mathrm{P}_1 (2,t)$}} \\ \tikzmarknode{fp3}{\highlight{violet}{$\tilde{c}^\mathrm{P}_1 (3,t) c^\mathrm{P}_1 (3,t)$}}
		\end{bmatrix}, \\\\
		x_2(t  + \Delta  t) &= \begin{bmatrix}
			\tikzmarknode{Irt}{\highlight{red}{$1$}} & 0 & 0 & 0 & 0 & 0 & 0 \\
			0 & 0 & 0 & \tikzmarknode{Amjt}{\highlight{olive}{$\tilde{a}_{24}$}} & 0 & 0 & 0 \\ 
			0 & 0 & \tikzmarknode{Atktkt}{\highlight{orange}{$\tilde{a}_{33}$}} & 0 & 0 & 0 & \tikzmarknode{Aptkt}{\highlight{orange}{$\tilde{a}_{37}$}} \\ 
			\tikzmarknode{Armt}{\highlight{blue}{$1$}} & 0 & 0 & 0 & 0 & 0 & 0 \\
			0 & \tikzmarknode{Ajpt}{\highlight{violet}{$\tilde{a}_{52}$}} & 0 & 0 & \tikzmarknode{App11t}{\highlight{violet}{$\tilde{a}_{55}$}} & \tikzmarknode{App12t}{\highlight{violet}{$\tilde{a}_{56}$}} & 0 \\
			0 & 0 & 0 & 0 & \tikzmarknode{App21t}{\highlight{violet}{$\tilde{a}_{65}$}} & \tikzmarknode{App22t}{\highlight{violet}{$\tilde{a}_{66}$}} & \tikzmarknode{App23t}{\highlight{violet}{$\tilde{a}_{67}$}} \\
			0 & 0 & \tikzmarknode{Atkpt}{\highlight{violet}{$\tilde{a}_{73}$}} & 0 & 0 & \tikzmarknode{App32t}{\highlight{violet}{$\tilde{a}_{76}$}} & \tikzmarknode{App33t}{\highlight{violet}{$\tilde{a}_{77}$}}
		\end{bmatrix}
		\begin{bmatrix}
			\tilde{c}^\mathrm{R}_1 (t) \\ \tilde{c}^\mathrm{J}_1 (t) \\ \tilde{c}^\mathrm{TK}_1 (t) \\ \tilde{c}^\mathrm{M}_1 (t) \\ \tilde{c}^\mathrm{P}_1 (1,t) \\ \tilde{c}^\mathrm{P}_1 (2,t) \\ \tilde{c}^\mathrm{P}_1 (3,t)
		\end{bmatrix} -k_r \begin{bmatrix}
			0 \\ 0 \\ 
			\tikzmarknode{ftkt}{\highlight{orange}{$\tilde{c}^\mathrm{TK}_1 (t) c^\mathrm{TK}_1 (t)$}} \\ 0 \\
			\tikzmarknode{fp1t}{\highlight{violet}{$\tilde{c}^\mathrm{P}_1 (1,t) c^\mathrm{P}_1 (1,t)$}} \\  \tikzmarknode{fp2t}{\highlight{violet}{$\tilde{c}^\mathrm{P}_1 (2,t) c^\mathrm{P}_1 (2,t)$}} \\ \tikzmarknode{fp3t}{\highlight{violet}{$\tilde{c}^\mathrm{P}_1 (3,t) c^\mathrm{P}_1 (3,t)$}}
		\end{bmatrix},
	\end{align*}
\end{subequations}

where
\begin{subequations}
	\begin{align*}
		\tikzmarknode{Amj2}{\highlight{olive}{$a_{24}$}}&=\tikzmarknode{Amjt2}{\highlight{olive}{$\tilde{a}_{24}$}} = \frac{q^\mathrm{M}_{1}(t+ \Delta t)}{q^{\mathrm{D}_\mathrm{J}}_1(t+ \Delta t)+q^\mathrm{P}_{1}(t+ \Delta t)}, \\
		\tikzmarknode{Atktk2}{\highlight{orange}{$a_{33}$}}&= \frac{(1-k^\mathrm{TK}_1 \Delta t)V_1^\mathrm{TK}(t)-q^{\mathrm{D}_\mathrm{TK}}_1(t)\Delta t}{V_1^\mathrm{TK}(t + \Delta t)},\\
		\tikzmarknode{Atktkt2}{\highlight{orange}{$\tilde{a}_{33}$}}&= \frac{V_1^\mathrm{TK}(t)-q^{\mathrm{D}_\mathrm{TK}}_1(t)\Delta t}{V_1^\mathrm{TK}(t + \Delta t)},\\
		\tikzmarknode{Aptk2}{\highlight{orange}{$a_{37}$}}&=\tikzmarknode{Aptkt2}{\highlight{orange}{$\tilde{a}_{37}$}} = \frac{q^\mathrm{P}_1(t) \Delta t}{V_1^\mathrm{TK}(t + \Delta t)} \\ 
		\tikzmarknode{Ajp2}{\highlight{violet}{$a_{52}$}}&=\tikzmarknode{Ajpt2}{\highlight{violet}{$\tilde{a}_{52}$}}  = \tikzmarknode{App212}{\highlight{violet}{$a_{65}$}}=\tikzmarknode{App21t2}{\highlight{violet}{$\tilde{a}_{65}$}} = \tikzmarknode{App322}{\highlight{violet}{$a_{76}$}}=\tikzmarknode{App32t2}{\highlight{violet}{$\tilde{a}_{76}$}} = \underline{\lambda}_1(t),\\
		\tikzmarknode{App112}{\highlight{violet}{$a_{55}$}}&= \tikzmarknode{App222}{\highlight{violet}{$a_{66}$}}= \tikzmarknode{App332}{\highlight{violet}{$a_{77}$}} = \lambda_1(t) - k^\mathrm{P}_1 \Delta t, \\
		\tikzmarknode{App11t2}{\highlight{violet}{$\tilde{a}_{55}$}}&= \tikzmarknode{App22t2}{\highlight{violet}{$\tilde{a}_{66}$}}= \tikzmarknode{App33t2}{\highlight{violet}{$\tilde{a}_{77}$}} = \lambda_1(t),\\
		\tikzmarknode{App122}{\highlight{violet}{$a_{56}$}}&=\tikzmarknode{App12t2}{\highlight{violet}{$\tilde{a}_{56}$}}  = \tikzmarknode{App232}{\highlight{violet}{$a_{67}$}}=\tikzmarknode{App23t2}{\highlight{violet}{$\tilde{a}_{67}$}} = \tikzmarknode{Atkp2}{\highlight{violet}{$a_{73}$}}=\tikzmarknode{Atkpt2}{\highlight{violet}{$\tilde{a}_{73}$}} = \overline{\lambda}_1(t),\\
		\tikzmarknode{Btk2}{\highlight{orange}{$b_{3}$}} & = \frac{V^\mathrm{B}(t + \Delta t)}{V_1^\mathrm{TK}(t + \Delta t)}.
	\end{align*}
\end{subequations}

    \begin{tikzpicture}[overlay,remember picture,>=stealth,nodes={align=left,inner ysep=1pt},<-]
	\path (Ir.north) ++ (0,1em) node[anchor=south east,color=red!67] (scalep){\textbf{$I_{n_\mathrm{R}}$}};
	\draw [color=red!87](Ir.north) |- ([xshift=-0.3ex,color=red]scalep.south west);
	
	\path (Amj.north) ++ (-3.3,0.7em) node[anchor=south east,color=olive!100] (scalep){\textbf{$A^\mathrm{M}_\mathrm{J}$}};
	\draw [color=olive!100](Amj.north) |- ([xshift=-0.3ex,color=olive]scalep.south west);
	
	\path (Atktk.north) ++ (-3,0.45em) node[anchor=south east,color=orange!100] (scalep){\textbf{$A^\mathrm{TK}_\mathrm{TK}$}};
	\draw [color=orange!100](Atktk.north) |- ([xshift=-0.3ex,color=orange]scalep.south west);
	
	\path (Aptk.north) ++ (-6.3,0.7em) node[anchor=south east,color=orange!100] (scalep){\textbf{$A^\mathrm{P}_\mathrm{TK}$}};
	\draw [color=orange!100](Aptk.north) |- ([xshift=-0.3ex,color=orange]scalep.south west);
	
	\path (Btk.north) ++ (1.5,0.3em) node[anchor=north east,color=orange](scalep){\textbf{$B_\mathrm{TK}$}};
	\draw [color=orange!100](Btk.east) -- ([xshift=0.3ex, xshift=0.3ex,color=orange!100]scalep.south east);
	
	\path (ftk.north) ++ (-1.2,0.5em) node[anchor=south east,color=orange!100] (scalep){\textbf{$f(c^\mathrm{TK}(t),\tilde{c}^\mathrm{TK}(t))$}};
	\draw [color=orange!100](ftk.north) |- ([xshift=-0.3ex,color=orange]scalep.south west);
	
	\path (Arm.north) ++ (-0.5,0.7em) node[anchor=south east,color=blue!67] (scalep){\textbf{$A^\mathrm{R}_\mathrm{M}$}};
	\draw [color=blue!87](Arm.north) |- ([xshift=-0.3ex,color=blue]scalep.south west);
	
	\draw [-, color=violet] (App11.north west) --  ([xshift=1.6ex, color=violet] App13.north east);
	
	\draw [-, color=violet] ([xshift=-1.5ex, yshift=-3.5, color=violet] App31.south west) --  (App33.south east);
	
	\draw [-, color=violet] ([color=violet] App11.north west) --  ([xshift=-1.5ex, yshift=-3.5, color=violet] App31.south west);
	
	\draw [-, color=violet] ([xshift=7, color=violet] App13.north east) --  ( App33.south east);
	
	\path (Ajp.north) ++ (-1.2,0.3em) node[anchor=north east,color=violet] (scalep){\textbf{$A^\mathrm{J}_\mathrm{P}$}};
	\draw [color=violet](Ajp.north) |- ([xshift=-0.3ex,color=violet]scalep.north west);
	
	\path (Atkp11.north) ++ (-3,0.7em) node[anchor=north east,color=violet] (scalep){\textbf{$A^\mathrm{TK}_\mathrm{P}$}};
	\draw [color=violet](Atkp11.north) |- ([xshift=-0.3ex,color=violet]scalep.north west);
	
	\path (App32.south) ++ (-6,1.1em) node[anchor=north east,color=violet] (scalep){\textbf{$A^\mathrm{P}_\mathrm{P}$}};
	\draw [color=violet](App32.south) |- ([xshift=-0.3ex,color=violet]scalep.south west);
	
	\path (fp3.south) ++ (-3.7,0.9em) node[anchor=north west,color=violet] (scalep){\textbf{$f(c^\mathrm{P}(t),\tilde{c}^\mathrm{P}(t))$}};
	\draw [color=violet](fp3.south) |- ([xshift=0.3ex,color=violet]scalep.south west);
	
	\draw [-, color=violet] (fp1.north east) -- (fp1.north west);

	\draw [-, color=violet] (fp3.south east) -- (fp3.south west);	
	
	\draw [-, color=violet] (fp1.north west) -- (fp3.south west);	
	
	\draw [-, color=violet] (fp1.north east) -- (fp3.south east);	
	
	\draw [-, color=violet] (Ajp.north west) -- (Ajp.north east);	
	
	\draw [-, color=violet] ([xshift=-1.5ex, yshift=-1ex] Ajp33.south west) -- ([xshift=1.5ex, yshift=-1ex] Ajp33.south east);	
	
	\draw [-, color=violet] ([color=violet] Ajp.north west) --  ([xshift=-1.5ex, yshift=-1ex, color=violet] Ajp33.south west);

	\draw [-, color=violet] ([color=violet] Ajp.north east) --  ([xshift=1.5ex, yshift=-1ex, color=violet] Ajp33.south east);
	
	\draw [-, color=violet] (Atkp.south west) -- (Atkp.south east);	
	
	\draw [-, color=violet] ([xshift=-1.5ex] Atkp11.north west) -- ([xshift=1.5ex] Atkp11.north east);	
	
	\draw [-, color=violet] ([color=violet] Atkp.south west) --  ([xshift=-1.5ex, color=violet] Atkp11.north west);
	
	\draw [-, color=violet] ([color=violet] Atkp.south east) --  ([xshift=1.5ex, color=violet] Atkp11.north east);

    \end{tikzpicture}
    \vspace{\baselineskip}
\end{figure}

%% file: TechniquesExample.tex

\begin{table}[h]
	\centering
	\caption{General form for the state-space matrices for the three-node network example}~\label{tab:SSGnrlMt}
	{\small\begin{tabular}{c|c|c c}
		\hline
		$E(t)$ &  $A(t)$ & $f(x_1,x_2,t)$ & \\
		\hline 
		& & & \\
		$\begin{bmatrix} \tikzmarknode{Ir}{\highlight{red}{$1$}} & 0 & 0 & 0 & 0 & 0 & 0 \\
			0 & \tikzmarknode{Ij}{\highlight{olive}{$1$}} & 0 & 0 & 0 & 0 & 0 \\ 
			0 & 0 & \tikzmarknode{Itk}{\highlight{orange}{$1$}} & 0 & 0 & 0 & 0 \\ 
			0 & 0 & 0 & \tikzmarknode{Im}{\highlight{blue}{$1$}} & 0 & 0 & 0 \\
			0 & \tikzmarknode{Ejp}{\highlight{violet}{$e_{52}$}} & \tikzmarknode{Etkp11}{$0$} & 0 & \tikzmarknode{Epp11}{\highlight{violet}{$e_{55}$}} & \tikzmarknode{Epp12}{\highlight{violet}{$e_{56}$}} & \tikzmarknode{Epp13}{$0$} \\
			0 & \tikzmarknode{AEp22}{$0$} & \tikzmarknode{Atkp22}{$0$} & 0 & \tikzmarknode{Epp21}{\highlight{violet}{$e_{65}$}} & \tikzmarknode{Epp22}{\highlight{violet}{$e_{66}$}} & \tikzmarknode{Epp23}{\highlight{violet}{$e_{67}$}} \\
			0 & \tikzmarknode{Ejp33}{$0$} & \tikzmarknode{Etkp}{\highlight{violet}{$e_{73}$}} & 0 & \tikzmarknode{Epp31}{$0$} & \tikzmarknode{Epp32}{\highlight{violet}{$e_{76}$}} & \tikzmarknode{Epp33}{\highlight{violet}{$e_{77}$}}
				\end{bmatrix}$ & $\begin{bmatrix} \tikzmarknode{Ir}{\highlight{red}{$1$}} & 0 & 0 & 0 & 0 & 0 & 0 \\
		0 & 0 & 0 & \tikzmarknode{Amj}{\highlight{olive}{$a_{24}$}} & 0 & 0 & 0 \\ 
		0 & 0 & \tikzmarknode{Atktk}{\highlight{orange}{$a_{33}$}} & 0 & 0 & 0 & \tikzmarknode{Aptk}{\highlight{orange}{$a_{37}$}} \\ 
		\tikzmarknode{Arm}{\highlight{blue}{$1$}} & 0 & 0 & 0 & 0 & 0 & 0 \\
		0 & \tikzmarknode{Ajp}{\highlight{violet}{$a_{52}$}} & \tikzmarknode{Atkp11}{$0$} & 0 & \tikzmarknode{App11}{\highlight{violet}{$a_{55}$}} & \tikzmarknode{App12}{\highlight{violet}{$a_{56}$}} & \tikzmarknode{App13}{$0$} \\
		0 & \tikzmarknode{Ajp22}{$0$} & \tikzmarknode{Atkp22}{$0$} & 0 & \tikzmarknode{App21}{\highlight{violet}{$a_{65}$}} & \tikzmarknode{App22}{\highlight{violet}{$a_{66}$}} & \tikzmarknode{App23}{\highlight{violet}{$a_{67}$}} \\
		0 & \tikzmarknode{Ajp33}{$0$} & \tikzmarknode{Atkp}{\highlight{violet}{$a_{73}$}} & 0 & \tikzmarknode{App31}{$0$} & \tikzmarknode{App32}{\highlight{violet}{$a_{76}$}} & \tikzmarknode{App33}{\highlight{violet}{$a_{77}$}}
		\end{bmatrix} $	 & 
		$\begin{bmatrix}
			0 \\ 0 \\ 
			\tikzmarknode{ftkt}{\highlight{orange}{$f^\mathrm{TK}_1$}} \\ 0 \\
			\tikzmarknode{fp1t}{\highlight{violet}{$f^\mathrm{P}_1(1)$}} \\  \tikzmarknode{fp2t}{\highlight{violet}{$f^\mathrm{P}_1(2)$}} \\ \tikzmarknode{fp3t}{\highlight{violet}{$f^\mathrm{P}_1(3)$}}
				\end{bmatrix}$ &
			$\begin{matrix}
				\begin{array}{c}
					$\textcolor{lightgray}{R1}$\\				$\textcolor{lightgray}{J1}$\\
					$\textcolor{lightgray}{TK1}$\\
					$\textcolor{lightgray}{M1}$\\
					\\
					$\textcolor{lightgray}{P1}$
					\\ \mathbb{ }
				\end{array}
			\end{matrix} $\\
			& & & \\
		\hline
		\hline
	\end{tabular}}
\end{table}

 \begin{tikzpicture}[overlay,remember picture,>=stealth,nodes={align=left,inner ysep=1pt},<-]
	
	\draw [-, color=violet] (Epp11.north west) --  ([xshift=1.6ex, color=violet] Epp13.north east);
	
	\draw [-, color=violet] ([xshift=-1.5ex, yshift=-3.5, color=violet] Epp31.south west) --  (Epp33.south east);
	
	\draw [-, color=violet] ([color=violet] Epp11.north west) --  ([xshift=-1.5ex, yshift=-3.5, color=violet] Epp31.south west);
	
	\draw [-, color=violet] ([xshift=7, color=violet] Epp13.north east) --  ( Epp33.south east);
	
	\path (Ejp.north) ++ (-1.2,0.3em) node[anchor=north east,color=violet] (scalep){\textbf{$E^\mathrm{J}_\mathrm{P}$}};
	\draw [color=violet](Ejp.north) |- ([xshift=-0.3ex,color=violet]scalep.north west);
	
	\path (Etkp11.north) ++ (-3,0.7em) node[anchor=north east,color=violet] (scalep){\textbf{$E^\mathrm{TK}_\mathrm{P}$}};
	\draw [color=violet](Etkp11.north) |- ([xshift=-0.3ex,color=violet]scalep.north west);
	
	\path (Epp32.south) ++ (-5.3,0.5em) node[anchor=north east,color=violet] (scalep){\textbf{$E^\mathrm{P}_\mathrm{P}$}};
	\draw [color=violet](Epp32.south) |- ([xshift=-0.3ex,color=violet]scalep.south west);
	
	\draw [-, color=violet] (Ejp.north west) -- (Ejp.north east);	
	
	\draw [-, color=violet] ([xshift=-1.5ex, yshift=-1ex] Ejp33.south west) -- ([xshift=1.5ex, yshift=-1ex] Ejp33.south east);	
	
	\draw [-, color=violet] ([color=violet] Ejp.north west) --  ([xshift=-1.5ex, yshift=-1ex, color=violet] Ejp33.south west);
	
	\draw [-, color=violet] ([color=violet] Ejp.north east) --  ([xshift=1.5ex, yshift=-1ex, color=violet] Ejp33.south east);
	
	\draw [-, color=violet] (Etkp.south west) -- (Etkp.south east);	
	
	\draw [-, color=violet] ([xshift=-1.5ex] Etkp11.north west) -- ([xshift=1.5ex] Etkp11.north east);	
	
	\draw [-, color=violet] ([color=violet] Etkp.south west) --  ([xshift=-1.5ex, color=violet] Etkp11.north west);
	
	\draw [-, color=violet] ([color=violet] Etkp.south east) --  ([xshift=1.5ex, color=violet] Etkp11.north east);
	
\end{tikzpicture}

\begin{table}[h!]
	\centering
	\caption{Different techniques implementation on the three-node network example}~\label{tab:TechsEx}
	{\small\begin{tabular}{c|c|c|c}
		Element & Tech. \#2 - Backward Euler Scheme & Tech. \#3 - Crank-Nicolson Scheme & Tech. \#4 - Method of Characteristics \\
		\hline
		\tikzmarknode{EjpT}{\highlight{violet}{$e_{52}$}} = \tikzmarknode{Epp21T}{\highlight{violet}{$e_{65}$}} = \tikzmarknode{Epp32T}{\highlight{violet}{$e_{76}$}} & $-0.5 \tilde{\lambda_1}$& $-0.25 \tilde{\lambda_1}$ & 0 \\
		\hline
		\tikzmarknode{Epp11T}{\highlight{violet}{$e_{55}$}} = \tikzmarknode{Epp22T}{\highlight{violet}{$e_{66}$}} = \tikzmarknode{Epp33T}{\highlight{violet}{$e_{77}$}} & 1 & 1 & 1 \\
		\hline
		\tikzmarknode{Epp12T}{\highlight{violet}{$e_{56}$}} = \tikzmarknode{Epp23T}{\highlight{violet}{$e_{67}$}} = \tikzmarknode{EtkpT}{\highlight{violet}{$e_{73}$}} & $0.5 \tilde{\lambda_1}$& $0.25 \tilde{\lambda_1}$ & 0 \\
		\hline
		\tikzmarknode{AjpT}{\highlight{violet}{$a_{52}$}} & 0 & $0.25 \tilde{\lambda_1}$ & $1$ \\
		\hline
		 \tikzmarknode{App21T}{\highlight{violet}{$a_{65}$}}& 0 & $0.25 \tilde{\lambda_1}$ & $\tilde{\lambda_i}(t) \exp(-(k_i^\mathrm{P}+k_r \tilde{c}^\mathrm{P}_1 (1,t)) \Delta t)$ \\
		\hline
		 \tikzmarknode{App32T}{\highlight{violet}{$a_{76}$}} & 0 & $0.25 \tilde{\lambda_1}$ & $\tilde{\lambda_i}(t) \exp(-(k_i^\mathrm{P}+k_r \tilde{c}^\mathrm{P}_1 (2,t)) \Delta t)$ \\
		\hline
		\tikzmarknode{App11T}{\highlight{violet}{$a_{55}$}}  & $1-k^\mathrm{P}_1$& $1-k^\mathrm{P}_1$ & $0$ \\
		\hline
		\tikzmarknode{App22T}{\highlight{violet}{$a_{66}$}} & $1-k^\mathrm{P}_1$& $1-k^\mathrm{P}_1$ & $(1-\tilde{\lambda_i}(t)) \exp(-(k_i^\mathrm{P}+k_r \tilde{c}^\mathrm{P}_1 (2,t)) \Delta t)$ \\
		\hline
		 \tikzmarknode{App33T}{\highlight{violet}{$a_{77}$}} & $1-k^\mathrm{P}_1$& $1-k^\mathrm{P}_1$ & $(1-\tilde{\lambda_i}(t)) \exp(-(k_i^\mathrm{P}+k_r \tilde{c}^\mathrm{P}_1 (3,t)) \Delta t)$ \\
		\hline
		\tikzmarknode{App12T}{\highlight{violet}{$a_{56}$}} = \tikzmarknode{App23T}{\highlight{violet}{$a_{67}$}} = \tikzmarknode{AtkpT}{\highlight{violet}{$a_{73}$}} & 0 & $-0.25 \tilde{\lambda_1}$ & 0 \\
		\hline	
		\tikzmarknode{fp1T}{\highlight{violet}{$f^\mathrm{P}_1(1)$}} & $\tilde{c}^\mathrm{P}_1 (1,t) c^\mathrm{P}_1 (1,t)$ & $\tilde{c}^\mathrm{P}_1 (1,t) c^\mathrm{P}_1 (1,t)$  & 0 \\
		\hline
		\tikzmarknode{fp2T}{\highlight{violet}{$f^\mathrm{P}_1(2)$}} & $\tilde{c}^\mathrm{P}_1 (2,t) c^\mathrm{P}_1 (2,t)$& $\tilde{c}^\mathrm{P}_1 (2,t) c^\mathrm{P}_1 (2,t)$ & 0 \\
		\hline
		\tikzmarknode{fp3T}{\highlight{violet}{$f^\mathrm{P}_1(3)$}} & $\tilde{c}^\mathrm{P}_1 (3,t) c^\mathrm{P}_1 (3,t)$&
		$\tilde{c}^\mathrm{P}_1 (3,t) c^\mathrm{P}_1 (3,t)$ & 0 \\
		\hline
			\hline
	\end{tabular}}
\end{table}